\documentclass[preprint,authoryear,12pt]{elsarticle}

\usepackage{amsmath,amsfonts,amssymb,amscd,amsthm,amsbsy}
\usepackage{bm}
\usepackage{mathabx}
\usepackage{mathrsfs}
\usepackage{comment}

\usepackage{graphicx}
\graphicspath{{graphics/}}
\usepackage[font= footnotesize]{caption}
\usepackage[font=footnotesize]{subcaption}

\usepackage[table]{xcolor}
\usepackage{booktabs}
\usepackage{tabu}
\usepackage{threeparttable}
\usepackage{multirow}

\usepackage{algorithm}
\usepackage{algpseudocode}
\usepackage{tikz}
\usetikzlibrary{shapes.geometric, arrows.meta, positioning, shapes, arrows, calc}

\usepackage{xcolor}
\usepackage[normalem]{ulem} 

\theoremstyle{remark}
\newtheorem*{remark*}{Remark} 

\usepackage{fullpage}

\usepackage[%
breaklinks=true,%
colorlinks=true,%
pdfauthor={First Author et al.},%
pdftitle={Template for manuscripts in Advances in Space Research}%
]{hyperref}

\journal{Advances in Space Research}

\begin{document}

\begin{frontmatter}

\title{Cislunar Resonant Transport and Heteroclinic Pathways: \\ From 3:1 to 2:1 to L1}

\author{Bhanu Kumar\corref{contrib}}
\cortext[contrib]{Authors contributed equally}
\ead{bhkumar@umich.edu}
\address{Department of Mathematics, University of Michigan, 530 Church St, Ann Arbor, MI, USA}
\address{Jet Propulsion Laboratory, California Institute of Technology, 4800 Oak Grove Dr, Pasadena, CA, USA}

\author{Anjali Rawat\corref{contrib}}
\ead{anjalirawat@vt.edu}
\address{Department of Aerospace and Ocean Engineering, Virginia Tech, 495 Old Turner St, Blacksburg, VA, USA}

\author{Aaron J. Rosengren}
\ead{arosengren@ucsd.edu}
\address{Department of Mechanical and Aerospace Engineering, UC San Diego, 9500 Gilman Dr, La Jolla, CA, USA}

\author{Shane D. Ross}
\ead{sdross@vt.edu}
\address{Department of Aerospace and Ocean Engineering, Virginia Tech, 495 Old Turner St, Blacksburg, VA, USA}

\begin{abstract}
Understanding the dynamical structure of cislunar space beyond geosynchronous orbit is critical for both lunar exploration and for high-Earth-orbiting trajectories. In this study, we investigate the role of mean-motion resonances and their associated heteroclinic connections in enabling natural semi-major axis transport in the Earth–Moon system. Working within the planar circular restricted three-body problem, we compute and analyze families of periodic orbits associated with the interior 4:1, 3:1, and 2:1 lunar resonances. These families exhibit a rich bifurcation structure, including transitions between prograde and retrograde branches and connections through collision orbits. We construct stable and unstable manifolds of the unstable resonant orbits using a perigee-based Poincar\'e map, and identify heteroclinic connections---both between resonant orbits and with lunar $L_1$ libration-point orbits---across a range of Jacobi constant values. Using a new generalized distance metric to quantify the closeness between trajectories, we establish operational times-of-flight for such heteroclinic-type orbit-to-orbit transfers. These connections reveal ballistic, zero-$\Delta v$ pathways that achieve major orbit changes within reasonable times-of-flight, thus defining a network of accessible semi-major axes. Our results provide a new dynamical framework for long-term spacecraft evolution and cislunar mission design, particularly in regimes where lunar gravity strongly perturbs distant circumterrestrial orbits.
\end{abstract}


\end{frontmatter}

\parindent=0.5 cm
\section{Introduction}

In recent years, cislunar space has emerged as a focal point for both ongoing and planned  space activities. 
Human exploration programs such as NASA's Artemis and Lunar Gateway, as well as robotic lunar exploration missions such as NASA/JPL's CADRE, the Commercial Lunar Payload Services (CLPS) initiatives, China's {\it Chang'e} series, India's {\it Chandrayaan} missions, and JAXA's SLIM lander, underscore the growing operational presence in the Earth--Moon environment. 
Beyond explicitly Moon-focused endeavors, cislunar space has also drawn increasing attention from broader strategic and  scientific perspectives. 
The 2022 U.S. {\it National Cislunar Science and Technology Strategy} explicitly identified space situational awareness in the cislunar domain as a national development priority \citep{cislunarWhiteHouse}. 
Recent policy-driven and application-oriented studies, such as those by \citet{bB24} and \citet{aW25}, further emphasize the importance of understanding the complex structure of the cislunar phase space for purposes including navigation, station-keeping, and governance, building on prior dynamical systems studies while expanding their strategic context. 
Moreover, even spacecraft not explicitly targeting the Moon may traverse regions where lunar gravity effects induces significant  orbital perturbations \citep{bSaR25}. 
This must therefore be taken into account during mission design. For instance, during its initial two-year nominal mission, NASA's Interstellar Boundary Explorer (IBEX) spacecraft was placed into an orbit whose evolution was found to be chaotic and unpredictable beyond 2.5 years due to lunar perturbations. For its extended mission, IBEX was transferred into a stable 3:1 mean-motion resonant orbit with the Moon to mitigate these effects \citep{ibex}. 

The above considerations highlight the pivotal role of the dynamical structure of cislunar space beyond geosynchronous orbit in shaping both lunar exploration and high-Earth-orbit mission design. A key aspect of these dynamics is the presence of mean-motion resonances (MMRs) with the Moon. Spacecraft such as the aforementioned IBEX and NASA's Transiting Exoplanet Survey Satellite (TESS) have operated in stable lunar resonant orbits \citep{tess}, while others, such as Russia's Spektr-R and China's Tiandu-1, appear to inhabit more unstable resonant regime. 
Although libration-point dynamics in the Earth--Moon system have been studied extensively (e.g., \citet{simoetal}), the detailed architecture of MMRs---especially their stable and unstable orbit families and overlapping heteroclinic connections---has received comparatively less attention. 
These heteroclinic connections can enable ballistic transitions between resonances, allowing a spacecraft’s semi-major axis to evolve naturally under lunar perturbations, without the need for propulsion.
While such transport mechanisms offer promising avenues for low-energy trajectory design, they can also pose hazards if not accounted for in mission planning.

While heteroclinic connections between resonances play a central role in enabling natural changes in a spacecraft's semi-major axis \citep{KoLoMaRo2000}, and have been investigated for trajectory design in the Jovian \citep{Anderson2010, Anderson2011, Anderson2021b} and Saturnian \citep{vaqueroThesis} systems, they remain markedly underexplored in the cislunar context. 
Recent studies by \citet{mVkH14a, mVkH14b}, \citet{yL20}, and \citet{cP24} have demonstrated the utility of resonant orbit families and their associated invariant manifolds for constructing long-duration transfers and infrastructural frameworks in the Earth--Moon system.
However, none of these studies fully investigate resonance-driven dynamics, nor do they find true, zero-$\Delta v$, heteroclinic connections in the dynamically sensitive region between Earth and Moon. 
The studies of \citet{mVkH14a, mVkH14b} focus primarily on  non-ballistic transfers tied to a single  lunar resonance or orbits exterior to the Moon's orbit, while \citet{yL20} and \citet{cP24} omit invariant manifolds and heteroclinics altogether. 
Nevertheless, their results hint at the potential richness of the cislunar phase space,  a richness that could be more fully revealed through modern computational approaches to manifold-based transport, offering deeper insight into the resonant and libration-driven pathways that shape access, stability, and maneuverability in this regime. 


This study revisits cislunar resonance structures through the lens of modern manifold-based transport theory.
Our first main contribution is the computation and analysis of key families of resonant periodic orbits in the Earth--Moon system, formulated  within the framework of the planar circular restricted three-body problem (PCR3BP). 
We focus on the interior 4:1, 3:1, and 2:1 mean-motion resonances, identifying bifurcations, symmetry transitions, and connectivity between stable and unstable prograde and retrograde branches. 
Our second contribution is the construction of the associated stable and unstable invariant manifolds, using a parameterization method \citep{haroetal,kumar2025tools} and a Poincar\'e map defined at osculating orbit perigee. 
This enables the identification of natural heteroclinic connections not only between  resonant orbits, but also with lunar $L_1$ libration-point orbits. A new distance metric we develop then allows us to precisely define and determine such heteroclinic transfers' times-of-flight.
By performing these analyses across a range of Jacobi constant values, we characterize the range of dynamically accessible semi-major axes for future ballistic transfers in cislunar space, offering both theoretical insights and practical implications for trajectory design. 

The remainder of this paper is structured as follows.
Section~\ref{backgroundsection} reviews the PCR3BP and provides background
on mean-motion resonances and resonance overlap.
Section~\ref{orbitComputeSection} details our computational and visualization methods. 
Section~\ref{poSection} presents the families of interior resonant periodic orbits, and 
Section~\ref{Transport} analyzes their manifolds and heteroclinic connections. Finally, Section~\ref{resToL1Section} demonstrates heteroclinic transfers between mean motion resonances and the vicinity of lunar $L_1$. 
We conclude by discussing the implications of our findings for 
zero-$\Delta v$ trajectory planning in the Earth--Moon system. 

\section{Background} \label{backgroundsection}

\subsection{Planar Circular Restricted Three-Body Problem} \label{modelSection}

\begin{figure}
	\centering
    \begin{tabular}{cc}
	\includegraphics[width=0.4\linewidth]{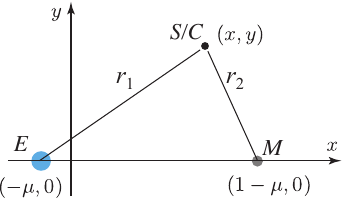} &
    \includegraphics[width=0.4\linewidth]{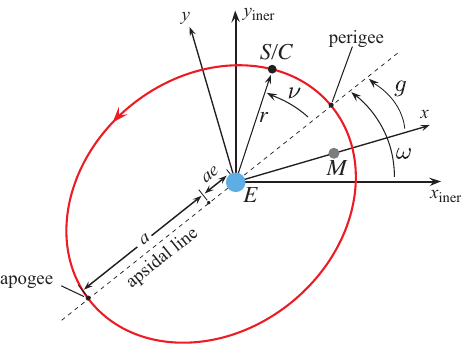}\\ 
        (a)  & (b) 
     \end{tabular}

    \caption{
     (a) PCR3BP barycentered rotating $(x,y)$ frame in normalized units.
     (b) The geocentric osculating orbital elements showing inertial longitude of perigee ($\omega$) and synodic longitude of perigee ($g$).
    }
	\label{frame}
\end{figure}

For the sake of completeness, we include the following description of the well-known PCR3BP, reproduced from \citet{kumar2022}, and use it to establish the nomenclature adopted in this work. The PCR3BP describes the motion of an infinitesimally small particle (thought of as a spacecraft) under the gravitational influence of two large bodies of masses $m_{1}$ and $m_{2}$, which revolve about their barycenter in a circular Keplerian orbit (e.g., a planet and a moon). Units are also normalized so that the distance between $m_{1}$ and $m_{2}$ becomes 1, their period of revolution becomes $2 \pi$, and $\mathcal{G}(m_{1}+m_{2})$ becomes 1. We define a mass ratio $\mu = \frac{m_{2}}{m_{1} + m_{2}}$, and use a synodic, rotating, non-inertial Cartesian coordinate system centered at the $m_{1}$--$m_{2}$ barycenter such that $m_{1}$ and $m_{2}$ are always on the $x$-axis, as shown in Figure \ref{frame}(a). The PCR3BP also assumes that the spacecraft moves in the same plane as $m_{1}$ and $m_{2}$. In this case, the equations of motion are Hamiltonian with form \citep{celletti},
\begin{equation} 
\label{pcrtbpH_EOM} 
\dot x = \frac{\partial H}{\partial p_{x}}, \quad \dot y = \frac{\partial H}{\partial p_{y}}, \quad \dot p_{x} = -\frac{\partial H}{\partial x}, \quad \dot p_{y} = -\frac{\partial H}{\partial y}, 
\end{equation}
\begin{equation}  
\label{pcrtbpH} 
H(x,y,p_x,p_{y}) = \frac{p_{x}^{2}+p_{y}^{2}}{2} + p_{x}y -p_{y}x - \frac{1-\mu}{r_{1}} - \frac{\mu}{r_{2}},
\end{equation}
where $r_{1} = \sqrt{(x+\mu)^{2} + y^{2}}$ and $r_{2} = \sqrt{(x-1+\mu)^{2} + y^{2}} $ are the distances from the spacecraft to $m_{1}$ and $m_{2}$, respectively. For $m_{1}$ and $m_{2}$ as the Earth and Moon, respectively, we use $\mu = 1.2150584270572 \times 10^{-2}$. The momenta $p_{x}$ and $p_{y}$ are the spacecraft velocity components in an inertial reference frame, and are related to the rotating (non-inertial) frame velocities $\dot x$ and $ \dot y$ as $\dot x = p_{x} + y$, $\dot y = p_{y} -x$. 

There are two important properties of Eqs.\ \eqref{pcrtbpH_EOM}--\eqref{pcrtbpH} to note. First of all, the Hamiltonian in Eq.\ \eqref{pcrtbpH} is autonomous and is thus an integral of motion. Hence, PCR3BP trajectories are restricted to 3D energy submanifolds of the state space satisfying $H(x,y,p_{x}, p_{y})=$ constant. The quantity $C = -2H$ is referred to as the \emph{Jacobi constant}, and is generally used in lieu of $H$ to specify energy levels; note that lower $C$ corresponds to higher energy $H$. The second property is that the equations of motion have a time-reversal symmetry: if $(x(t), y(t), t)$ is a solution of Eqs.\ \eqref{pcrtbpH_EOM}--\eqref{pcrtbpH} for $t > 0$, then $(x(-t), -y(-t), t)$ is a solution for $t < 0$. 

\subsubsection{Osculating Orbital Elements and Synodic Delaunay Coordinates} \label{delaunaySection}

While Eqs.\ \eqref{pcrtbpH_EOM}--\eqref{pcrtbpH} for the PCR3BP are written in Cartesian coordinates, other coordinate systems can also be used to express the Hamiltonian and equations of motion. In certain cases, one of the most beneficial sets of coordinates to use are the \emph{synodic Delaunay coordinates} $(L, G, \ell, g)$. These coordinates can be most easily defined in terms of the classical (osculating) orbital elements via the equations $L = \sqrt{(1-\mu)a}$, $G = L\sqrt{1-e^{2}}$, and the definitions $\ell =$ mean anomaly and $g=$ longitude of perigee \emph{with respect to the rotating frame positive $x$-axis}; as usual, $a$ and $e$ here denote osculating semi-major axis and eccentricity, respectively. Note that unlike the traditional inertial longitude of perigee $\omega$, in this case $\dot g \approx -1 \neq 0$ as the rotating frame $x$-axis is constantly moving with respect to an inertial frame; the definition of $g$ and difference with $\omega$ are illustrated in Figure \ref{frame}(b). 

The main advantage of using synodic Delaunay coordinates stems from the fact that they are action-angle coordinates \citep{celletti} for the $\mu=0$ PCR3BP; i.e., the Kepler problem in a rotating reference frame. This means that (1) when $\mu=0$, the Hamiltonian $H$ can be written solely as a function of the actions $L$ and $G$; and (2) the transformation between Cartesian and synodic Delaunay variables is canonical; i.e., the equations of motion in synodic Delaunay coordinates remain Hamiltonian. Note that the traditional osculating orbital elements satisfy above fact (1) but not (2). Combined, these two facts imply that when $\mu=0$, $\dot L = \dot G = 0$ and both $\dot \ell$ and $\dot g$ are constants. Furthermore, when $\mu>0$, the rich existing literature on Hamiltonian perturbation theory of near-integrable systems \citep[e.g.,][]{morbyBook} becomes applicable to the system written in synodic Delaunay variables.

While synodic Delaunay coordinates have advantages as just described, the geocentric osculating orbital elements $a$ and $e$ are arguably more physically intuitive than $L$ and $G$. Osculating elements also provide a natural connection to the literature on resonances \citep[e.g.,][]{malhotra2020divergence}, thereby clarifying the positioning and spacing of resonances and rendering the overall resonant structure more transparent. However, note that $L$ and $a$ have a simple relationship $L = \sqrt{(1-\mu)a}$, with each increasing in tandem with the other. Thus, the ability to qualitatively analyze results through the lens of Hamiltonian perturbation theory partially remains even if $a$ is used instead of $L$. This is the approach we will adopt; although we use Cartesian coordinates for most computations in this study, we will use osculating orbital elements $(g, a)$ to aid in the visualization and interpretation of results in line with perturbation theory. From a mission design perspective, we hope that this representation will also improve interpretability and accessibility for practicing astrodynamicists.

\subsection{Mean-Motion Resonances and Resonance Overlap} \label{mmrSection}

Resonant motions are ubiquitous in celestial systems.
Among the most important resonant phenomena, especially for astrodynamics, is that of \emph{mean-motion resonance} (MMR). Roughly speaking\footnote{More rigorously, MMRs are defined through studying the dynamics of the resonant angle $m g + n \ell$, $m, n \in \mathbb{Z}$; if this angle librates along a trajectory rather than circulating, the trajectory is contained within the $m$:$n$ MMR. See for instance the book of \citet{morbyBook} for more details.}, an $m$:$n$ MMR is a region of a celestial system's phase space where one body (in our case the spacecraft) makes approximately $m$ integer number of revolutions around some large central body (e.g., Earth) in the same time that another body (here, the Moon) makes $n$ integer revolutions around the same central body. For spacecraft--Moon MMRs, since this is a relation between orbital periods of two bodies with the second body's (Moon's) period being fixed and known, the different MMRs correspond to certain spacecraft semi-major axis values---one for each MMR. 

In multi-body celestial systems, MMR regions contain both stable and unstable resonant orbits with various topological properties. Of these, the unstable resonant orbits are of special interest for mission design and low-fuel-cost orbit transfers, as they possess attached \emph{stable and unstable invariant manifolds}. If the unstable manifold of an orbit contained in one MMR intersects the stable manifold of an orbit contained in another MMR, then one gets a zero-$\Delta v$ \emph{heteroclinic} trajectory from the first MMR to the second. This dynamical phenomenon, known as \emph{MMR overlap}, in turn yields a natural change of spacecraft semi-major axis. As described in \citet{morbyBook}, heteroclinic intersections provide a rigorous mathematical basis for the resonance overlap criterion\footnote{Note that semi-analytical approaches to Chirikov's overlap criterion, as commonly employed by dynamical astronomers, tend to underestimate the onset threshold for chaotic transport \citep{morbyBook} as compared to manifold-based analyses, in the sense of requiring stronger perturbations than necessary. This is in part due to the neglect of higher-order resonances in perturbation series approaches---an approximation which is not made when computing invariant manifolds using the original equations of motion.} of \citet{chirikov1960,chirikov1979}, which states that overlap of MMRs is the key driver of global transport across phase space in celestial systems, due to its destruction of topological dynamical barriers whose existence would otherwise prevent major changes in semi-major axis. MMR overlap thus determines the semi-major axis values a spacecraft can reach without using fuel. 

In the PCR3BP, at each mean-motion resonance, families of stable and unstable resonant orbits exist over a range of energy levels. The unstable resonant orbits correspond to 1-parameter families of unstable periodic orbits, where the parameter along the family of orbits can be taken as the Jacobi constant \citep{kumar2021journal}. Thus, except for at fold bifurcation points (where $C$ reaches an extremum along the family), locally there will be one unstable resonant periodic orbit for each Jacobi constant $C$ across some range of $C$ values. For the $m$:$n$ MMR, its resonant periodic orbits will have periods of \emph{approximately} but not exactly $2 \pi n$; the range of periods in fact will vary throughout the family as a function of Jacobi constant. Unstable orbits will encounter the Moon at apogee once every $m$ revolutions.

\section{Computational and Visualization Methodologies} 
\label{orbitComputeSection}

In the PCR3BP, to understand the structure of resonant orbit families and the heteroclinic dynamics induced by the unstable ones, one needs to compute the corresponding periodic orbits as well as their stable/unstable manifolds and intersections between them. To inform these computations and aid in visualization and dynamical analysis, an appropriate Poincar\'e section also needs to be selected and used. Finally, to determine practical times-of-flight (TOF) for theoretically infinite-duration heteroclinic transfers, a metric of ``closeness'' to an orbit must be defined. In this section, we summarize the methods used for these purposes. 

\subsection{Perigee Poincar\'e Map} \label{sectionSection}

The PCR3BP is a dynamical system on a 4D phase space $(x, y, p_{x}, p_{y})$ with an integral of motion given by the Jacobi constant $C$. Thus, any one trajectory will lie entirely within a single energy submanifold of form $\mathcal M_C = \{ (x,y,p_{x},p_{y}) : -2H(x,y,p_{x},p_{y}) = C\}$, where $C$ is the Jacobi constant associated with any point along that trajectory. Stable and unstable manifolds of a periodic orbit in $\mathcal M_C$ will also belong to $\mathcal M_{C}$, as must any heteroclinic connections to other periodic orbits. Thus, for the dynamical analysis we wish to carry out, one can restrict attention to studying the PCR3BP dynamics within $\mathcal M_C$ for a variety of $C$ values. 

$\mathcal M_C$ is a 3D submanifold, so if one now takes a Poincar\'e surface of section $\Sigma$ for the PCR3BP flow (with Poincar\'e map $P:\Sigma \rightarrow \Sigma$) and considers the fixed-energy section $\Sigma_C = \Sigma \cap \mathcal M_C$, then $\Sigma_C$ will be 2D. Thus, the dynamics of the resulting Poincar\'e map $P_C: \Sigma_C \rightarrow \Sigma_C$---the restriction of $P$ to $\Sigma_C$---will be much easier to study than those of the flow, as 2D map dynamics can be relatively easily visualized. Hence, as is standard in PCR3BP studies---see, e.g., \citet{KoLoMaRo, kumar2021journal}---we will follow this approach of studying the Poincar\'e map dynamics. However, ideally the PCR3BP flow should be transverse (i.e., not tangent) to the chosen Poincar\'e section $\Sigma$ at all points of the section, or otherwise have tangencies at as few points as possible. The most commonly used Poincar\'e sections in the literature, including the aforementioned studies, are those with fixed $x$ or $y$. These sections however generally experience many tangencies with unstable resonant periodic orbits' stable/unstable manifolds, leading to discontinuities when the manifolds are plotted on the section for analysis; see, e.g., \citet{kumar2021journal}. 

Given the disadvantages of fixed-$x$ or fixed-$y$ Poincar\'e sections in the PCR3BP, in this study, we instead utilize a perigee Poincar\'e surface of section $\Sigma = \{ (x, y, p_{x}, p_{y}) \in \mathbb{R}^{4}: \ell(x, y, p_{x}, p_{y}) = 0\}$ defined by Earth-relative osculating mean and true anomalies equal to zero. Such sections have been employed by, e.g., \citet{villacScheeres2003}, \citet{RoSc2007}, and \citet{howellDavisHaapala}, and they offer significantly better transversality to the PCR3BP flow compared to the commonly used sections mentioned earlier. This is because the mean anomaly generally increases continuously with time, meaning that orbits typically do not experience tangencies with the section. The exception to this occurs in the region around the Moon, where Earth's gravity is no longer the dominant influence on the spacecraft. In this case, the geocentric osculating true and mean anomalies may begin to decrease. However, for the interior resonant orbits and their stable or unstable manifolds studied in this paper, this region is not the primary focus, as intersections with the perigee section occur away from the Moon.

\subsubsection{Identifying Crossings of the Perigee Section}

To identify crossings of PCR3BP trajectories with the aforementioned perigee Poincar\'e surface of section $\Sigma$ during orbit propagation, we adopt an alternative approach rather than directly detecting when the osculating true or mean anomaly reaches zero. 
Specifically, it is well-known that at both perigee and apogee, the geocentric position and velocity vectors in an inertial reference frame become perpendicular, resulting in their dot product becoming zero. Furthermore, this dot product is positive when $\ell \in (0, \pi)$ and negative for $\ell \in (\pi, 2\pi)$. 
Assuming the mean anomaly increases monotonically, perigee occurs when the dot product crosses zero in the positive direction (i.e., from negative to positive), while apogee occurs when it crosses in the opposite direction. Thus, to detect perigee crossings and distinguish them from (most) apogees, we look for when, 
\begin{equation} \sigma(x, y, p_{x}, p_{y})= [(x+\mu), y] \cdot [p_{x}, (p_{y}+\mu)] \label{dotProd} \end{equation} 
crosses from negative to positive during a trajectory propagation. This can easily be computationally implemented in MATLAB or Julia using those programming languages' \texttt{odeEvent} and \texttt{ContinuousCallback} functionalities, respectively. Note that the additions of $\mu$ to $x$ and $p_{y}$ in Eq.\ \eqref{dotProd} are due to the change from barycentered to geocentric coordinates. 

In the Kepler problem, where mean anomaly $\ell$ is always increasing, the aforementioned strategy will reliably detect every perigee without generating any false detections. 
However, if the assumption of $\dot \ell> 0$ is violated, there may be instances where the test erroneously identifies a ``false perigee'' in the PCR3BP, particularly when the true mean anomaly is actually $\pi$ instead of zero. Such a situation can arise in particular when the geocentric $\ell$ briefly begins to decrease near apogee, which can occur when the trajectory closely approaches the Moon. If $\dot \ell$ becomes negative near apogee, then the trajectory may transition from $\ell \in (\pi, 2\pi)$ to $\ell \in (0, \pi)$ by passing through $\ell = \pi$. In this case, the dot product in Eq.\ \eqref{dotProd} also crosses from negative to positive, mirroring the behavior observed when $\ell$ crosses zero with $\dot{\ell} >0$. To mitigate these false detections, an additional check can be incorporated into the perigee event detection algorithm. Namely, if the previous dot product method detects a potential perigee crossing during propagation, then the mean anomaly $\ell$ of the resulting state is computed; only if $\ell$ is very near 0 or $2\pi$ is the perigee detection then confirmed and the propagation stopped. Both MATLAB and Julia's event detection capabilities can implement such functionality to decide whether or not to stop a numerical integration based on an additional test. 

\subsection{PCR3BP Periodic Orbits} 
\label{periodicOrbitCompSection}

\begin{figure}
\centering
\includegraphics[width=0.5\columnwidth]{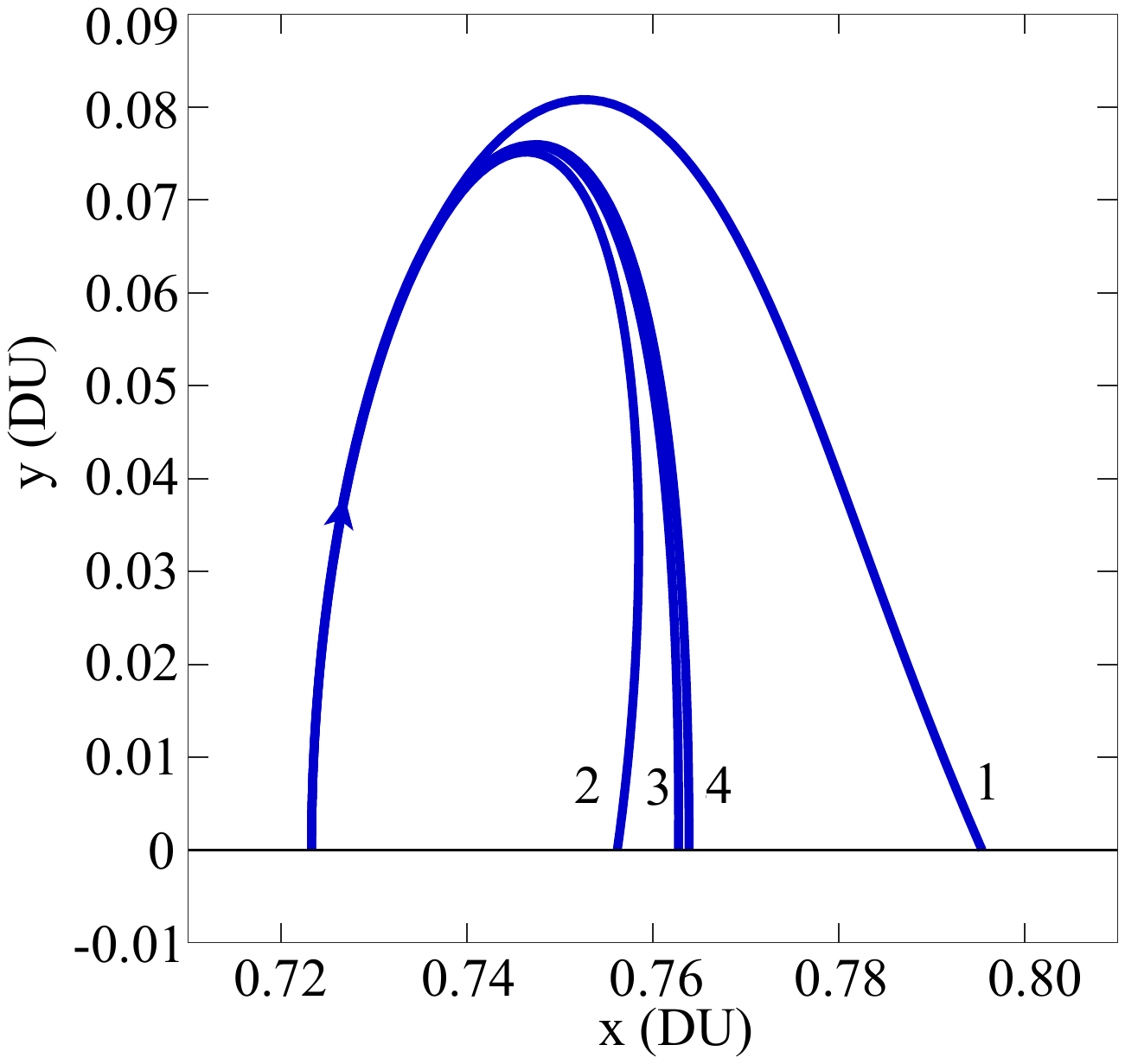}
\caption{ \label{fig::singleshoot} Illustration of perpendicular $x$-axis crossings method for computing PCR3BP periodic orbits \citep{perpCrossing}.}
\vspace{-15pt}
\end{figure}

To compute a family of $m$:$n$ unstable periodic orbits in the Earth--Moon PCR3BP, we start with an orbit state from the Earth Kepler problem having semi-major axis $a = \left(\frac{n}{m}\right)^{2/3}$ and initial longitude of perigee and true anomaly both $\pi$ (for the interior MMRs considered in this study); here, the longitude of perigee is measured with respect to the positive $x$-axis of the rotating frame. The previously mentioned Keplerian orbit will be symmetric about the $x$-axis and will also be periodic in the rotating Kepler problem (i.e., PCR3BP with $\mu = 0$). Thus, the method of perpendicular $x$-axis crossings can be used to numerically continue this Keplerian orbit to the true value of $\mu =  1.2150584270572 \times10^{-2}$ for the Earth--Moon system; see, for example, Section 2.6.6.2 of \citet{parkerAnderson} for details of this method. A schematic illustration of the method is also given in Figure \ref{fig::singleshoot}. 
The same method is then used to continue the resulting PCR3BP orbit through (most of) the remainder of its orbit family, using the perpendicular orbit $x$-intercept as the continuation parameter. 

Occasionally, a turning point may be encountered where this orbit $x$-intercept reaches an extremum along the family; in this case, using the Jacobi constant $C$ as the continuation parameter instead of $x$ can allow the computation to proceed past such points. Also, as will be seen in Section \ref{poSection}, it can occur that a periodic orbit family passes \emph{through} the singularity at the Earth ($r_{1}=0$), where the standard PCR3BP Equations \eqref{pcrtbpH_EOM}--\eqref{pcrtbpH} are undefined. In both cases, a Kustaanheim-Stiefel (KS) \citep{ksRegularization} regularization-based continuation method is used to overcome the difficulty. This method, developed by \citet{kumar2025aug} as a modification of that of \citet{howellBreakwell}, allows continuation of CR3BP periodic orbits by Jacobi constant $C$---including through singularities. {$C$ changes rapidly along the orbit family near singularities, and thus one can in fact use relatively \emph{large} continuation step sizes (e.g., $\Delta C = 0.1$) when passing through them.} We refer the reader to the aforementioned papers for details of the KS-based method, but note that it also relies on finding initial states that lead to perpendicular $xz$-plane crossings when propagated. In this paper, we consider only planar periodic orbits that are symmetric about the $x$-axis; both previously described (non-KS and KS-based) methods are valid for this case. 

As a final remark, note that an $m$:$n$ resonant periodic orbit will pass through perigee $m$ times during one orbital period in the PCR3BP. Thus, when using the perigee section $\Sigma$ described in Section \ref{sectionSection}, the $m$:$n$ periodic orbit will have $m$ intersection points with the section; moreover, these points will not be fixed points of the section's Poincar\'e map, but become $m$-iteration periodic orbits under the map. That is, one will have $m$ points $X(k) \in \Sigma \subset \mathbb{R}^{4}$, $k = 0, 1, \dots, m-1$, satisfying the equation,
\begin{equation} P(X(k)) = X(k+1 \mod m)  \end{equation}
for all $k$, where $P: \Sigma \rightarrow \Sigma$ is the perigee Poincar\'e map. Furthermore, by definition of the points $X(k)$, there exist times-of-flight $\tau(k) \in \mathbb{R}^+$ such that a trajectory starting at $X(k)$ reaches $X(k+1)$ in time $\tau(k)$ under the PCR3BP flow; in other words, $\tau(k)$ is the first-return time of the point $X(k)$ under the PCR3BP perigee Poincar\'e map. 

\subsection{Stable and Unstable Manifolds} \label{paramSection} 

Once the periodic orbits in a family $m$:$n$ have been computed as well as their intersection points $X(k)$ with the perigee section $\Sigma$, the computation of the orbits' stable/unstable manifolds is carried out. In particular, we compute the intersection of these manifolds with the perigee section. 

The portions of the periodic orbit stable/unstable manifolds lying in the chosen Poincar\'e section will correspond to 1D curves---one curve attached to each of the $m$ periodic orbit intersection points $X(k)$, $k = 0, 1, \dots, m-1$ with the section. To accurately compute these manifold curves, we used a parameterization method developed by the first author \citep{kumar2025tools} for calculating Taylor-series approximations of periodic orbit stable and unstable manifolds. The algorithm, an example of a general class of methods for computing various types of invariant manifolds \citep{haroetal}, extends the first author's related work \citep{kumar2021journal} on stable/unstable manifolds of Poincar\'e map fixed points to the $m$-iteration map-periodic orbit case.
Although a complete treatment lies beyond the scope of this paper (see \citet{kumar2025tools} for details), we note briefly that for each manifold we solve for a function
$W: \{ 0, \dots, m-1\} \times \mathbb{R} \rightarrow \mathbb{R}^{4}$ such that for all $k \in \{0, 1, \dots, m-1\} $,
\begin{equation}  \label{invariancequationpcrtbp}  
\Phi_{\tau(k)}(W(k,s)) = W(k+1 \mod m, \lambda s).
\end{equation}
Here, $\Phi_{t}(\bold{x})$ is the PCR3BP flow map of a point $\bold{x} \in \mathbb{R}^{4}$ by time $t$, $\tau(k)$ are the first return times of the points $X(k)$ as described in Section \ref{periodicOrbitCompSection}, and $\lambda$ is the $m$th root of the monodromy matrix eigenvalue corresponding to the stable or unstable manifold. The parameter $s \in \mathbb{R}$ is 0 at the base periodic orbit points $X(k)$ and grows larger in absolute value as the manifold moves further away from them (in terms of time-of-flight); the manifold will have two branches  corresponding to $s<0$ and $s>0$ here. Note that $\Phi_{\tau(k)}(\bold{x})$ is \emph{not} the perigee Poincar\'e mapping $P$, as the flow propagation time $\tau(k)$ is fixed to the first return time of $X(k)$, not to the return time of the arbitrary point $\bold{x} \in \mathbb{R}^{4}$. 

Equation \eqref{invariancequationpcrtbp} can be solved recursively by expressing $W$ as a set of $m$ Taylor series depending on the integer $k \in \{0, 1, \dots, m-1\} $,
\begin{equation}  \label{taylorSeries}   W(k, s) = X(k) + \sum_{i=1}^{\infty} W_{i}(k) s^{i} \quad k \in \{0, 1, \dots, m-1\}, \end{equation}
where $W_{1}(k) \in \mathbb{R}^{4}$ are appropriately scaled stable or unstable eigenvectors of the periodic orbit monodromy matrix at each of its perigee passages $X(k)$. The $W_i(k)s^i$ for $i\ge2$ correspond to higher-order (nonlinear) terms in the stable/unstable manifold approximation that are solved recursively. Due to both finite truncation and radius of convergence, these series representations of $W$ will be valid only in a finite domain of validity $s \in (-D, D)$, where $D>0$ is the largest  number for which the error in Equation \eqref{invariancequationpcrtbp} remains within a chosen tolerance for all $s \in (-D, D)$ (we chose $10^{-5}$). Although Equation \eqref{taylorSeries} is thus still a \emph{local} rather than global representation of the stable/unstable manifold, it yields several orders of magnitude improvement in accuracy as compared to traditional linear eigenvector-based approximations, as shown in \cite{kumar2025tools}; this in turn gives us a much larger local manifold segment requiring less propagation for globalization.

As $\Phi_{\tau(k)}(\bold{x})$ was not the perigee Poincar\'e map, the $m$ curves parameterized by $W$ lie near but not on the perigee section of interest $\Sigma$. Thus, to finally compute the manifolds on the section, one simply numerically integrates fine grids of points from the domains of validity of those curves either backwards or forwards to the section. Then, further applications of the perigee Poincar\'e map $P$ either forwards or backwards in time are used to  globalize the full unstable and stable manifolds respectively. Using the series $W$ in combination with $P$, one can in fact define a function $W_{p}(k,s): \{ 0, \dots, m-1\} \times \mathbb{R} \rightarrow \Sigma$ which is valid for \emph{all} $s \in \mathbb{R}$ and parameterizes the  \emph{global} stable or unstable manifold of the periodic orbit on $\Sigma$. 

As usual, for each fixed Jacobi constant value $C$, one can plot these Poincar\'e map manifolds of various orbits at that $C$ value using just 2D coordinates for $\Sigma_C = \Sigma \cap \mathcal M_C$. Intersections of 1D manifold curves from different orbits on this 2D plot yield heteroclinic trajectories between those orbits. In this study, though manifolds are computed using Cartesian coordinates, we use the osculating elements $(g,a)$ of Section \ref{delaunaySection} for plotting them.

\subsection{Heteroclinic Connections} 
\label{heteroComputeSection}

As described earlier, an intersection in $\Sigma_C$ of 1D unstable and stable manifold curves belonging to two different orbits indicates the presence of a zero-$\Delta v$ heteroclinic transfer between them. While the corresponding trajectory can be approximately found by visually estimating the coordinates of the intersection point, the global parameterizations previously described for the stable/unstable manifolds enable a much more accurate and automatic  search and calculation. Such a heteroclinic computation algorithm was developed by \citet{kumar2025tools} to accompany the parameterization method of the previous section. As most of the heteroclinics of this paper were calculated by that algorithm, we briefly summarize its main steps and ideas here; full details can be found in the aforementioned reference. 

Assume that one has found parameterizations $W^{u}_{1p}(k,s)$ and $W^{s}_{2p}(k,s)$ for the \emph{global} unstable and stable manifolds respectively of two periodic orbits, as described in the second-to-last paragraph of Section \ref{paramSection}. The goal is then to find $(k_{1},s_{1}, k_{2},s_{2}) \in \mathbb{Z} \times \mathbb{R} \times \mathbb{Z} \times \mathbb{R}$ such that 
\begin{equation} \label{intersectcondition} W^{u}_{1p}(k_{1},s_{1}) = W^{s}_{2p}(k_{2},s_{2}),
\end{equation}
in which case either side of Eq. \eqref{intersectcondition} gives the desired heteroclinic intersection point. We also assume that, e.g., during the manifold visualization process, many points of the manifolds represented by $W^{u}_{1p}$ and $W^{s}_{2p}$ on $\Sigma_C$ have been computed. Then, to solve Eq. \eqref{intersectcondition}, we:

\begin{enumerate} 
 
\item Use the dynamics to restrict the solution search to only subsets of $(k_{1},s_{1}, k_{2},s_{2})$ space. 
 
\item For each subset of step 1, check if any line segment between precomputed, consecutive points of $W^{u}_{1p}$ intersects any segment between $W^{s}_{2p}$ points in $(x,y)$ coordinates for $\Sigma_C$.
\item For each intersection of line segments from step 2, use the segment endpoints' $(k_{1}, s_{1})$ and $(k_{2}, s_{2})$ values to refine $s_{1}$ and $s_{2}$ to high accuracy by bisection. 
\end{enumerate}
Assuming that the precomputed manifold curve points used to define the line segments of step 2 are spaced closely enough together, the resulting segment endpoint $(k_{1},s_{1}, k_{2},s_{2})$ values should be sufficiently accurate for step 3 to quickly yield a highly accurate refined solution. Note that though we previously discussed using $(g,a)$ coordinates for visualization on $\Sigma_C$, the above segment-intersection and bisection calculations are performed in 2D Cartesian $(x,y)$ coordinates instead, which form a valid coordinate system for $\Sigma_C$ as well.


As a final remark, note that the above process does not actually require any plots to be created or visualized; as long as the manifold parameterizations and enough manifold points have been computed and given to the algorithm, the search and calculation of heteroclinics is completely automatic and highly parallelizable for very fast performance. This is in addition to providing a significantly more accurate heteroclinic intersection point than a purely visual inspection can hope to achieve. However, visualization of stable/unstable manifolds can still provide additional information on dynamical behaviors that are not captured by computing heteroclinics alone, so in this paper, we will present $(g,a)$ plots of these manifolds as well. 

\subsection{Distance Metric between Orbits} 
\label{DistanceMetric}
A heteroclinic connection between two orbits is a trajectory that converges to an initial orbit backwards in time, and to a final orbit forwards in time. Mathematically, convergence is an infinite-time concept: the distance between the heteroclinic trajectory and the initial or final orbit goes to zero as $t \rightarrow \pm \infty$. However, to define and calculate a practical, finite transfer time for heteroclinics between orbits, one needs a way of determining when the spacecraft is ``close enough'' to its initial or final orbit for practical purposes. Thus, in this section, we present a framework for quantifying the distance between two orbits as well as between a state and an orbit in the PCR3BP. 

Traditional distance metrics in unperturbed 2-body Keplerian dynamics have previously been developed for the design of controllers for orbital transfers. For instance, \cite{marsden2001} introduced a distance metric based on two vector constants of motion in the 2-body problem---angular momentum and the Laplace-Runge-Lenz vector (hereafter referred to as the Laplace vector)---that uniquely define non-degenerate, bounded Keplerian orbits. Initially developed in the context of a Lyapunov function (for a Lyapunov-based controller), this metric facilitated global, low-thrust transfers between arbitrary elliptical orbits. 
Similarly, \cite{GAO2010117} proposed a Lyapunov-based guidance scheme leveraging the dynamic evolution of mean equinoctial orbital elements to optimize low-thrust, many-revolution transfers. However, these metrics are limited to unperturbed Keplerian orbits, where orbital elements along with angular momentum and Laplace vector remain invariant over time.

To address these limitations, we propose a generalized distance metric suitable for 
a broader range of orbit types, including both perturbed and unperturbed orbits. 
Notably, our metric is applicable to scenarios involving at least two 3-body solutions---perturbed Keplerian orbits like resonant orbits and Lyapunov orbits about the libration point $L_1$.

For the set of states 
under consideration, we assume that the instantaneous 3-body state $X$ is uniquely 
represented in terms of 
the angular momentum and Laplace vector, effectively constituting a change of variables.
For 3-body trajectories, the angular momentum and Laplace vector may exhibit osculatory behavior. 
This necessitates defining a distance metric that extends from point-to-point comparisons to accommodating the relationship between an instantaneous state (point) and a set of states (e.g., those defining a periodic orbit).  

For a 3-body state $X$ in the Earth--Moon rotating frame, transform to the Earth-centered frame. Let $\textbf{r}$ and $\dot{\mathbf{r}}$ denote the position and velocity vectors in the Earth-centered inertial frame. The Earth's gravitational parameter in non-dimensional units can be expressed as  $\mu_e = 1-\mu$.
The angular momentum vector ($\textbf{L}$) and Laplace vector ($\textbf{A}$) are, respectively, 
\begin{align}
\label{eq:angular_momentum}
    \mathbf{L}(\mathbf{r},\dot{\mathbf{r}}) &= \mathbf{r} \times \dot{\mathbf{r}}, \\
\label{eq:laplace_vector}
    \mathbf{A}(\mathbf{r},\dot{\mathbf{r}}) &= \dot{\mathbf{r}} \times (\mathbf{r} \times \dot{\mathbf{r}})
    - (1-\mu) \frac{\mathbf{r}}{\lVert \mathbf{r} \rVert}.
\end{align}

\begin{figure}
\centering
\includegraphics[width=0.6\linewidth]{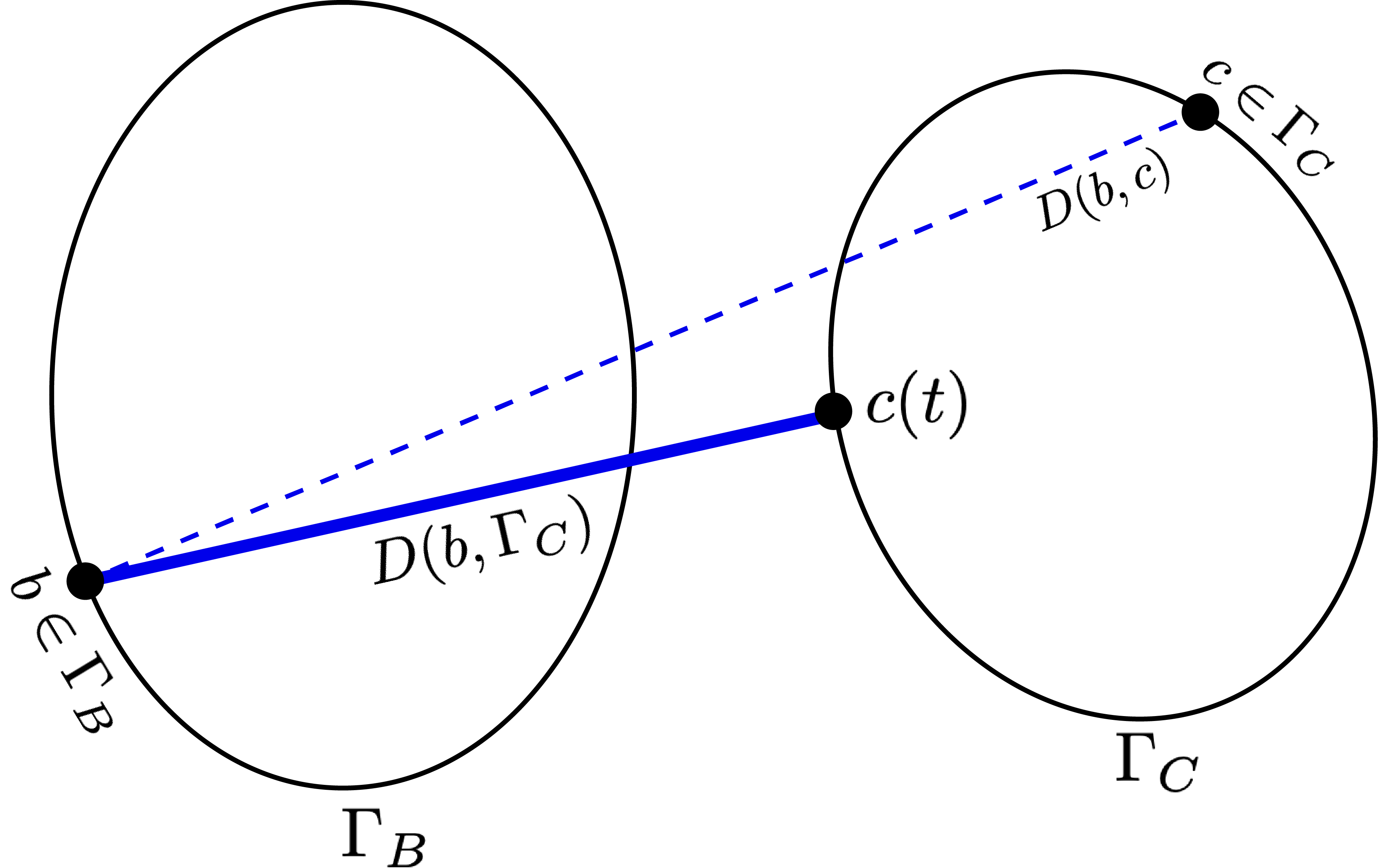}
\caption{\label{fig:orbit_orbit} Illustration of the definition of distance metric between two orbits $\Gamma_B$ and $\Gamma_C$.}
\end{figure} 

Consider two orbits at the same energy level in the inertial frame, $\Gamma_B$ and $\Gamma_C$ as shown schematically in Figure \ \ref{fig:orbit_orbit}, 
where $b \in \Gamma_B$ represents a state in orbit $\Gamma_B$ and $c \in \Gamma_C$ a state in orbit $\Gamma_C$. 
The distance between states $b$ and $c$ can be quantified by the difference between their corresponding angular momentum and Laplace vectors, $(\textbf{L}_{b},\textbf{A}_{b})$ and $(\textbf{L}_{c},\textbf{A}_{c})$. 
We define the distance metric $D(b,c)$ between states $b \in \Gamma_B$ and $c \in \Gamma_C$ as,
\begin{equation}
\label{eq:D(a,b)}
    D(b,c) = \sqrt{||\textbf{L}_{b} - \textbf{L}_{c}||^2 + ||\textbf{A}_{b} - \textbf{A}_{c}||^2}.
\end{equation}
One can also naturally extend the above definition to define the distance between a single state $b$ and the set $\Gamma_C$ as the minimum distance between $b$ and all points of $\Gamma_C$,
\begin{equation}
    D(b,\Gamma_C) = \min_{c\, \in \Gamma_C} D(b,c).
    \label{eq:point_set_distance}
\end{equation}
This point-to-set distance is a type of Hausdorff distance \citep{abraham2012manifolds}. 

While vectors $(\textbf{L},\textbf{A})$ were used to define the distance above, one can also use 
other variables, such as equinoctial elements, or even the Euclidean metric of the synodic state vector $(x,y, p_x, p_y)$ of two states ${b}$ and ${c}$  (see Appendix \ref{Alt_dist}). However, equinoctial elements are scalar quantities that have singularities for an inclination of $180^{\circ}$ when considering orbits in 3D space. 
For geocentric orbits, synodic state vectors tend to exhibit stronger short-period variations than equinoctial elements or the angular momentum and Laplace vectors; this contrast is especially pronounced for near-Keplerian orbits.
Therefore, we choose to define the distance metric in terms of angular momentum and Laplace vectors as they have less variability for the orbits of interest.

\begin{figure}
\centering
\includegraphics[width=0.6\linewidth]{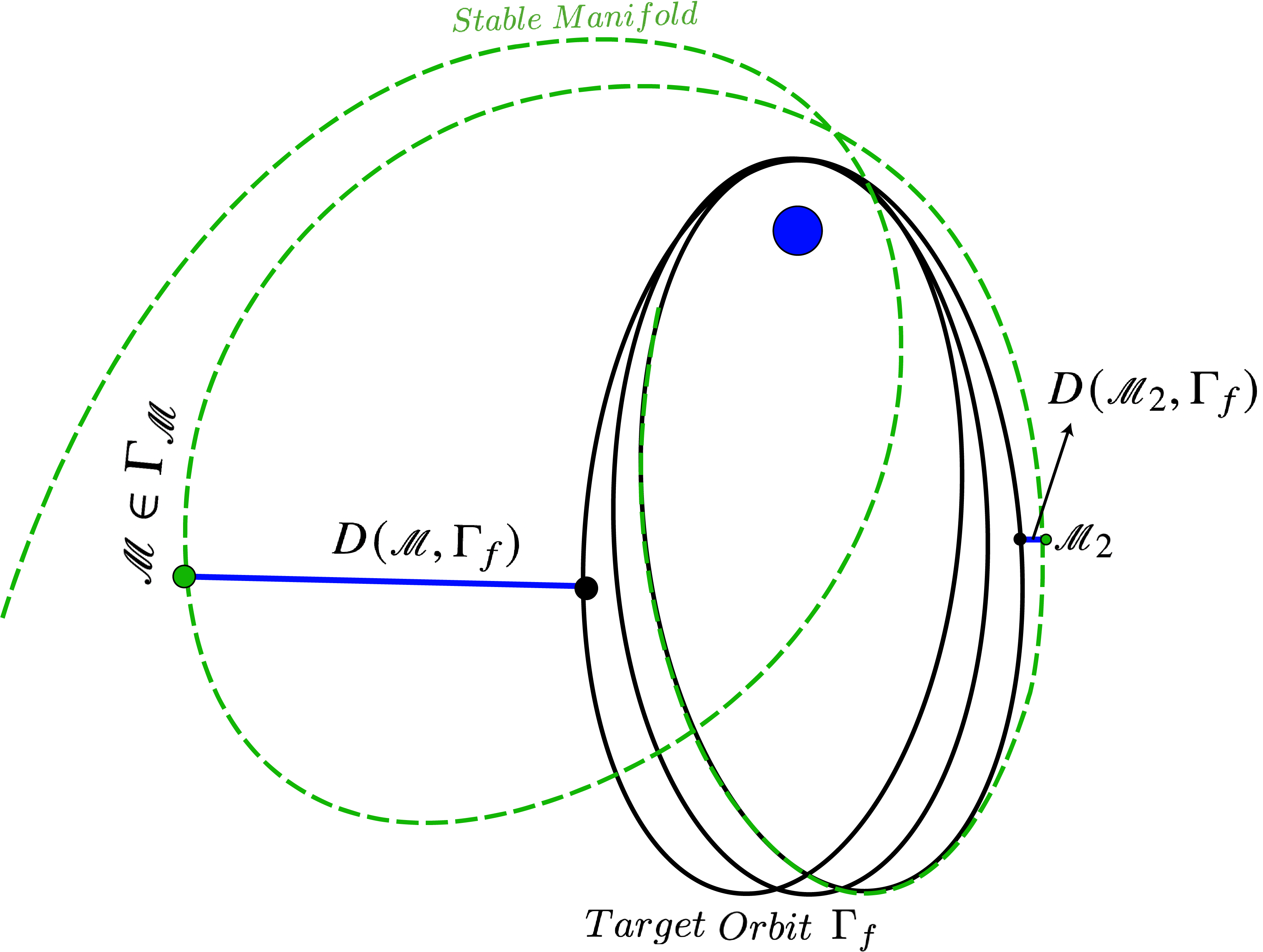}
\caption{\label{fig:manifold_orbit} Illustration of definition of distance metric $D(\mathscr{M},\Gamma_f)$ between a state $\mathscr{M}$ on the stable manifold $\Gamma_{\mathscr{M}}$ of an orbit $\Gamma_f$ and the orbit $\Gamma_f$ itself.}
\end{figure} 

Heteroclinic connections exist between unstable resonant periodic orbits, as we will discuss in Section \ref{heteroclinicSection}. As mentioned earlier, heteroclinic connections theoretically require infinite time-of-flight (TOF) due to asymptotic convergence onto initial and final orbits; in real life, though, one can consider the transfer as complete once the spacecraft is ``close enough'' to an orbit. 
Thus, for practical operational applications and computation of real-world transfer durations, certain distance tolerances ($\epsilon_i$ and $\epsilon_f$) are defined for the initial and final unstable resonant periodic orbits ($\Gamma_i$ and $\Gamma_f$). The transfer time is then calculated from the instant when the unstable manifold is within $\epsilon_i$ of $\Gamma_i$ as propagated backwards in time and the stable manifold is within $\epsilon_f$ of $\Gamma_f$ as propagated forwards in time. The distance metric defined in Equation \eqref{eq:point_set_distance} proves useful in this case. 

Figure \ref{fig:manifold_orbit} illustrates our distance metric for a stable manifold trajectory and a target orbit. Let $\Gamma_{\mathscr{M}}$ be the set of points on a stable manifold trajectory asymptotic to $\Gamma_f$. The distance between a state {\footnotesize $\mathscr{M}$} $\in \Gamma_\mathscr{M}$ and the final orbit $\Gamma_f$ is $D(${\footnotesize $\mathscr{M}$},$\Gamma_f)$ via \eqref{eq:point_set_distance}. By the definition of the stable manifold of an orbit $\Gamma_f$, we expect  $D(${\footnotesize $\mathscr{M}$}$(t),\Gamma_f) \rightarrow 0$ as $t \rightarrow \infty$. 
By setting a tolerance $\epsilon_f$ we can compute the state {\footnotesize $\mathscr{M}$}$_2 \in \Gamma_{\mathscr{M}}$ such that $D(${\footnotesize $\mathscr{M}$}$_2,\Gamma_f) = \epsilon_f$, and the corresponding time $t_2$, such that 
the manifold stays within a distance $\epsilon_f$, i.e., $D(${\footnotesize $\mathscr{M}(t)$}$,\Gamma_f) < \epsilon_f$, for all $t>t_2$. The time $t_1$ when the unstable manifold of $\Gamma_i$ reaches a distance  $\epsilon_i$-close to  $\Gamma_i$ can be found similarly. The effective heteroclinic transfer time is then given by $t_2 - t_1$.

\section{Cislunar Resonant Periodic Orbit Families} \label{poSection}

As described in Section \ref{mmrSection}, the overlap of mean-motion resonances allows for natural changes of  semi-major axis without use of spacecraft propulsion. This overlap, which corresponds to heteroclinics between unstable resonant orbits, can thus be useful for understanding and designing orbit changes for missions in celestial systems. Thus, a program for determining the extent of such phenomena in cislunar space involves computing the unstable resonant orbits, followed by their stable and unstable manifolds, and finally calculating heteroclinics or detecting barriers to transport between the various resonances in the system. In this section, we describe the first step: the computation of unstable interior resonant orbits, which are periodic in our Earth--Moon PCR3BP model. In addition, to try to gain a more complete understanding of how various resonant periodic orbit families are linked with each other, we will also compute stable resonant periodic orbits. 

\begin{figure}
\begin{centering}
\includegraphics[width=0.65\columnwidth]{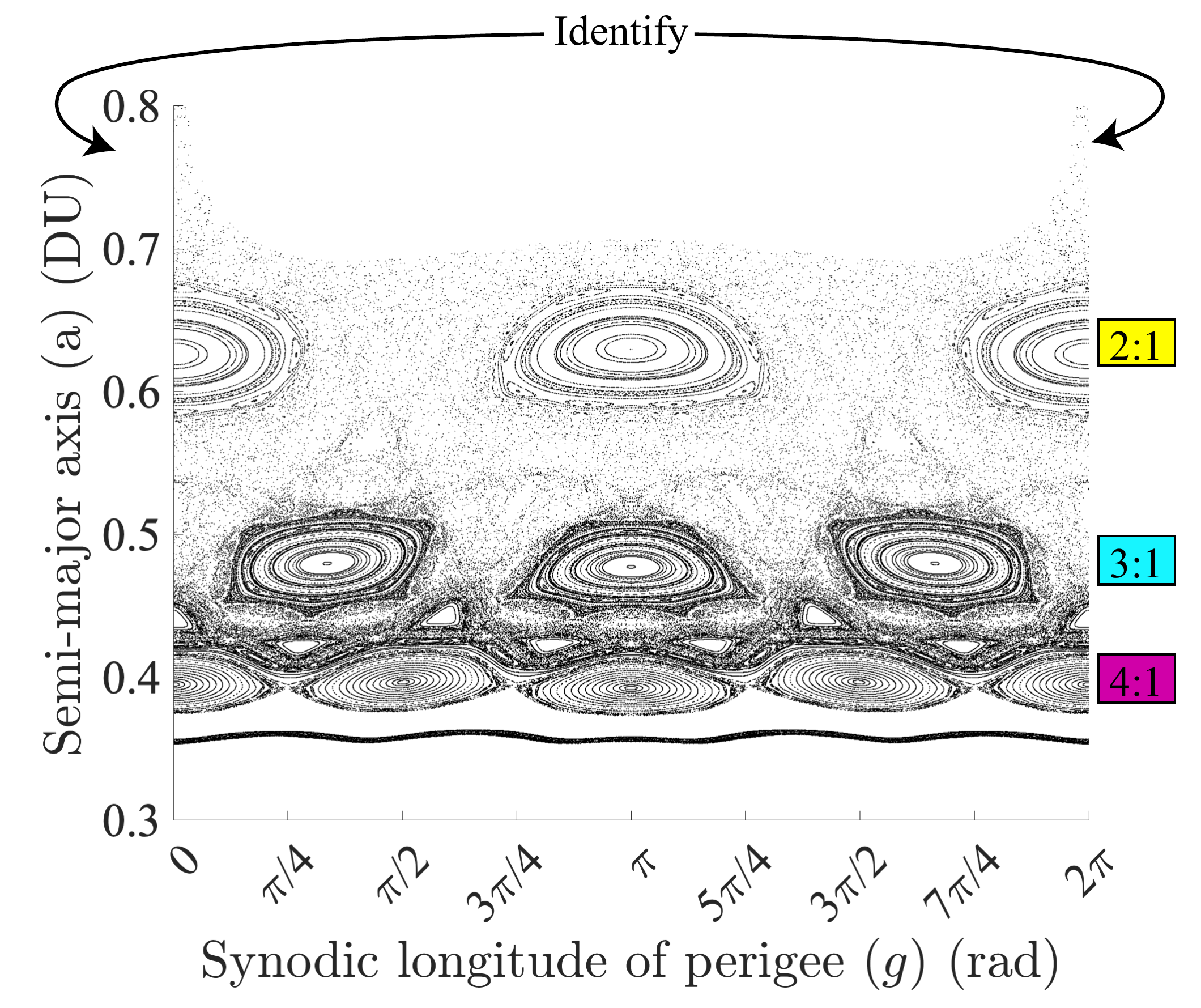}
\vspace{-5pt}
\caption{ \label{fig:poincareMap310} Poincar\'e section plot at perigee ($\ell=0$) for $C = 3.10$. The “identify” indicates that the synodic longitude of perigee is an angle
that maps back to itself.}
\end{centering}
\vspace{-10pt}
\end{figure}

As outlined in the introduction, the 2:1 and 3:1 lunar 
MMRs have been utilized in previous space missions such as TESS and IBEX, respectively---both of which operate in stable resonant orbits. Therefore, we will focus our study on these two MMRs. Additionally, Poincar\'e section plots generated by propagating massive numbers of points to our perigee section indicate that the 4:1 MMR is also prominent, so we will include it in our analysis as well.
For instance, Figure \ref{fig:poincareMap310} 
shows such a plot generated for the Jacobi constant $C=3.10$. In this plot, wide stability islands generated by the 2:1 MMR are prominently centered around 
$a=0.63$.
Additionally, significant 3:1 islands are observed near 
$a=0.48$, while notable 4:1 islands are clearly visible around 
$a=0.40$. Note that an $m$:$n$ MMR displays $m$ ``eye'' shaped regions in a row at its corresponding value of 
semi-major axis ($a$).

\subsection{Periodic Orbit Families for 4:1 MMR} \label{41section}

\begin{figure*}
\begin{centering}
\includegraphics[width=0.99\textwidth]{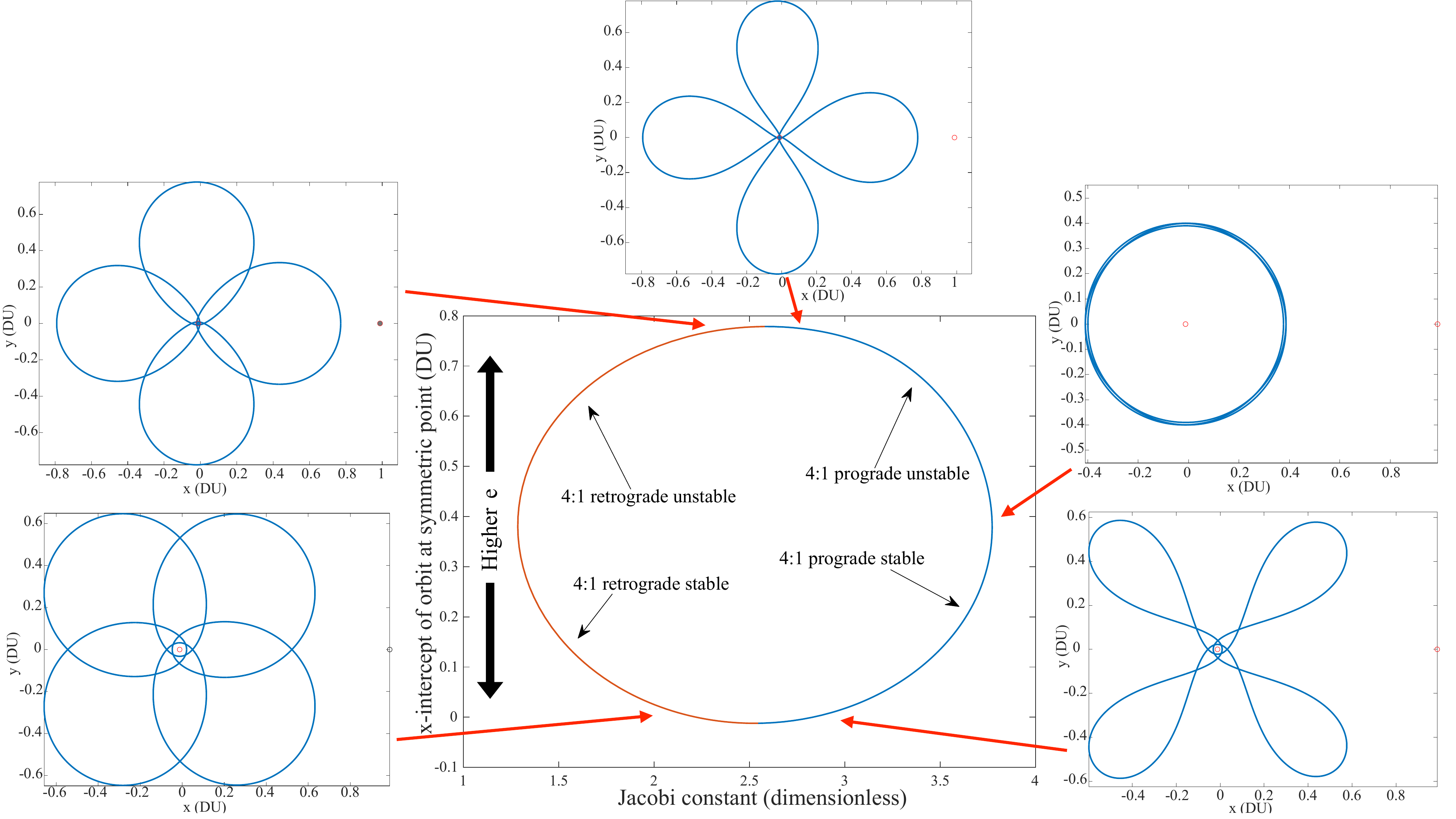}
\caption{ \label{fig:bifurcation41} Bifurcation diagram for 4:1 periodic orbits with example orbit plots. {Bottom left/right plotted orbits are stable, top left/center are unstable, and top right is very mildly unstable (nearly stable).} Change of curve color in bifurcation diagram indicates Earth collision. Earth/Moon shown as red circles in orbit plots.}
\end{centering}
\vspace{-15pt}
\end{figure*}

The 4:1 resonant periodic orbits were computed using the methods of Section \ref{periodicOrbitCompSection}, including stable and unstable periodic orbits. Both prograde and retrograde orbits were computed and numerically continued throughout their families. Figure \ref{fig:bifurcation41} displays a bifurcation diagram for the resulting orbit family, plotting the perpendicular positive $x$-axis crossing value of each orbit versus its Jacobi constant, along with a selection of representative orbits. First of all, it should be noted that the unstable and stable prograde 4:1 periodic orbits belong to a single continuous periodic orbit family, so that the stable orbits are continuations of the unstable ones. The same is true of the unstable and stable retrograde 4:1 periodic orbits, which also belong to the same family. The Jacobi constant reaches a maximum or minimum at points where the periodic orbit Floquet multipliers pass through 1, i.e., at fold bifurcation points which---as expected \citep{rimmer1978}---mark the dividing points between stable and unstable orbits in each family.

Furthermore, using the KS-based method described in Section \ref{periodicOrbitCompSection} to continue these orbits through the singularity at the Earth, we found that the prograde and retrograde orbit families in fact connect to each other to form a \emph{single, unified} 4:1 family. The orbits joining the families together correspond to collisions and very close approaches with the Earth at perigee (the $x$-intercepts near 0.8 correspond to apogee); the points in Figure  \ref{fig:bifurcation41} where the curve color changes correspond to Earth collision orbits, and mark the dividing line between prograde and retrograde motion. All orbits shown are in the 4:1 MMR, nevertheless. The retrograde orbits occur for lower Jacobi constants (higher energies) than the prograde orbits. 

\begin{figure}
\begin{centering}
\includegraphics[width=0.49\columnwidth]{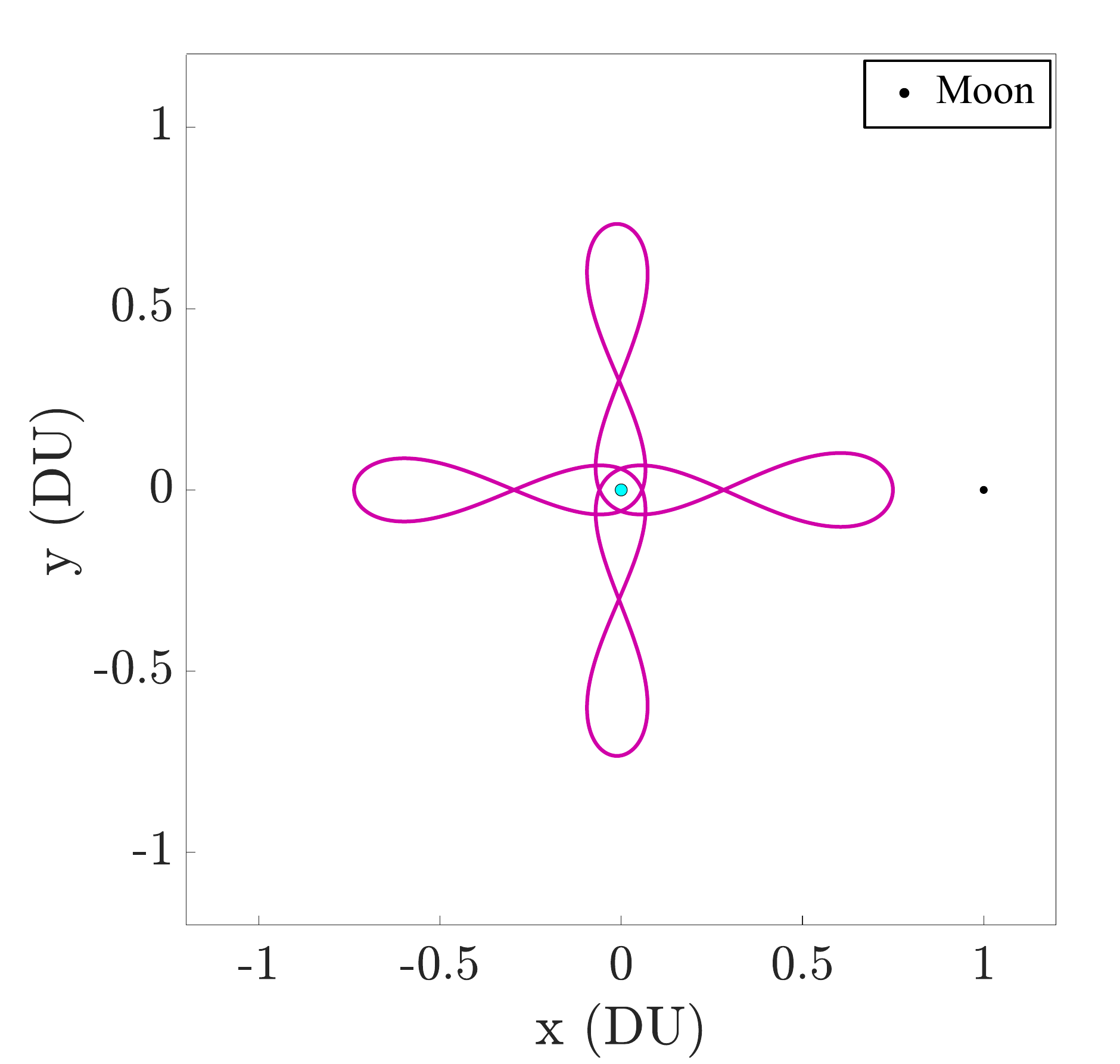}
\includegraphics[width=0.48\columnwidth]{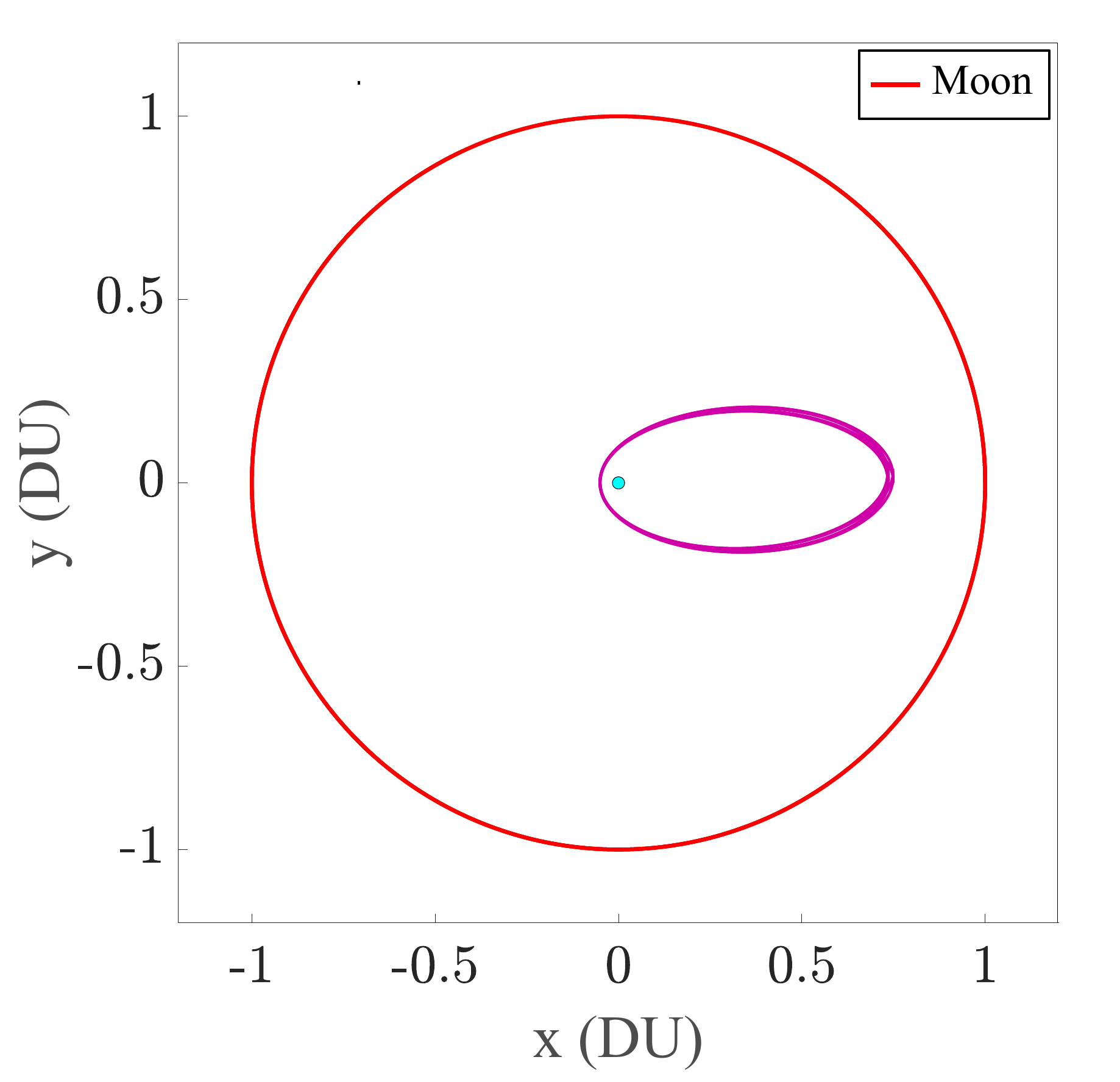}
\caption{ \label{fig:41orbit} Example 4:1 unstable resonant orbit, $C=3.15$ and period $6.3089 \approx 2.0082\pi$ TU, propagated for two periods in rotating (left) and inertial (right) frames.}
\end{centering}
\vspace{-15pt}
\end{figure}

The 4:1 orbits exhibit a wide range of eccentricities, from near-circular (as shown in the top-right unstable orbit plot) to highly eccentric (depicted in the top center orbit plot). An example of a moderate-eccentricity 4:1 unstable orbit is illustrated in Figure \ref{fig:41orbit}, propagated for twice its period and plotted in both rotating and inertial frames; note that the orbit is periodic only in the rotating frame---not in the inertial frame---as is usual. It is important to also note that the semi-major axis of the 4:1 MMR is too small to reach or exceed the Moon's orbit; even the highest-eccentricity 4:1 orbits, shown in Figure \ref{fig:bifurcation41}, have an apogee well below the Moon (shown in Figure \ref{fig:bifurcation41} orbit plots as a red circle at $x=1-\mu \approx 1$). The perigee altitudes of the 4:1 resonant orbits range from around 149,000 km down to Earth surface collision at 6378 km; the orbit shown in Figure \ref{fig:41orbit} has a perigee of approximately $19,000$ km---well within GEO, as is the case for all higher-eccentricity 4:1 resonant orbits. 

\subsection{Periodic Orbit Families for 3:1 MMR} \label{31section}

\begin{figure}
\begin{centering}
\includegraphics[width=0.6\columnwidth]{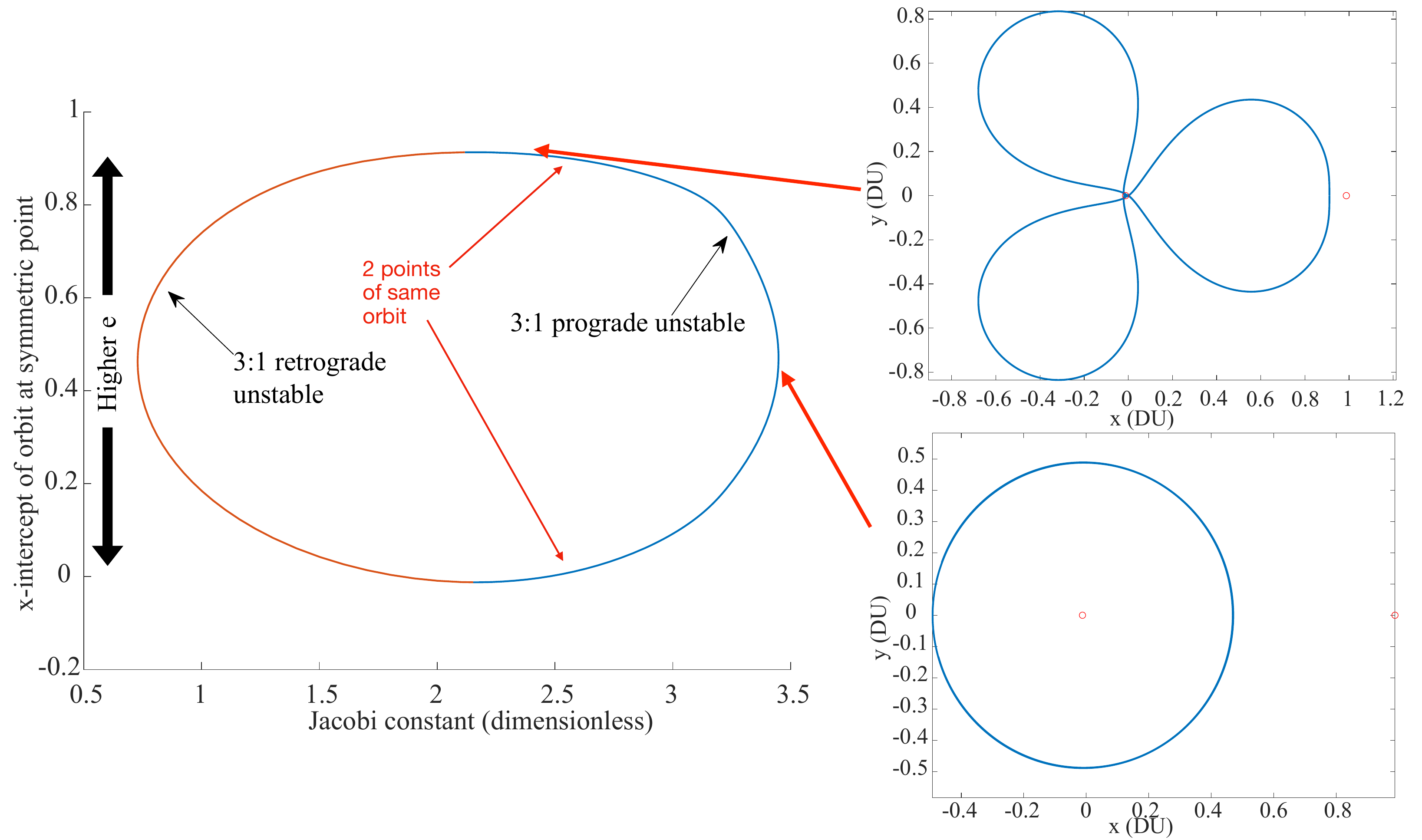}
\includegraphics[width=0.39\columnwidth]{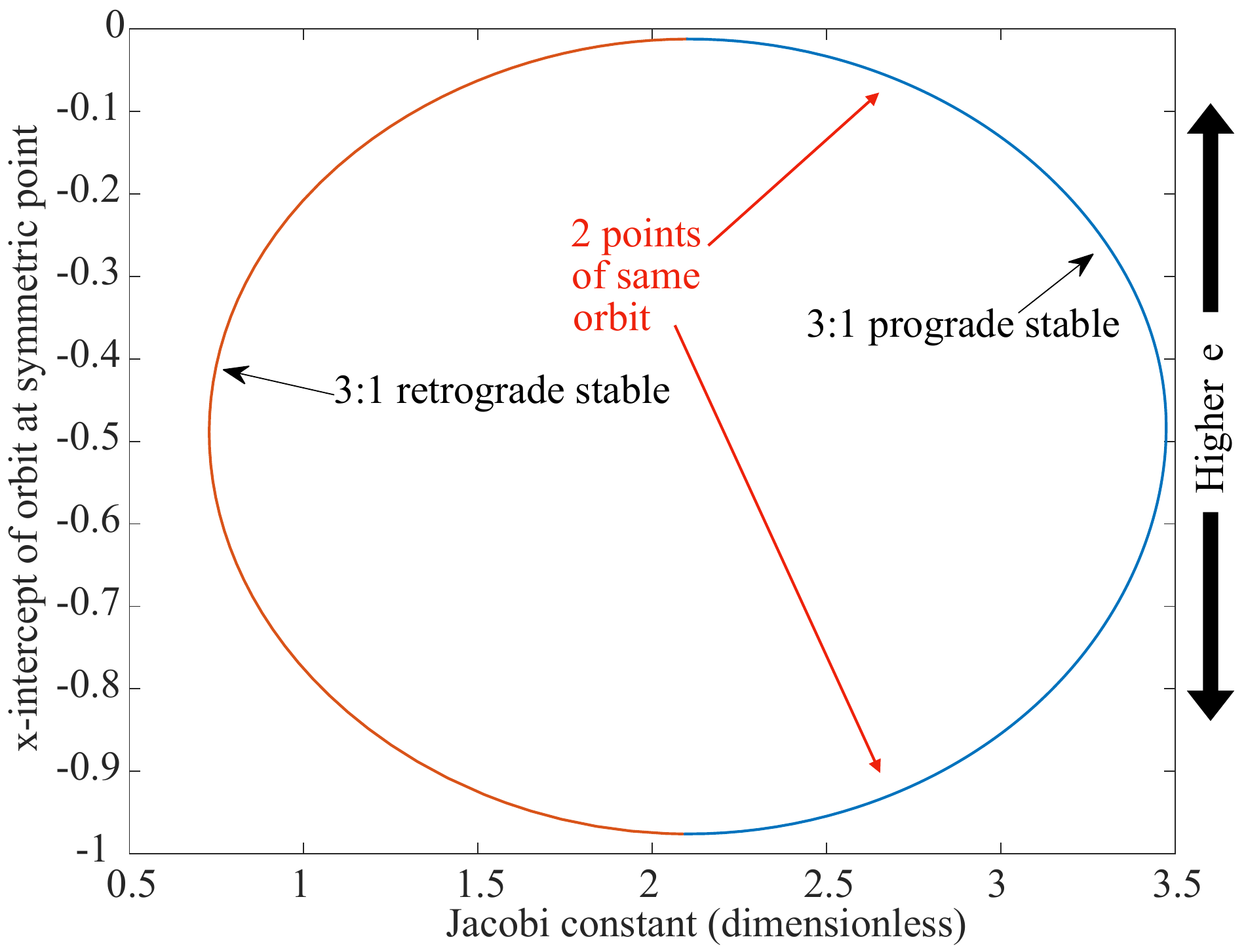}
\caption{ \label{fig:bifurcation31} Bifurcation diagrams for 3:1 unstable (left) and stable (right) periodic orbits with example orbit plots. {Top orbit is highly unstable, bottom orbit is very mildly unstable (nearly stable).} Change of curve color in bifurcation diagram indicates Earth collision. Earth and Moon shown as red circles in orbit plots.}
\end{centering}
\vspace{-12pt}
\end{figure}

For the 3:1 resonance, again both prograde and retrograde stable and unstable periodic orbits were computed and numerically continued throughout their families. In this case, the stable and unstable orbits are found to belong to separate families, and so we get two separate bifurcation diagrams, one for each family. Figure~\ref{fig:bifurcation31} displays the diagrams for the resulting orbit families, again as a perpendicular $x$-axis crossing vs Jacobi $C$ plot; the unstable diagram is accompanied by a few unstable periodic orbit plots. 
In this diagram, note that though each curve corresponding to a prograde/retrograde orbit family crosses a given Jacobi $C$ twice, both of these crossings correspond to the same orbit; these two replicated portions of a given orbit family both come together at fold bifurcation points where the Floquet multipliers go to 1 and the Jacobi constant reaches a maximum or minimum along the family. 

\begin{figure}[t!]
\begin{centering}
\includegraphics[width=0.49\columnwidth]{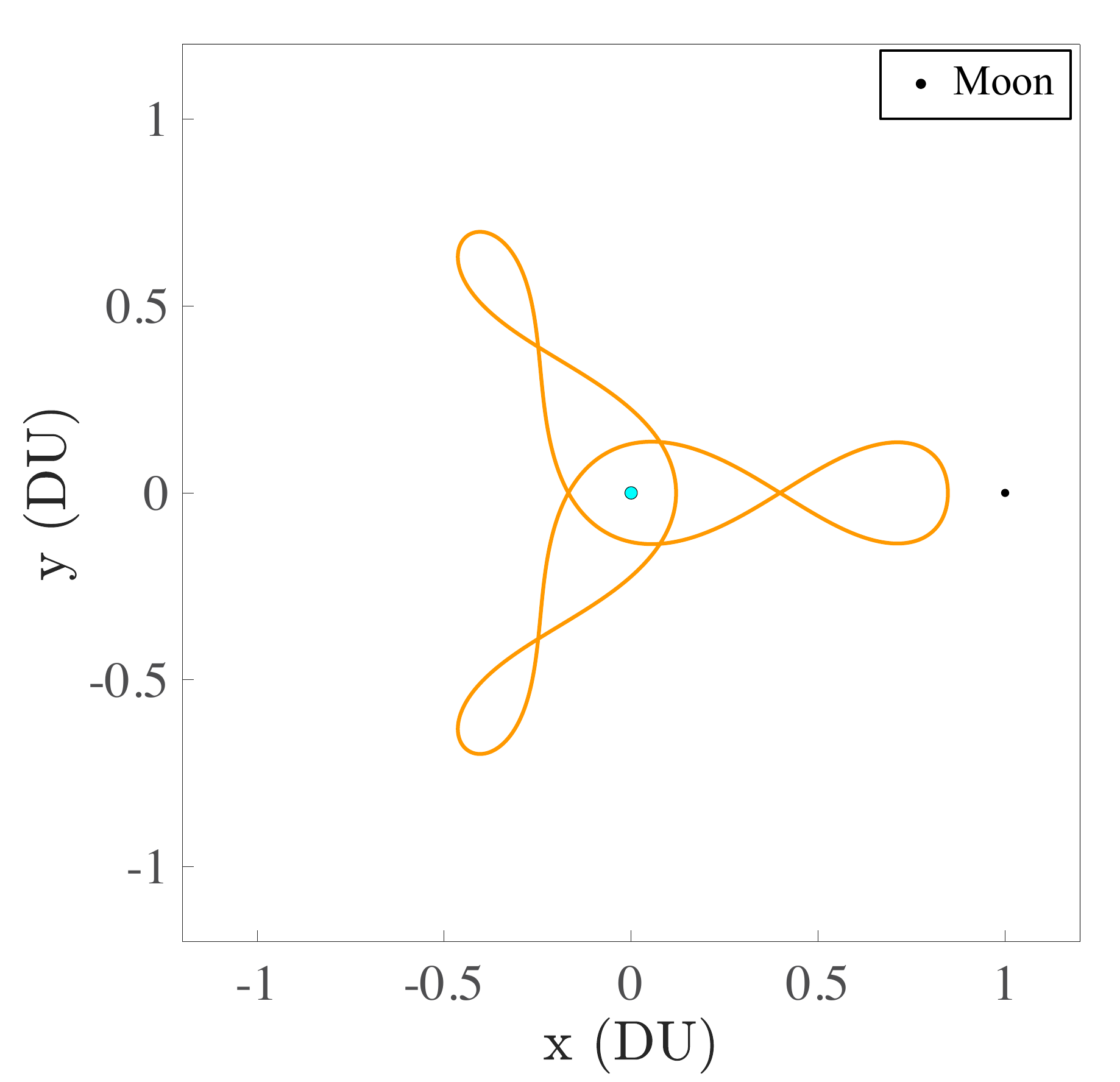}
\includegraphics[width=0.49\columnwidth]{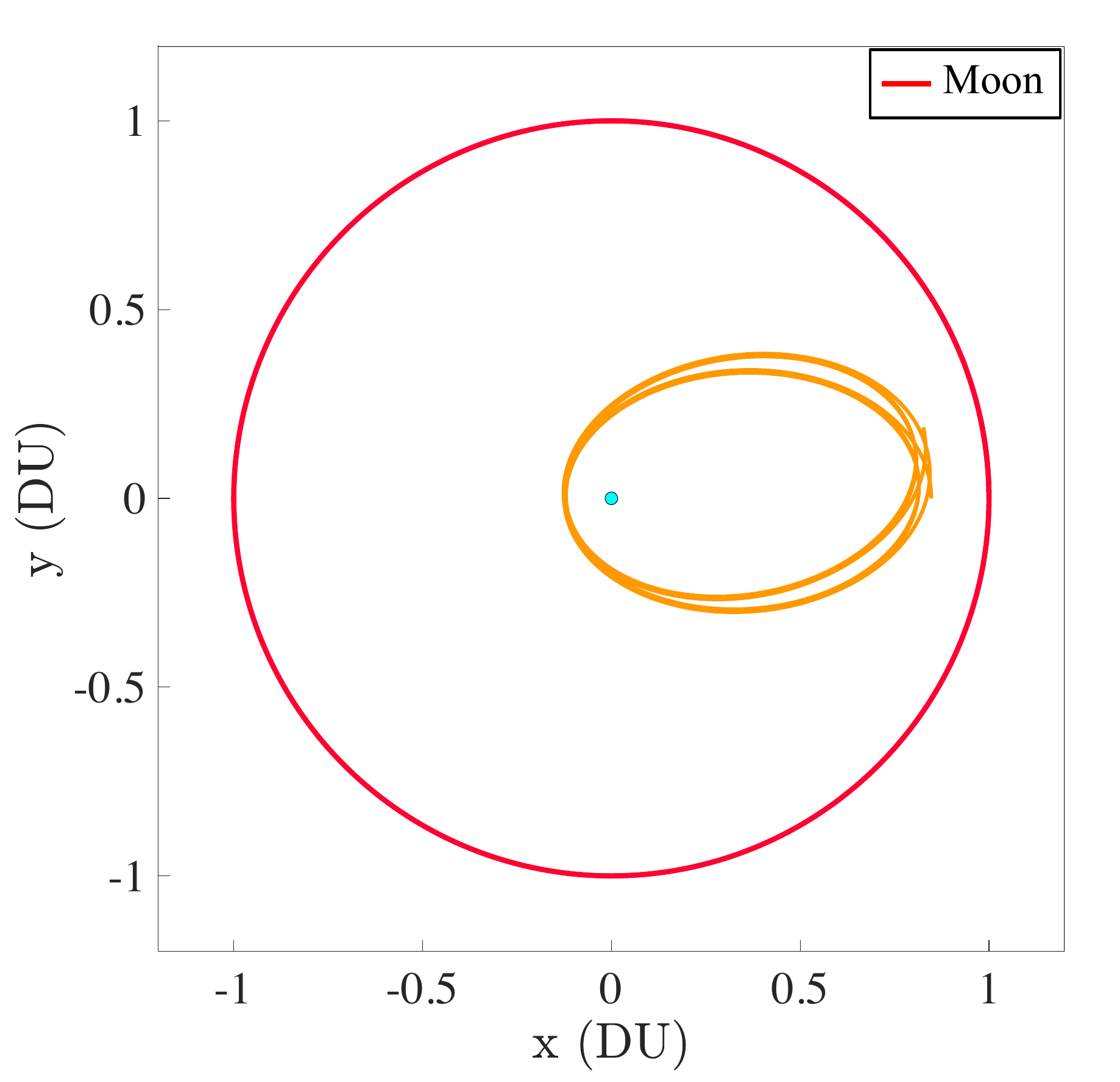}
\caption{ \label{fig:31orbit} Example 3:1 unstable resonant orbit, $C=3.05$ and period $6.3952 \approx 2.0357\pi$ TU, propagated for two periods in rotating (left) and inertial (right) frames.}
\end{centering}
\vspace{-12pt}
\end{figure}

Similar to the 4:1 case described earlier, we discovered that the prograde and retrograde unstable orbits are again part of a single continuous family, joined together by orbits that pass very close to and then through the Earth singularity. The same is also true of the stable prograde and retrograde orbits. Additionally, retrograde orbits again occur at lower Jacobi constants and higher energies than prograde orbits. The 3:1 orbits also exist across a wide range of eccentricities, from near-circular to highly eccentric, as shown in the example orbit plots in the top right of Figure~\ref{fig:bifurcation31}. An example of a moderately eccentric 3:1 unstable orbit propagated for twice its period is displayed in Figure \ref{fig:31orbit}. 
Once again, we observe that 3:1 unstable orbits never reach or exceed the Moon's orbit. Even the highest-eccentricity 3:1 unstable orbit shown in the top right of Figure~\ref{fig:bifurcation31} has an apogee well below the Moon (again shown in that figure as a red circle at $x=1-\mu \approx 1$). 

\subsection{Periodic Orbit Families for 2:1 MMR} \label{21section}

\begin{figure*}[!t]
\begin{centering}
\includegraphics[width=0.999\textwidth]{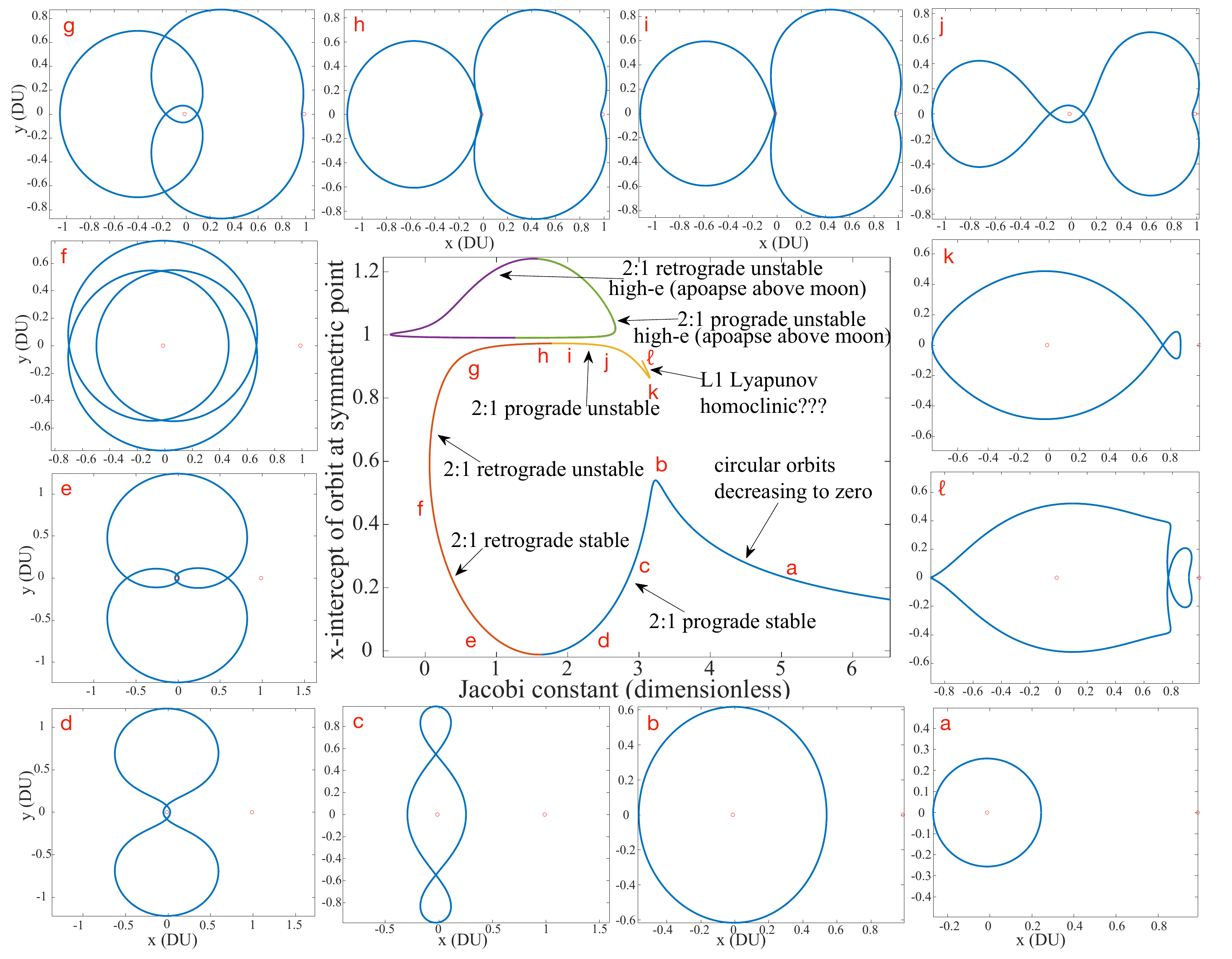}
\caption{ \label{fig:bifurcation21} Bifurcation diagram for orbit family containing 2:1 periodic orbits with example orbit plots. {Displayed orbits a--f are stable, g--$\ell$ are unstable.} Change of curve color in bifurcation diagram indicates Earth collision. Earth and Moon shown as red circles in orbit plots.}
\end{centering}
\vspace{-12pt}
\end{figure*}

Whereas the 4:1 and 3:1 orbit families possess many common characteristics, the 2:1 resonant periodic orbit family demonstrates significant distinctions within the Earth–Moon PCR3BP.
A bifurcation diagram of the perpendicular $x$-axis crossing value versus the Jacobi constant $C$ for the periodic orbit families containing the 2:1 orbits is presented in Figure~\ref{fig:bifurcation21}. This diagram is constructed similarly to those provided earlier for the 4:1 and 3:1 families.

First of all, we discuss the orbits with apogee below the Moon ($x$-intercept below $1-\mu$ in Figure~\ref{fig:bifurcation21}), examples of which are labeled with letters a-$l$ and plotted in the figure. This orbit family actually begins as a set of \emph{non-resonant} stable circular orbits around Earth, characterized by a vanishingly small radius as the Jacobi constant approaches infinity; one such circular orbit is orbit (a) in Figure~\ref{fig:bifurcation21}. These circular orbits then \emph{continuously} morph into the 2:1 prograde stable resonant periodic orbits, represented by orbits b, c, and d, listed in order of increasing eccentricity and decreasing perigee. These stable prograde orbits then pass through the singularity at Earth and become stable \emph{retrograde} 2:1 orbits, as also occurred in the 4:1 and 3:1 cases; examples of this are retrograde stable orbits e and f. These retrograde stable orbits then continuously join with retrograde unstable 2:1 orbits, exemplified by g and h, after passing through a fold bifurcation corresponding to a minimum in Jacobi $C$. The retrograde unstable orbits' perigee then passes through the singularity at Earth again, with the orbits continuously joining with the prograde unstable 2:1 orbits exemplified by orbits i, j, and k in order of decreasing eccentricity. These orbits eventually encounter another fold bifurcation point where $C$ reaches a local maximum, after which the unstable prograde 2:1 orbits become briefly stable, and then revert to instability through a pitchfork bifurcation---see \citet{kumar2025aug} for full details on these bifurcations and the asymmetric orbit families that emerge from the pitchfork.
Finally, these last symmetric 2:1 orbits continuously join with a family of \emph{non-resonant} unstable orbits exemplified by orbit $\ell$. This last orbit visually seems to be generated by a homoclinic trajectory to an $L_1$ planar Lyapunov orbit, although further investigation is needed to confirm this. 

Immediately, some key differences with 3:1 and 4:1 orbit families become apparent, First of all, focusing on the chain of orbits discussed above, the stable prograde and unstable prograde 2:1 resonant periodic orbit families do not end by joining each other at a fold bifurcation point, as occurred in the 4:1 case. Instead, they continuously connect to families of non-resonant orbits. In the 4:1 case, this fold bifurcation point corresponded to a near-circular, low-eccentricity orbit. In contrast, the 2:1 case reveals a significant vertical ``gap'' between orbits b and k. 
This gap appears where one otherwise might have expected to find low-eccentricity 2:1 unstable prograde orbits, which we believe do not exist in the Earth--Moon system. Indeed, we also tried continuation by $\mu$ of several low-eccentricity 2:1 prograde unstable resonant periodic orbits from the $\mu=0$ rotating Kepler problem; however, all of these attempts encountered fold bifurcations, where $\mu$ would reach a maximum along the curve of continuation solutions and then would start decreasing seemingly to negative infinity, thus never reaching the Moon's $\mu$ value. Other non-continuation based methods of searching for periodic orbits directly in the PCR3BP, e.g., that of \citet{broucke1968}, also failed to locate any low-eccentricity 2:1 prograde unstable orbits. 

\begin{figure} [t!]
\centering
\includegraphics[width=0.49\columnwidth]{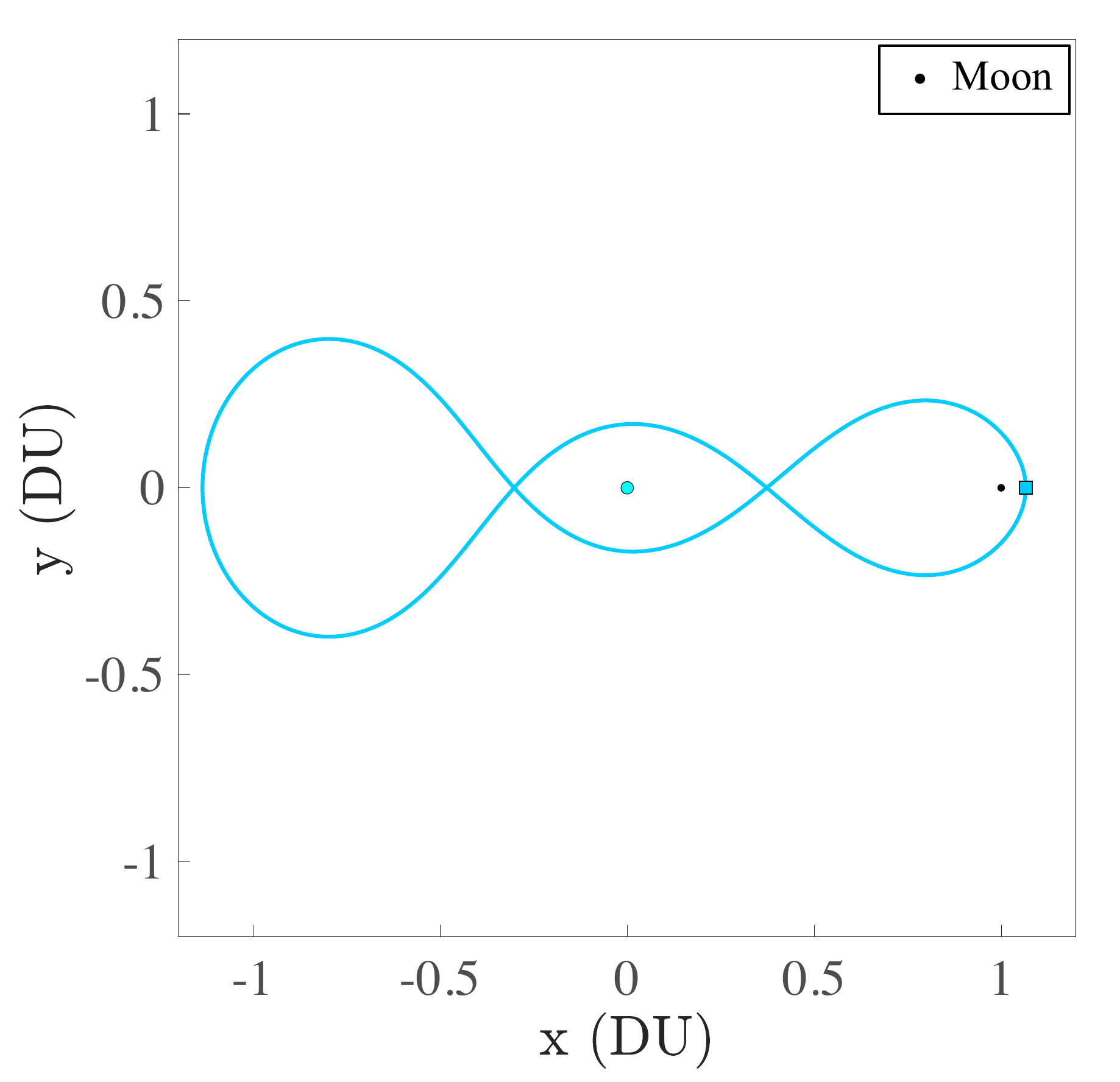}
\includegraphics[width=0.49\columnwidth]{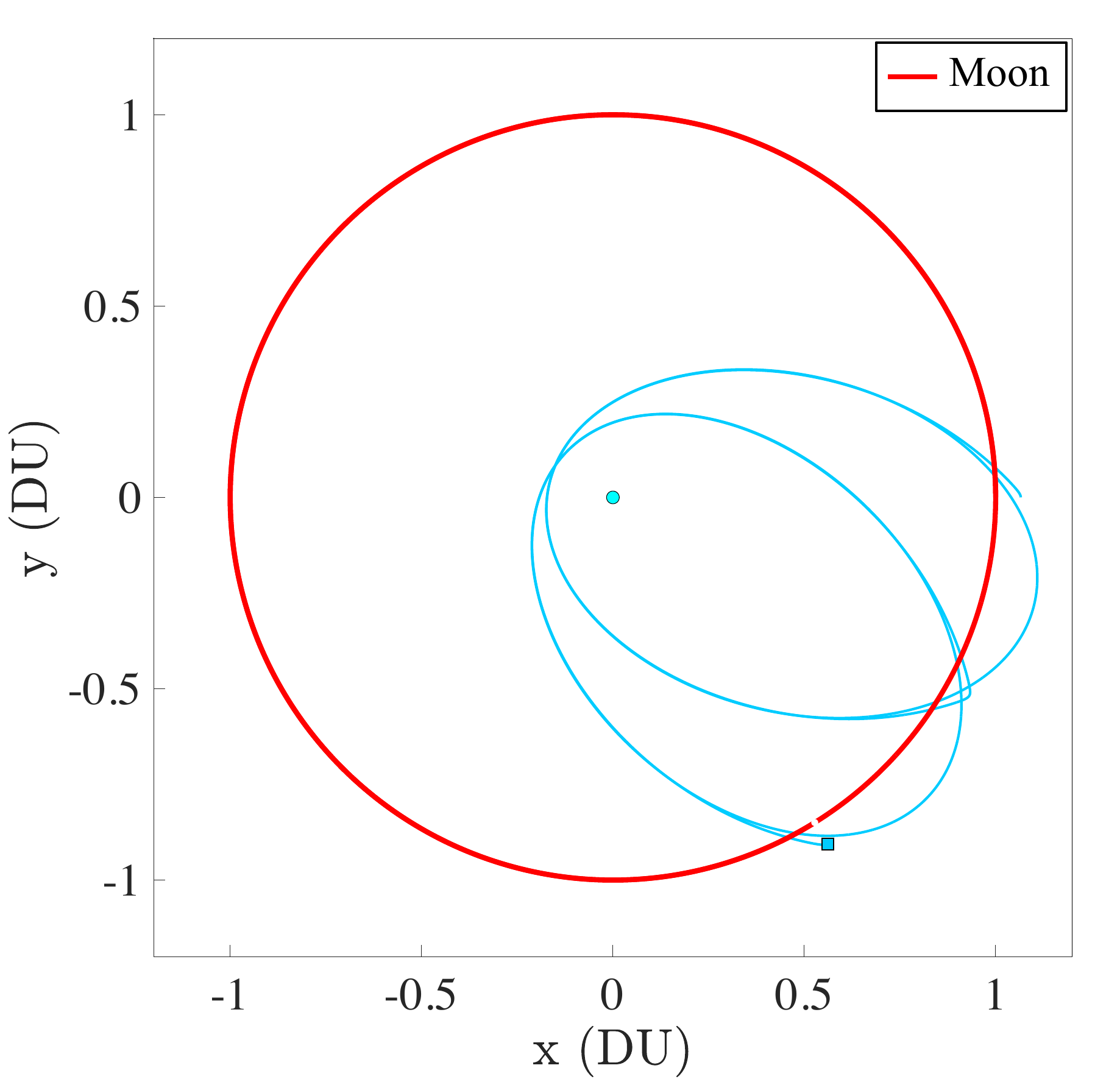}
\caption{ \label{fig:highe21} 2:1 prograde unstable orbit with apogee above the Moon, $C=2.628$ and period $5.7755 \approx 1.8384\pi$ TU.}
\end{figure}

Another significant difference between the 2:1 resonant periodic orbits and the previously discussed 3:1 and 4:1 families is that there exist 2:1 orbits with apogees extending beyond the Moon. These orbits correspond to the prograde and retrograde unstable orbit family curves in Figure \ref{fig:bifurcation21} that have $x$-intercept greater than $1-\mu$; they only occur for relatively high energies (low Jacobi constants), having a maximum Jacobi constant of 2.670 along the family as compared to 3.152 for the apogee-below-moon 2:1 unstable orbits. An example of such an orbit with apogee above the Moon is shown in Figure \ref{fig:highe21}. 
For these orbits with apogee beyond the Moon, the retrograde and prograde orbit families again join continuously with each other after passing through the singularity at Earth.

\begin{figure}[t!]
\begin{centering}
\includegraphics[width=0.49\columnwidth]{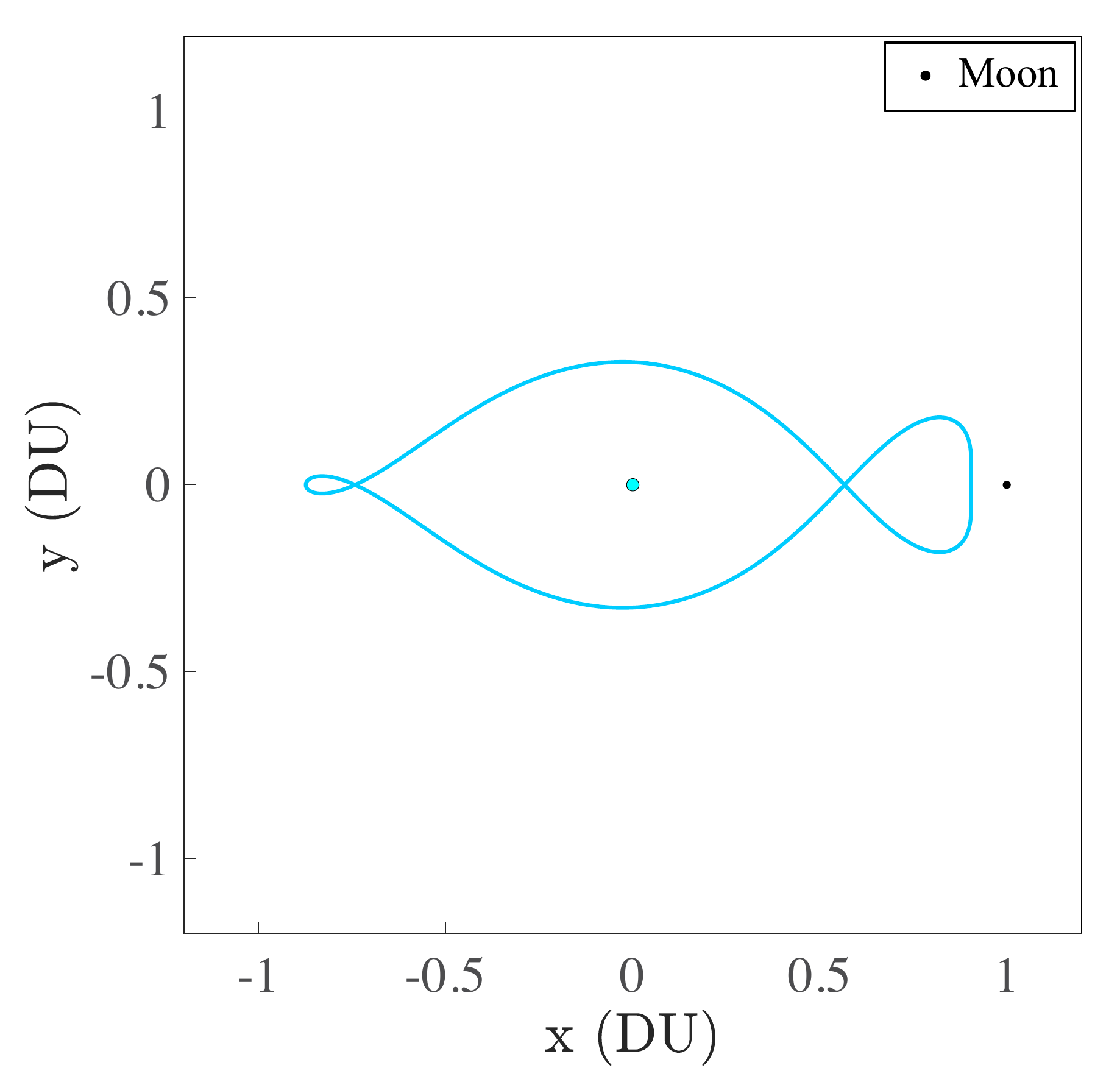}
\includegraphics[width=0.49\columnwidth]{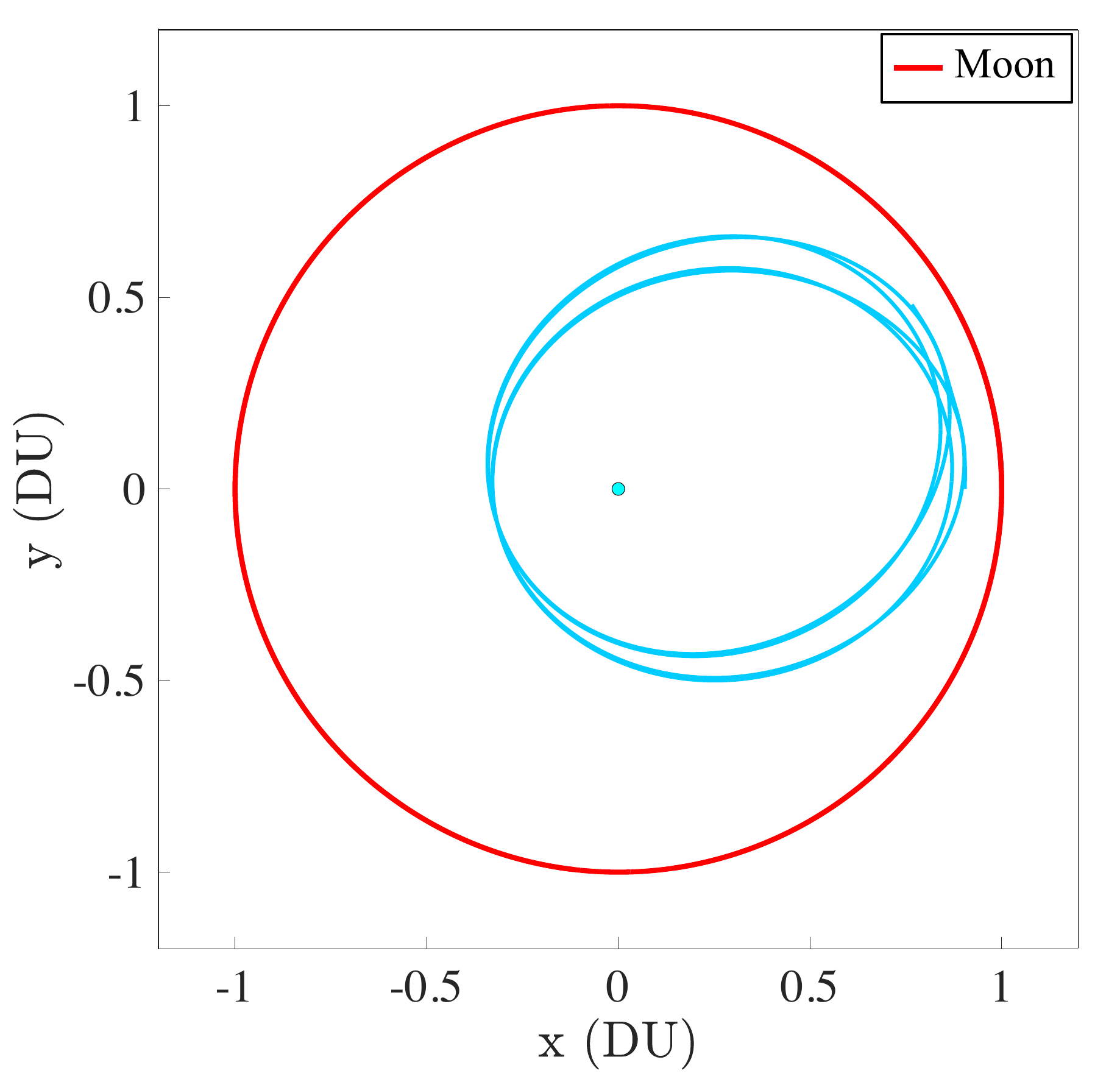}
\caption{ \label{fig:21orbit} Example 2:1 unstable resonant orbit, $C=3.05$ and period $6.5636 \approx 2.0893\pi$ TU, propagated for two periods in rotating (left) and inertial (right) frames.}
\end{centering}
\vspace{-15pt}
\end{figure}

As a final example, a 2:1 prograde unstable orbit of moderate eccentricity propagated for twice its period is shown in Figure \ref{fig:21orbit}. 
We would like to note that the continuous family of orbits joining orbits a through $\ell$ together in Figure \ref{fig:bifurcation21} was computed by \citet{broucke1968} as well, where he refers to it as as ``Family BD''. However, in that work, these orbits are not identified as belonging to the 2:1 MMR, nor are the unstable 2:1 orbits with apogee above the Moon found either. 

\section{Resonance Overlap and Natural Transport}
\label{Transport}

With the families of 4:1, 3:1, and 2:1 resonant periodic orbits computed and characterized, we are now ready to compute and visualize the orbits' stable and unstable manifolds as well as heteroclinics between them. The methodology described in Sections \ref{paramSection}--\ref{heteroComputeSection} is used for this purpose. As a reminder, we plot all manifolds on the 2D fixed-energy perigee Poincar\'e sections $\Sigma_C = \Sigma \cap \mathcal{M}_C$ using osculating orbital elements $(g,a)$ 
for visualization, where  $g$ is the longitude of perigee relative to the synodic frame $x$-axis and 
$a$ is the semi-major axis. We will compute manifolds and plot them on the perigee section for a variety of Jacobi constants, starting with a discussion of the 3:1 and 2:1 MMRs and then investigating the 4:1 resonance as well. From this point onward, we will analyze only prograde orbits. 

\subsection{Stable/Unstable Manifolds of Very Low-Energy 3:1 Orbits} \label{veryLowEnergySection}


As noted in Section~\ref{21section}, 2:1 unstable prograde resonant orbits occur only at higher eccentricities, corresponding to higher energies and lower Jacobi constants $C$. In contrast, 3:1 and 4:1 unstable prograde orbits persist across a broad range of eccentricities. For example, 3:1 unstable orbits are found even at $C = 3.45$, whereas the lowest-energy 2:1 unstable orbit appears only at $C = 3.1518$. Consequently, there exists a substantial interval of energy values in which 3:1 (and 4:1) unstable orbits are present, while 2:1 unstable orbits are absent.

\begin{figure}[b!]
    \begin{centering}
    \includegraphics[width=0.49\columnwidth]{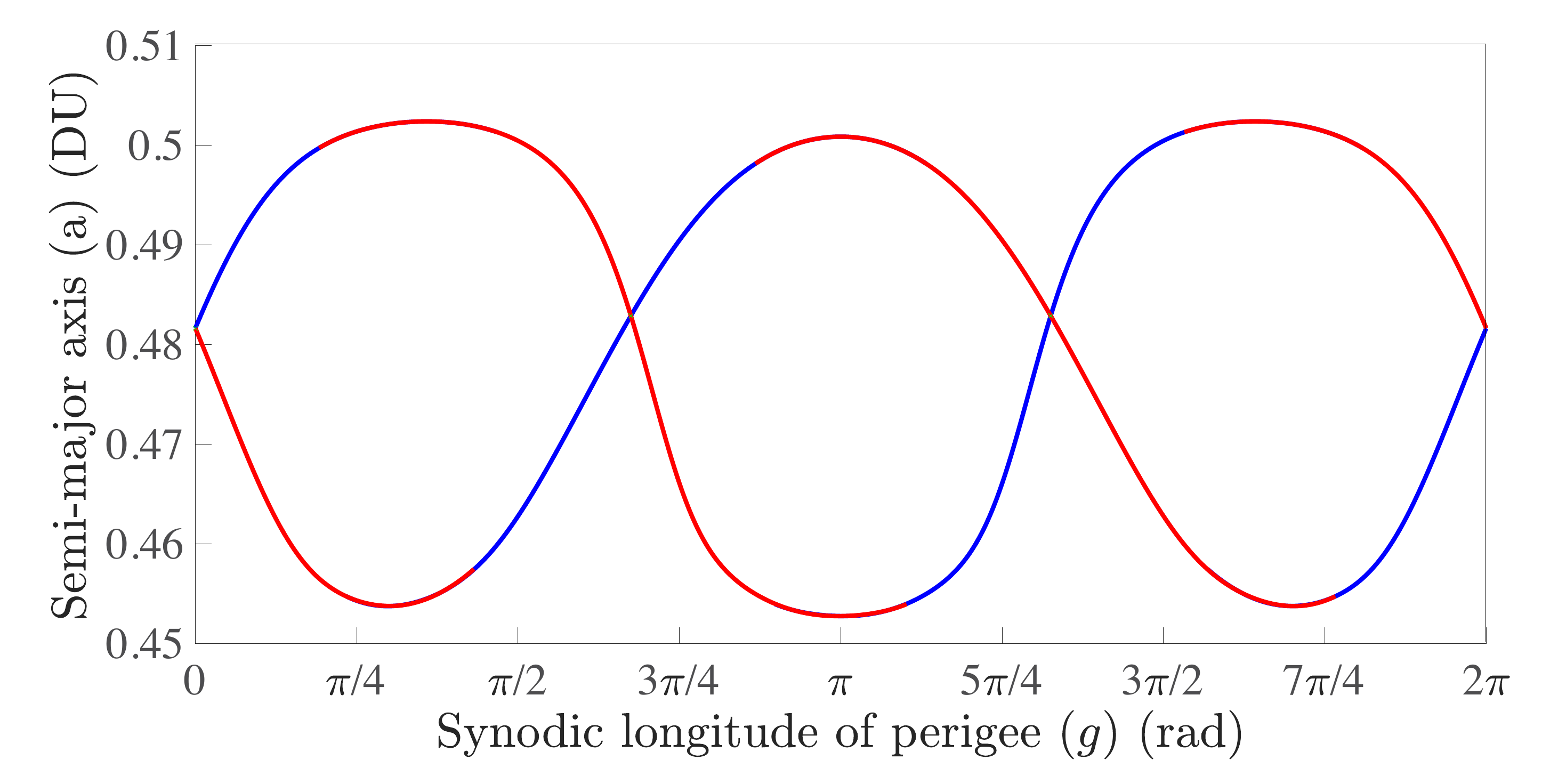} \\
    \includegraphics[width=0.49\columnwidth]{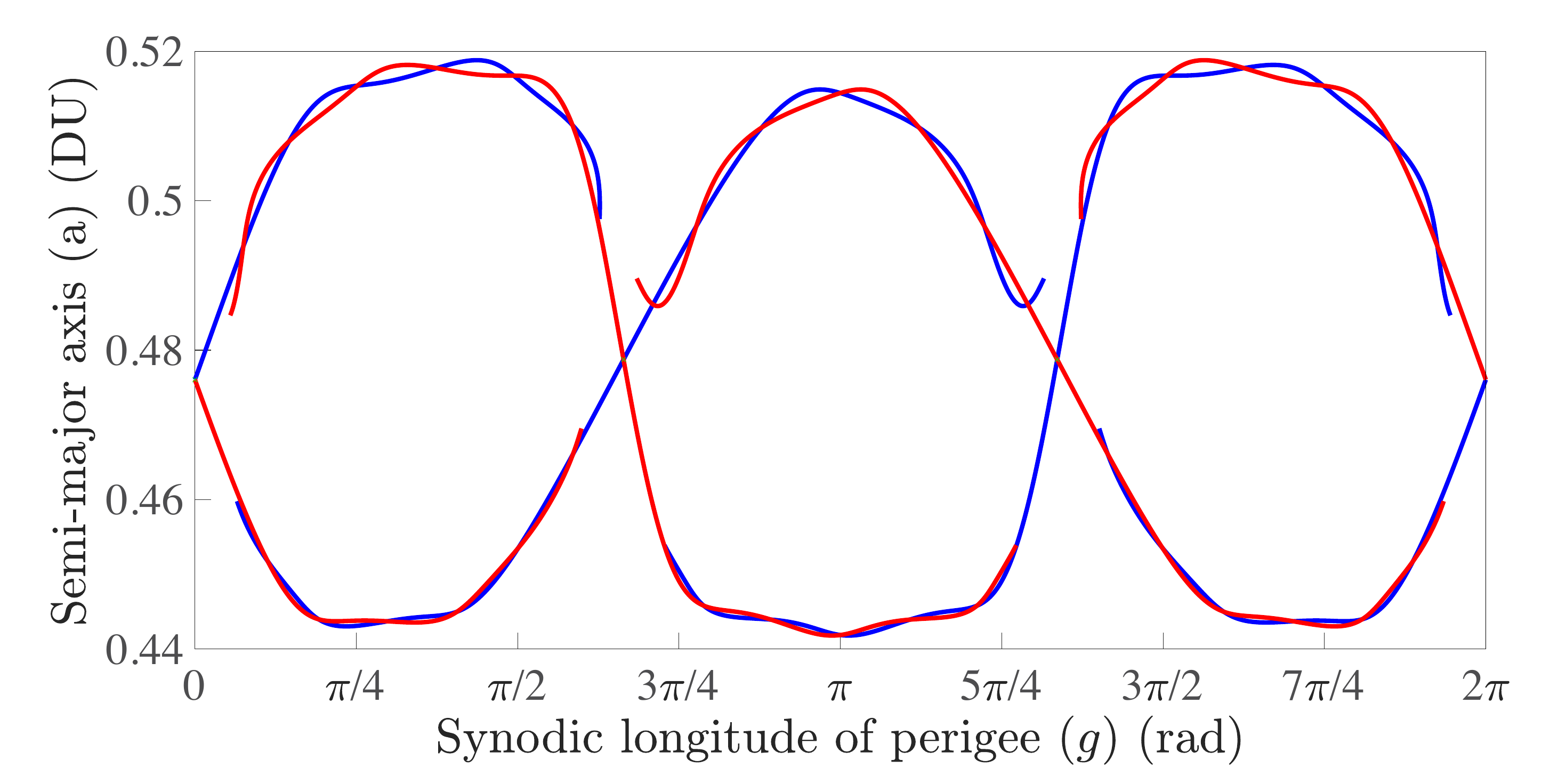} \\
    \includegraphics[width=0.49\columnwidth]{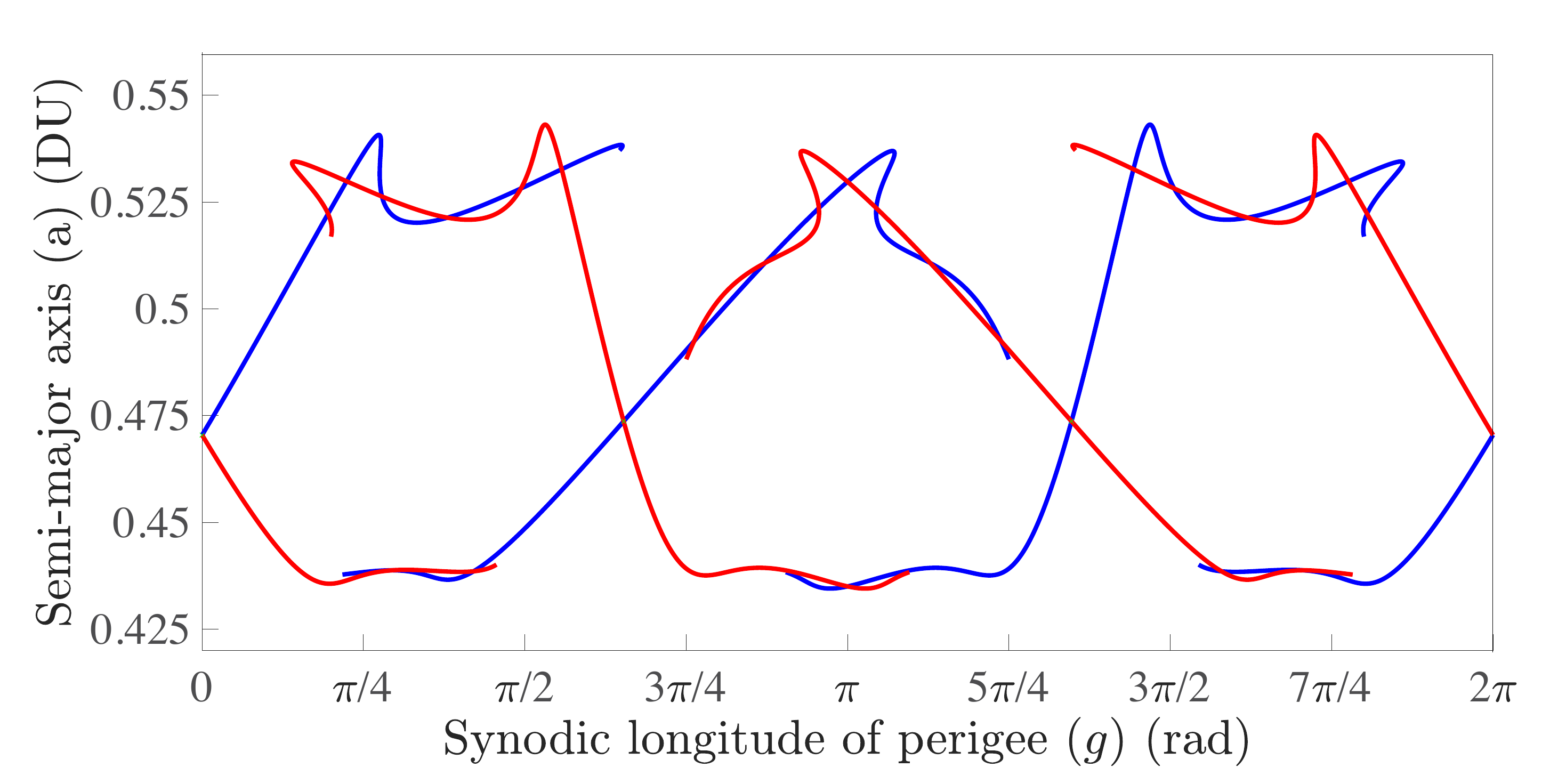}
    \caption{ 
    \label{fig:31manifolds} 
    3:1 stable/unstable manifolds (blue/red), perigee section, from top to bottom: $C = 3.363, 3.251, 3.158$.
    }
    \end{centering}
\end{figure}


Consider first the range of Jacobi constants $C \in [3.16, 3.45]$, spanning the emergence of the 3:1 unstable orbits through the onset of the 2:1 unstable orbits. The computed manifolds, projected onto our perigee sections $\Sigma_C$ (Figure~\ref{fig:31manifolds}), show that for the higher values of $C$ in this interval (top and middle panels), the structures closely resemble the separatrices of mathematical pendula. This behavior is consistent with perturbation theory \citep{morbyBook}, which predicts pendulum-like dynamics in regions dominated by a single resonance without overlap from others. Even as the system approaches the appearance of the 2:1 orbits at $C = 3.152$, the 3:1 manifolds largely preserve this pendulum-like character. The principal change is an increase in the density and transversality of the homoclinic intersections between stable and unstable branches of the 3:1 manifolds, as illustrated in the bottom panel of Figure~\ref{fig:31manifolds} for $C = 3.158$. At this value, the top lobe of the 3:1 orbit manifolds extends noticeably farther in 
semi-major axis than the bottom lobe.

\subsection{Heteroclinics Between 3:1 and 2:1 Orbits} \label{heteroclinicSection}
 
\begin{figure}[t!]
    \begin{centering}
    \includegraphics[width=0.495\textwidth]{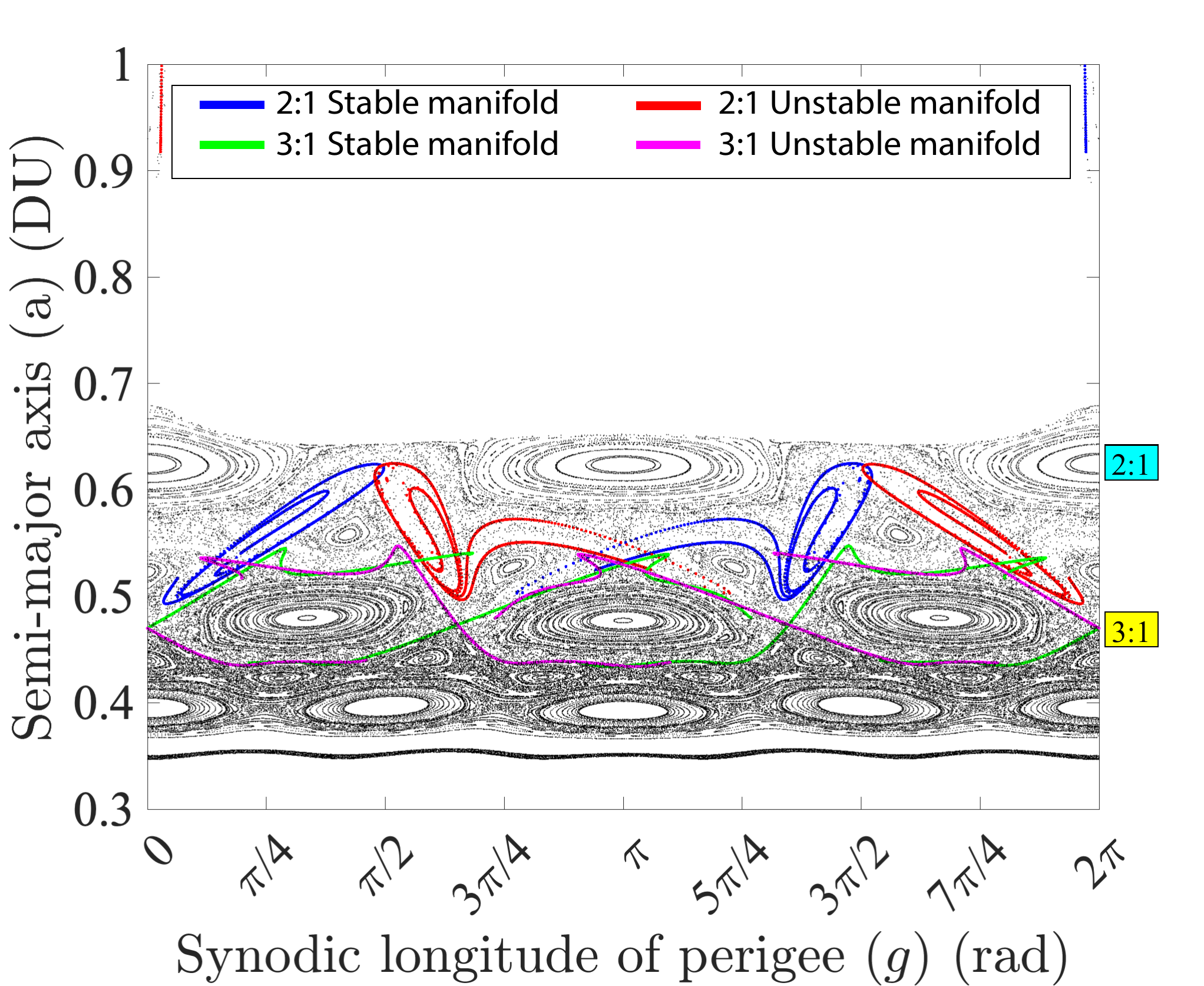}
    \includegraphics[width=0.495\textwidth]{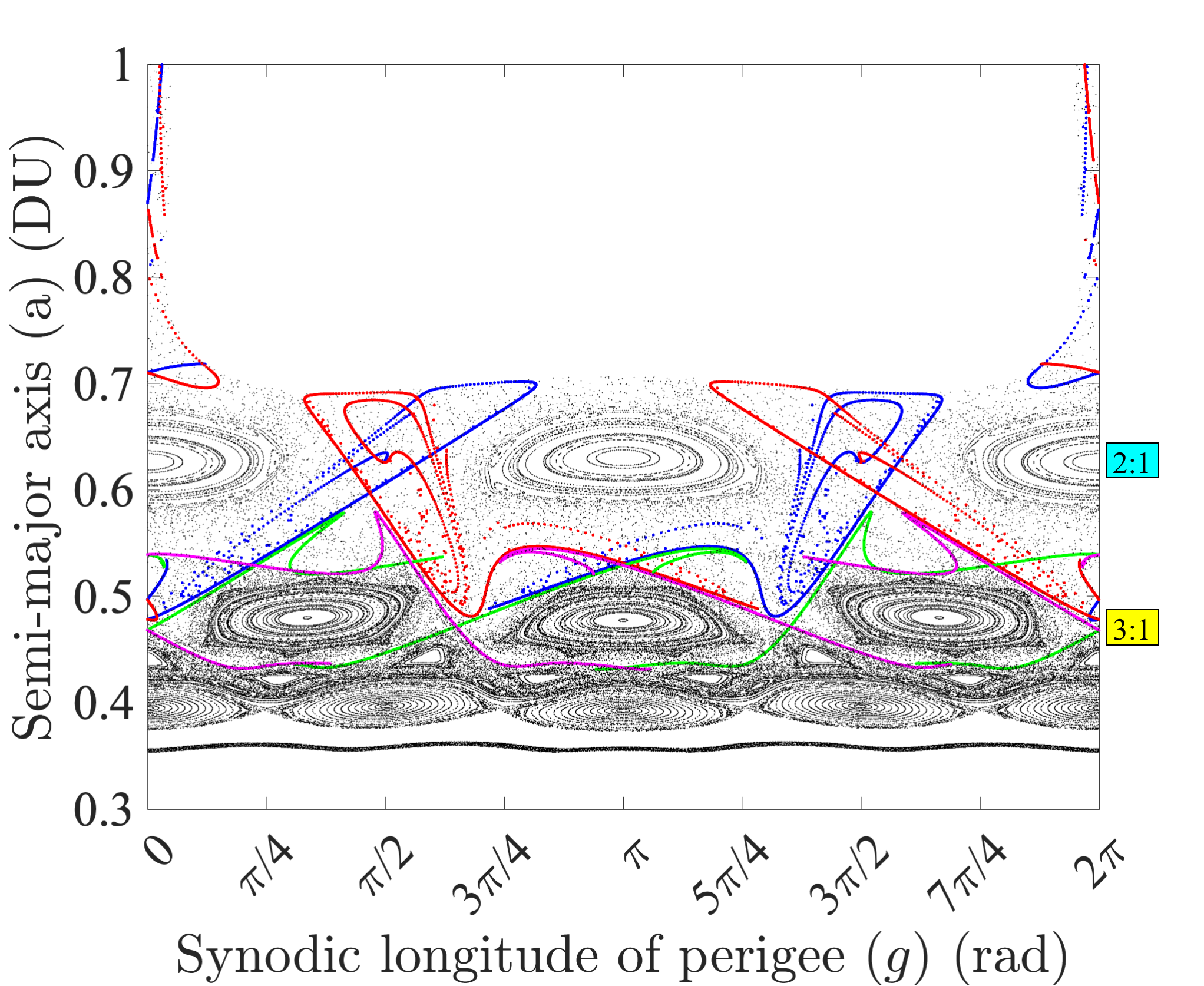}
    \includegraphics[width=0.495\textwidth]{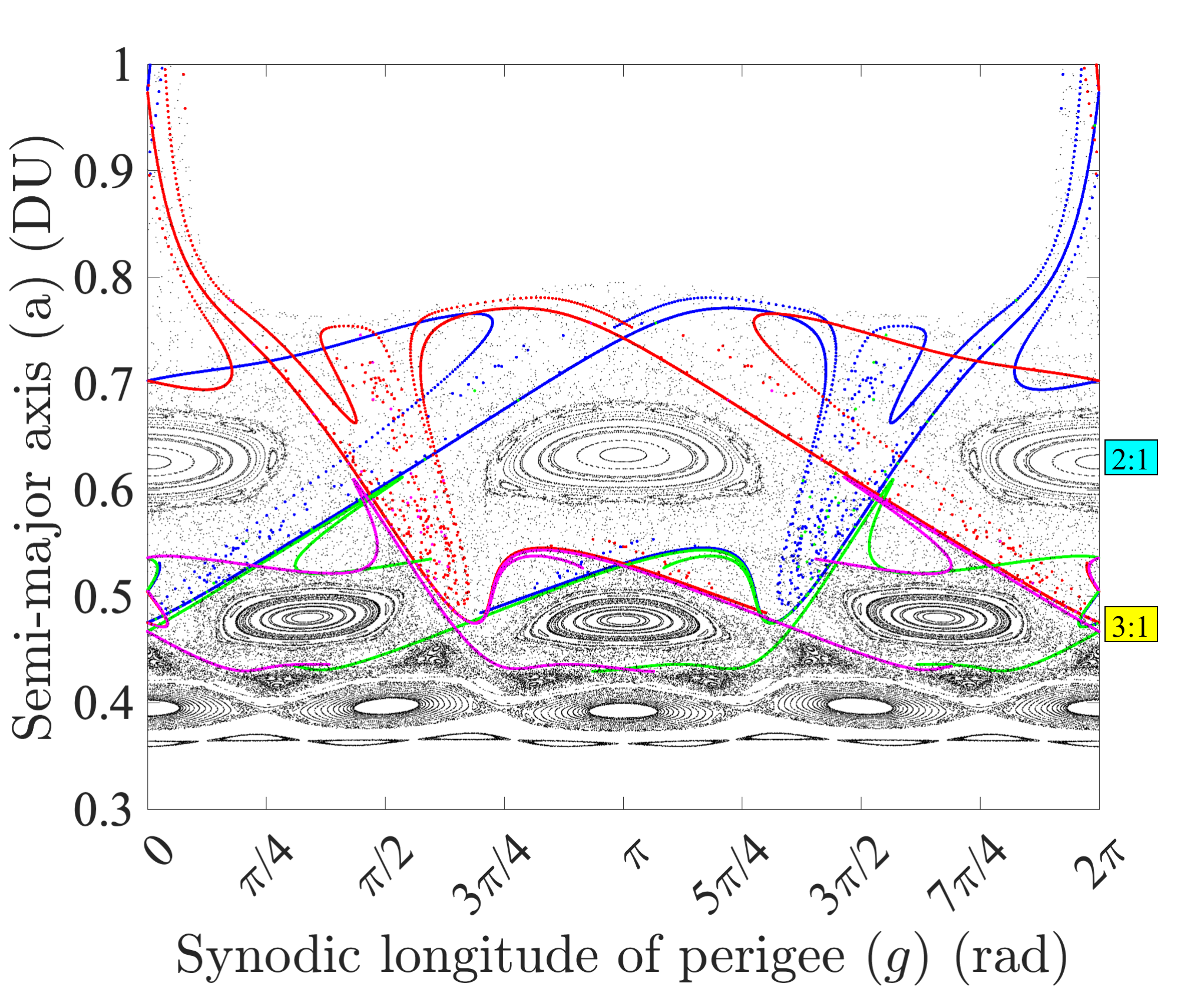}
    \includegraphics[width=0.495\textwidth]{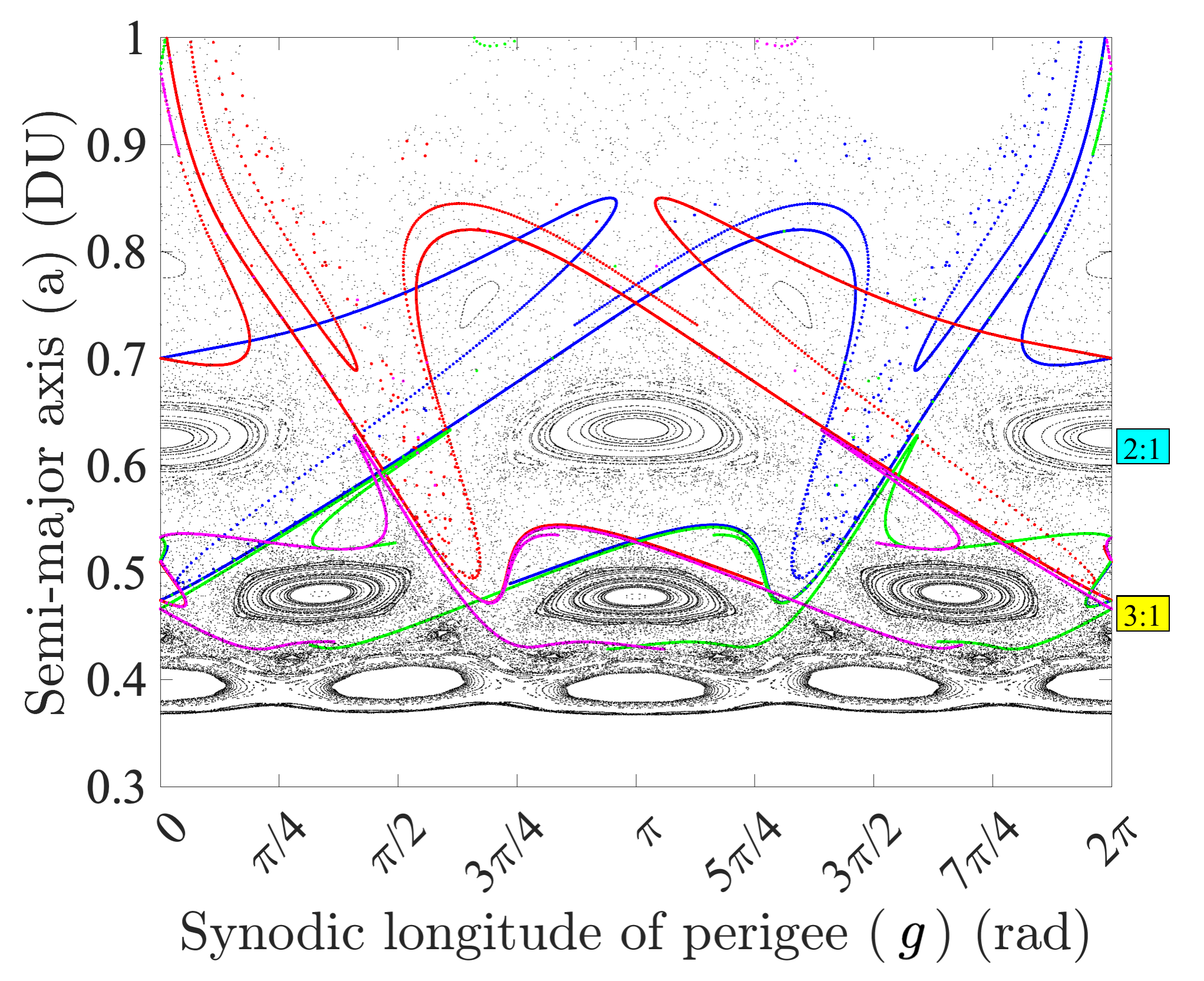}
    \caption{ \label{fig:3121manifolds} 3:1 (green/magenta) and 2:1 (blue/red) orbit stable/unstable manifolds on perigee Poincar\'e sections along with other propagated trajectories in gray, for Jacobi constants $C = 3.15$ (top left), $3.10$ (top right), $3.05$ (bottom left), $3.00$ (bottom right).}
    \end{centering}
\end{figure}

We now turn to the range of Jacobi constants for which both 2:1 and 3:1 unstable periodic orbits coexist, namely for $C \leq 3.1518$. Figure~\ref{fig:3121manifolds} shows the corresponding stable/unstable manifolds of the 3:1 and 2:1 orbits, plotted in green/magenta and blue/red, respectively, on our perigee Poincar\'e sections $\Sigma_C$ for $C = 3.15, 3.10, 3.05,$ and $3.00$. These cases remain within the ``low-energy'' regime of lunar trajectory design \citep{parkerAnderson}, though they are at higher energies than the orbits examined in Section~\ref{veryLowEnergySection}. The gray background points in Figure~\ref{fig:3121manifolds} correspond to non-manifold trajectories, obtained by propagating a large ensemble of initial conditions to our section, analogous to the earlier Figure~\ref{fig:poincareMap310}. These points provide additional context for the regions of phase space through which the manifolds drive motion.
 
In particular, once the 2:1 unstable orbits emerge for $C$ just above 3.15, heteroclinic intersections between the manifolds of the 3:1 and 2:1 orbits are immediately generated. This can be seen in the top left panel of Figure~\ref{fig:3121manifolds}, corresponding to $C = 3.15$. Such behavior contrasts with the more standard resonance-overlap dynamics observed in systems with smaller mass parameters $\mu$, such as Jupiter--Europa, where a range of low energies (high $C$ values) exists in which two unstable resonant orbits coexist without any mutual overlap. We suspect that, in the present case, the 2:1 resonance lies in a regime where standard predictions from Hamiltonian perturbation theory begin to lose validity. A further observation is that in both top-row plots of Figure~\ref{fig:3121manifolds}, at $C = 3.15$ and $3.10$, the upper branches of the 2:1 manifolds (above the resonance center at $a \approx 0.63$) 
deviate significantly from the pendulum-like separatrix structure. In fact, at $C = 3.15$, the upper stable and unstable branches of the 2:1 manifolds intersect only at the periodic orbit points themselves and at two other  points just above them---very much unlike the upper branch intersections seen along the $g=\pi$ line for $C=3.05$ and 3.00. As will be discussed in Section~\ref{resToL1Section}, this unexpected behavior for $C = 3.10$ and $3.15$ is tied to tube dynamics induced by the manifolds of planar Lyapunov orbits.


For all $C \leq 3.15$, heteroclinic intersections arise between the stable and unstable manifolds of the 3:1 and 2:1 orbits. At $C = 3.10$ and lower, several points from the 2:1 manifolds also escape toward $a = 1$, corresponding to the Moon’s semi-major axis; a smaller number of points do so at $C = 3.15$ as well, though transport toward the Moon generally strengthens as $C$ decreases (higher energies), as expected. For $C = 3.05$ and $3.00$, the upper branches of the 2:1 manifolds also develop a homoclinic intersection, rendering the manifolds more similar to the typical pendulum separatrix shape than at $C = 3.15$ or $3.10$. The overall structure of the manifolds and their heteroclinic connections does not change appreciably between $C = 3.05$ and $3.00$. Although we present Poincar\'e sections and heteroclinic intersections only for $3.00 \leq C \leq 3.15$ here, analogous 3:1–2:1 heteroclinic trajectories exist for $2.44 \leq C \leq 3.00$ as well. Representative examples will be shown in Section~\ref{heteroclinicTrajsSection}.


In summary, for $C \leq 3.15$, where 2:1 unstable orbits exist, zero-$\Delta v$ paths connect the 3:1 and 2:1 MMRs. For Jacobi constants below $C = 3.10$, a spacecraft in a 2:1 unstable resonant orbit can also reach the Moon with ease---a topic we examine further in Section~\ref{resToL1Section}. Thus, by inserting into the stable manifold of a sufficiently high-energy 3:1 orbit, a spacecraft can then surf heteroclinic connections and manifolds all the way to the Moon, saving substantial propellant compared to a direct injection.

\subsubsection{Examples of 3:1 to 2:1 Heteroclinic Trajectories} \label{heteroclinicTrajsSection}

As was discussed previously and shown in Figure \ref{fig:3121manifolds}, the stable and unstable manifolds of 3:1 and 2:1 unstable periodic orbits intersect each other when $C \leq 3.15$. For these energies, ballistic heteroclinic transfer trajectories must thus exist between 3:1 and 2:1 resonances, enabling a change in semi-major axis value without any deterministic $\Delta v$. 

Given their importance, we calculated these heteroclinic trajectories over a range of Jacobi constants, using the algorithm summarized in Section \ref{heteroComputeSection} to compute manifold intersections. 
Once an intersection point between stable and unstable manifolds is found, it is propagated backwards in time along the unstable manifold and forwards in time along the stable manifold until both propagations reach within a certain distance of their respective unstable periodic orbits. The distance here is defined by the metric of Section \ref{DistanceMetric}, with propagation stopped when this distance becomes less than some heuristically-chosen (small) tolerance. Though we will discuss heteroclinic trajectories \emph{from} 3:1 \emph{to} 2:1 orbits, note that each such transfer corresponds to a 2:1 to 3:1 heteroclinic as well, which can be calculated by applying the PCR3BP's time-reversal symmetry (see end of Section \ref{modelSection}) to any 3:1 to 2:1 transfer. 

\begin{figure}
\centering
\includegraphics[width=0.85\linewidth]{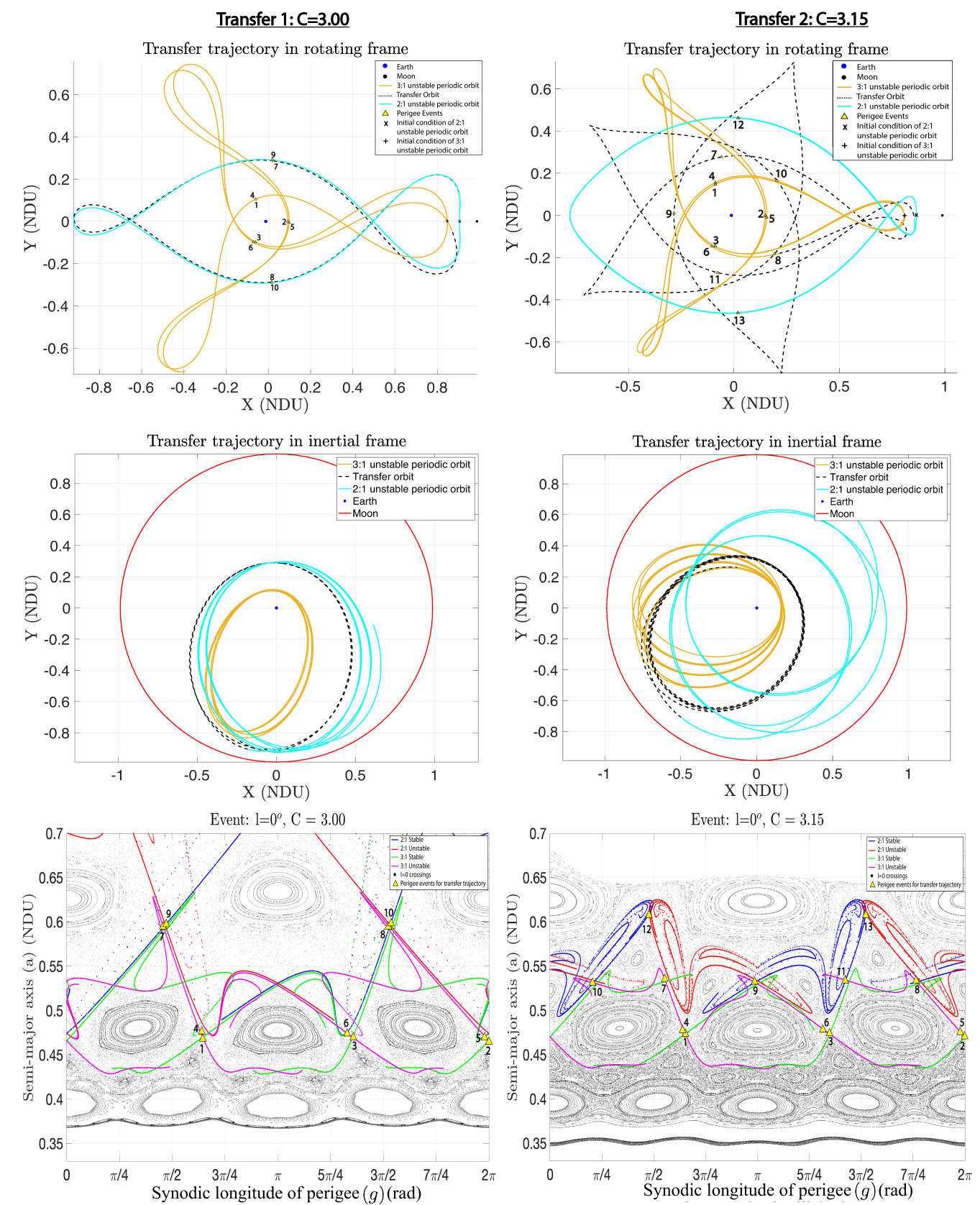}
\caption{\label{fig:transfer_comb} Two types of heteroclinic transfer trajectories between 3:1 and 2:1 resonances: Type 1 (short duration) for $C=3.00$ (left column) and type 2 (long duration) for $C=3.15$ (right column).}
\end{figure}

The selection of the intersection point of 3:1 unstable and 2:1 stable manifolds determines the type of transfer found, although in general we will seek transfers with the lowest time-of-flight (TOF). We examine two types of transfer trajectories: type 1 (short-duration) and type 2 (long-duration). An example of a type 1 transfer is given by intersection point 9 in the bottom left $C=3.00$ Poincar\'e section plot of Figure \ref{fig:transfer_comb}; 
this determines a direct, short-duration transfer between 3:1 and 2:1 resonances. This heteroclinic trajectory is plotted in both rotating and inertial reference frames on the left of Figure \ref{fig:transfer_comb} as well. However, such type 1 intersection points cease to exist for $C\geq3.09$ due to lower orbital energy, which results in reduced instability and consequently the disappearance of direct 3:1-2:1 transfers. Nevertheless, other transfers can still exist; an example of a type 2 transfer is given by heteroclinic intersection point 10 in the bottom right $C=3.15$ plot of Figure \ref{fig:transfer_comb}. It yields a longer-duration transfer that transitions from a 3:1 unstable resonant orbit to a 2:1 orbit via an intermediary 5:2 resonant segment, whose outline can clearly be seen in the rotating frame type 2 trajectory plot (the ``five-petaled flower'' shaped portion of the trajectory). An inertial frame plot of the type 2 transfer is also given in the right column of Figure \ref{fig:transfer_comb}. 

\begin{figure*}
\begin{centering}
\includegraphics[width=0.43\columnwidth]{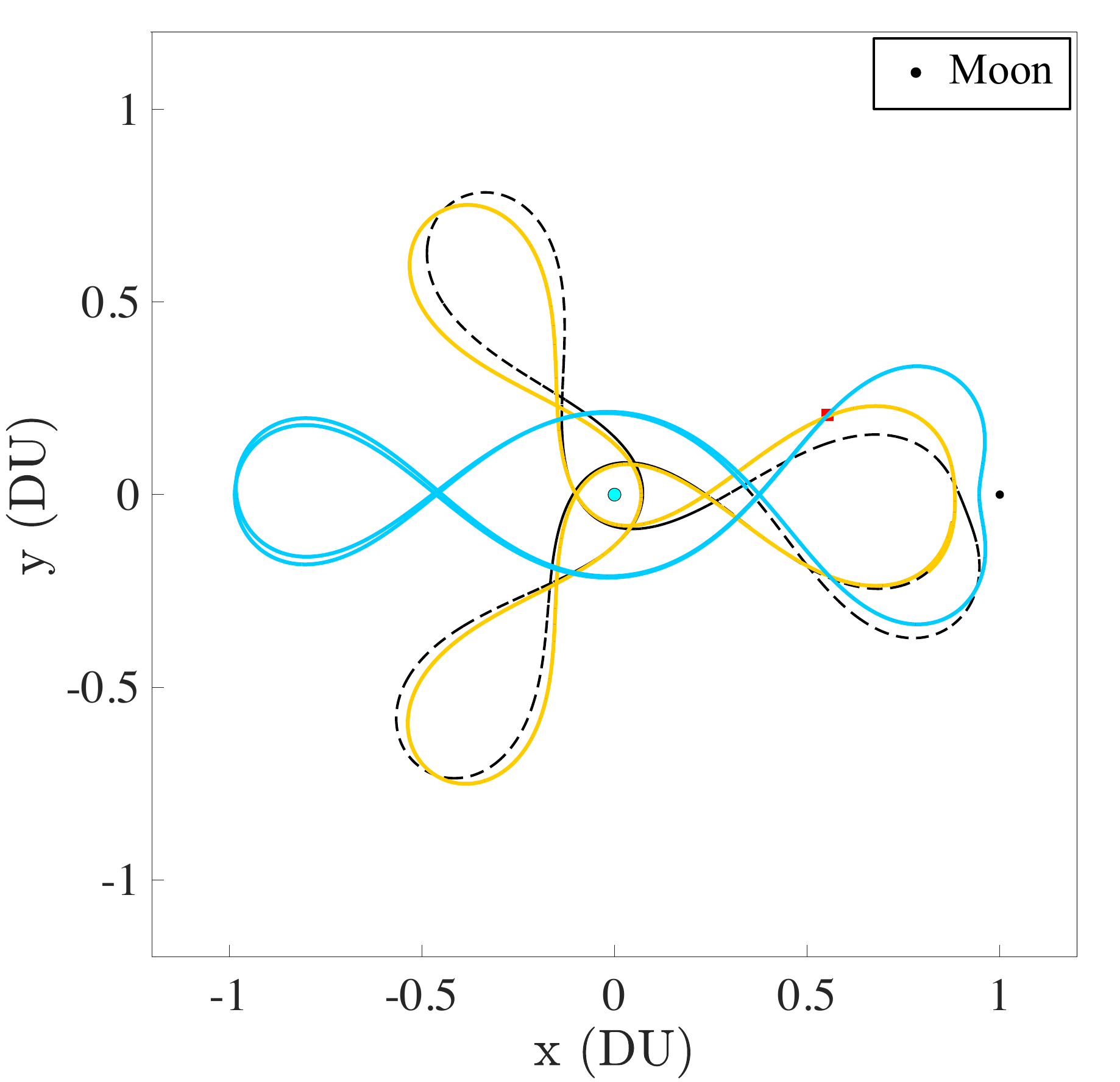}
\includegraphics[width=0.43\columnwidth]{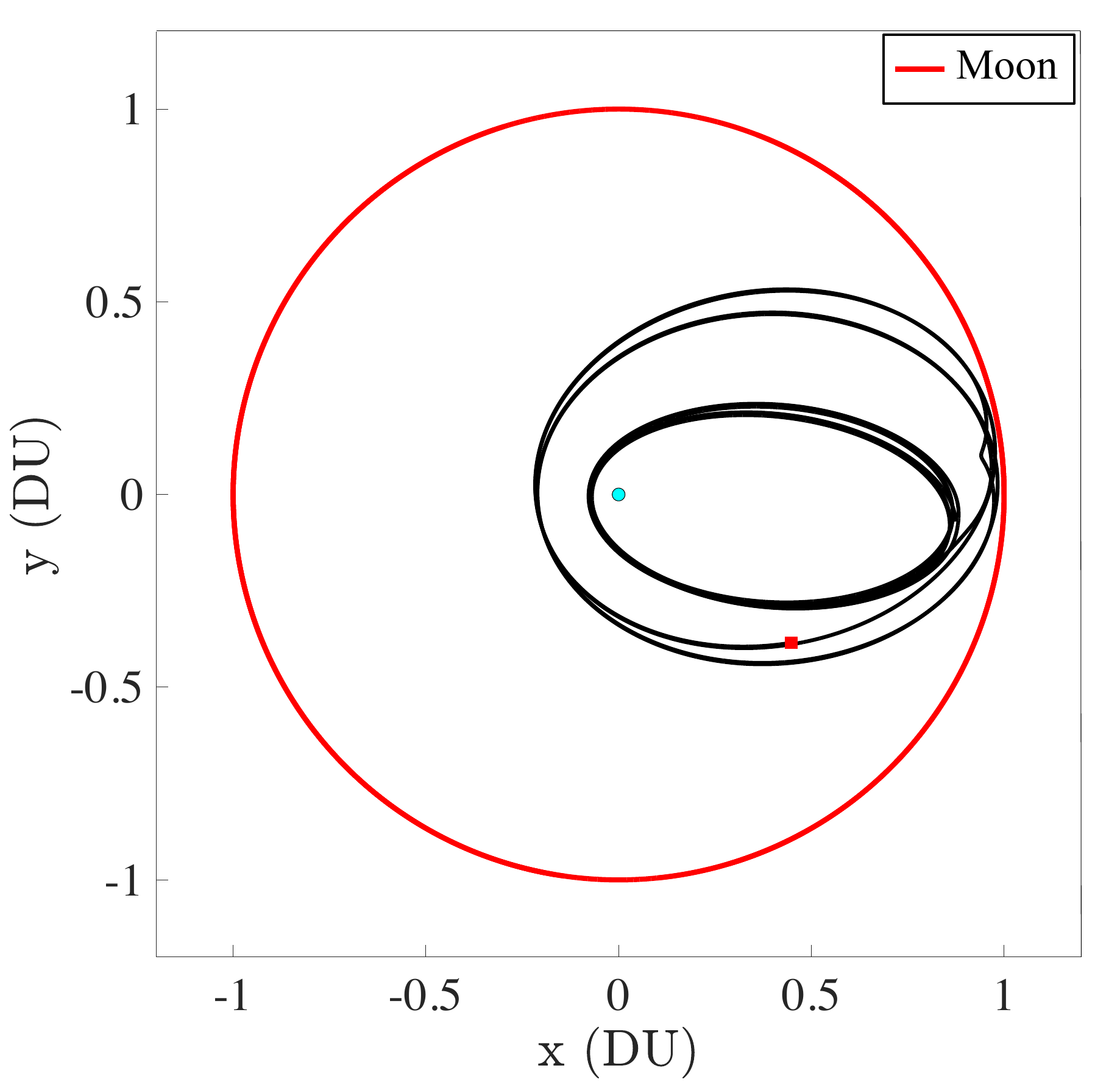}
\includegraphics[width=0.43\columnwidth]{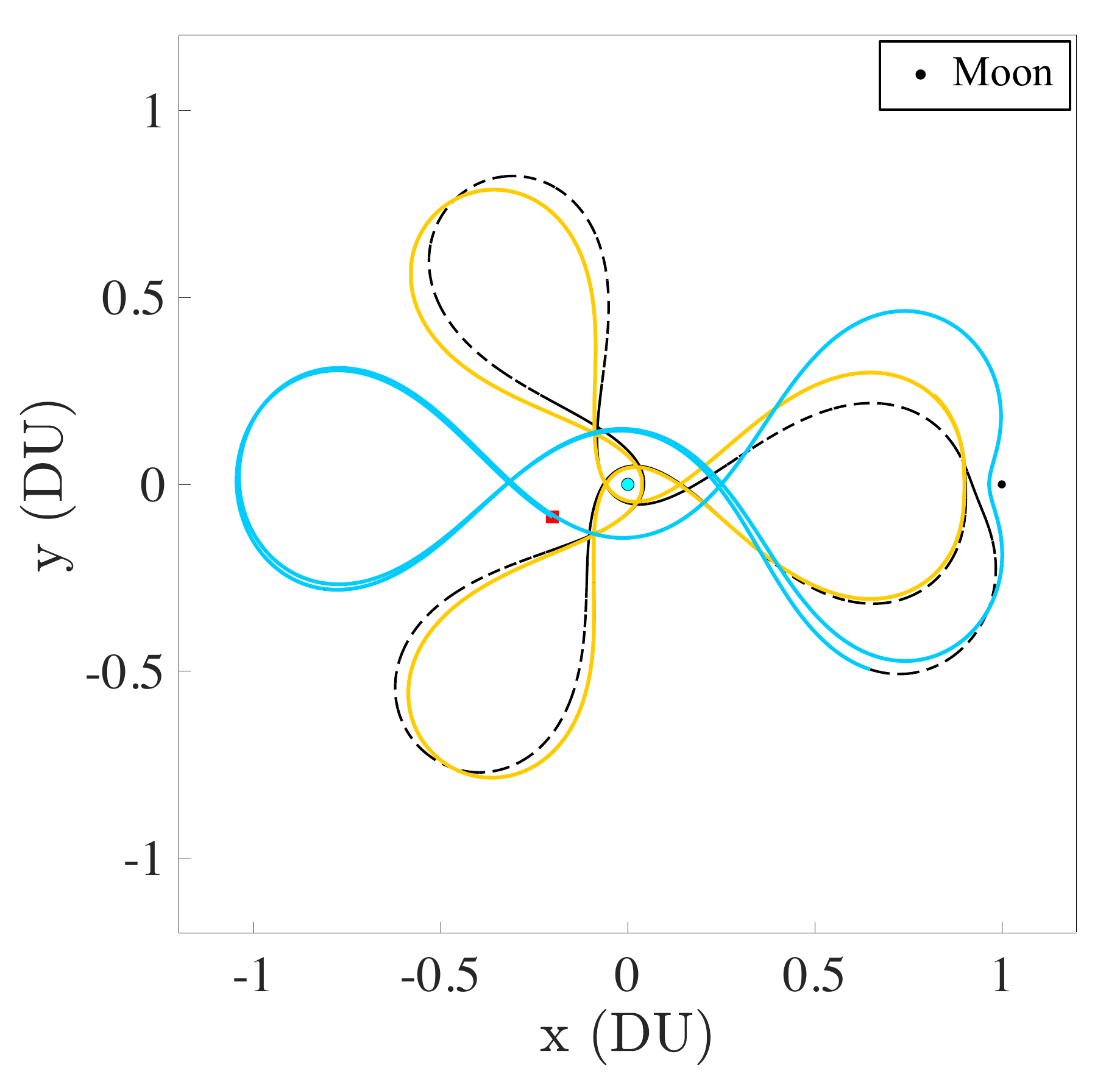}
\includegraphics[width=0.43\columnwidth]{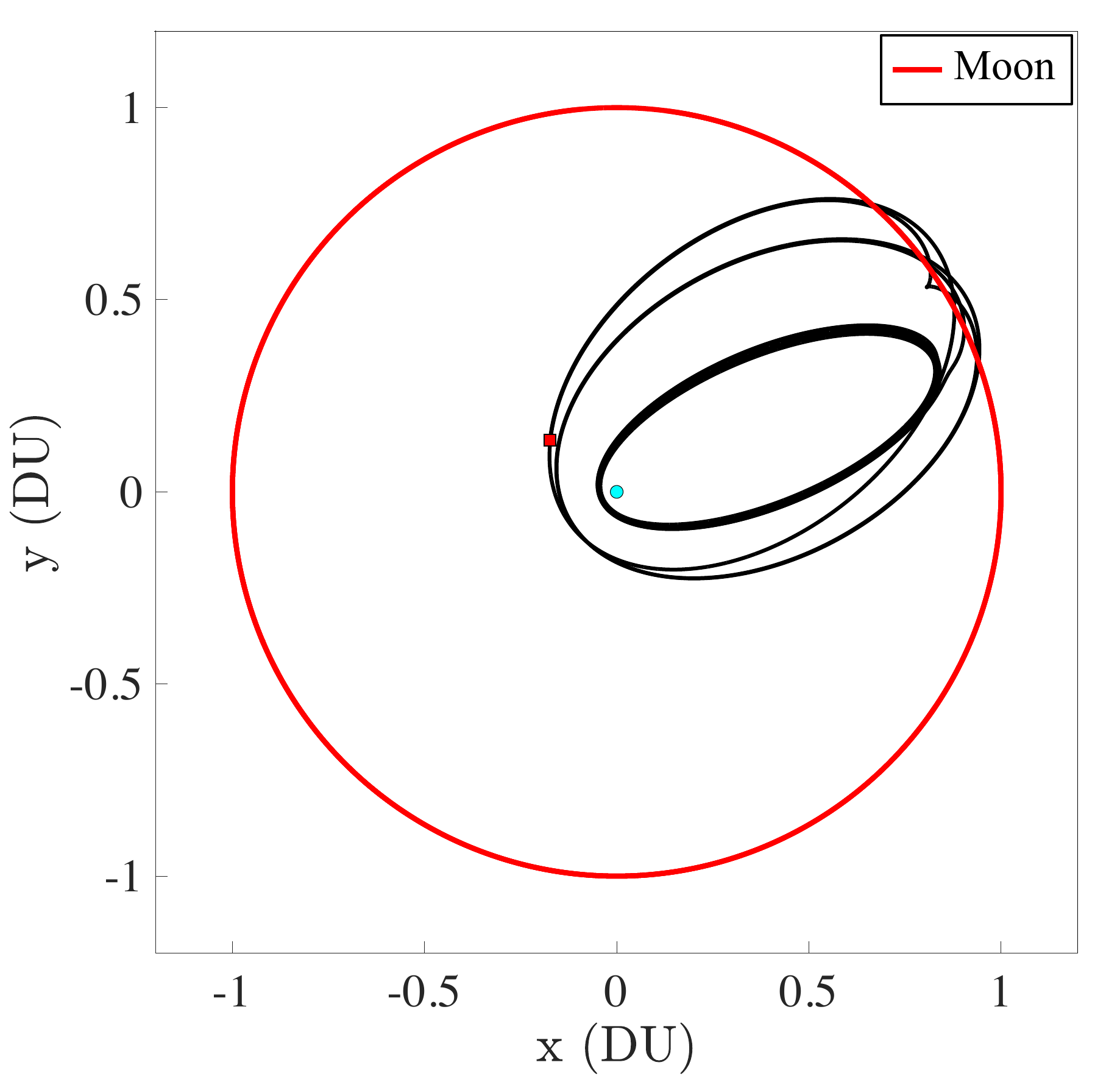}
\includegraphics[width=0.43\columnwidth]{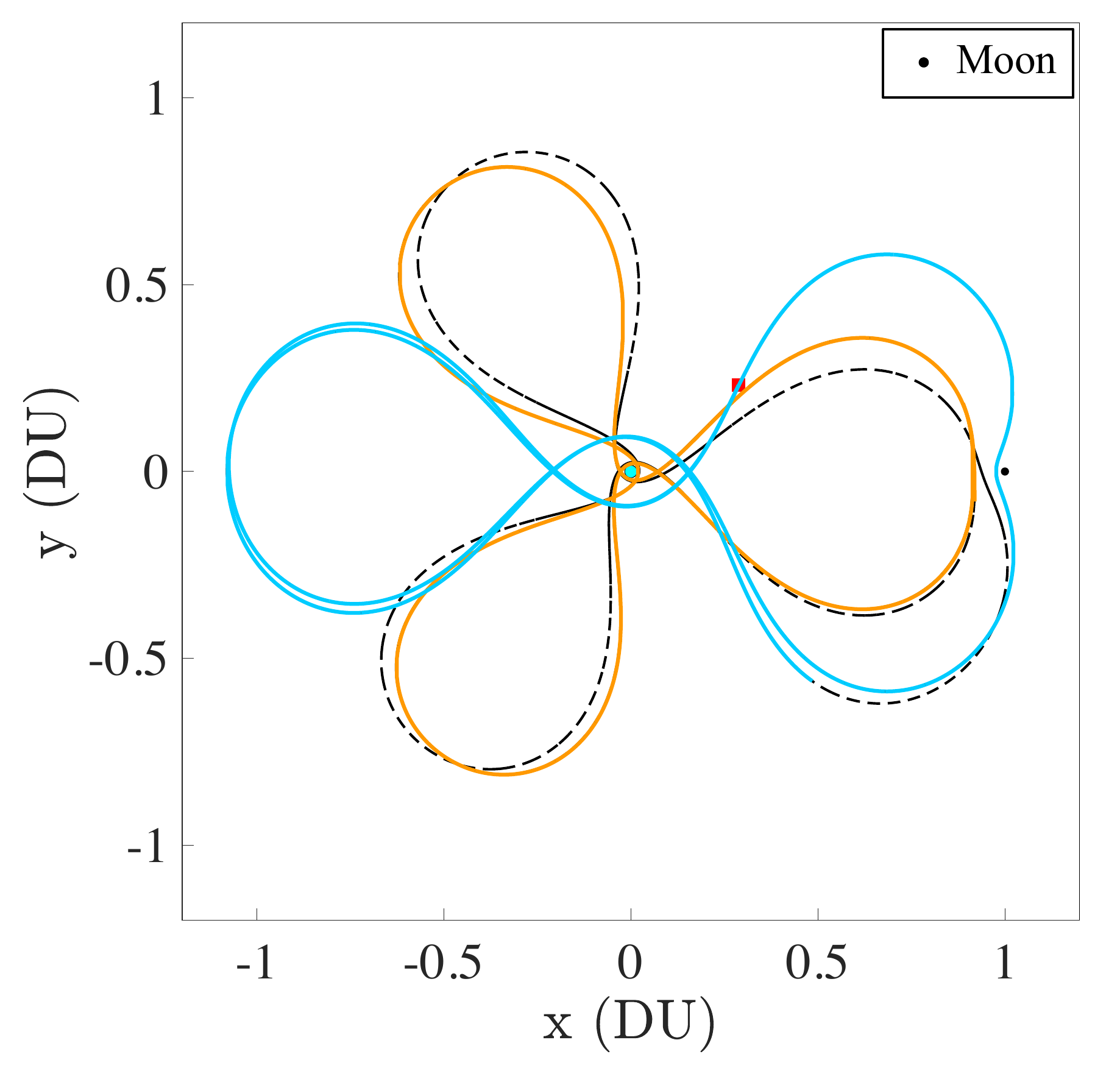}
\includegraphics[width=0.43\columnwidth]{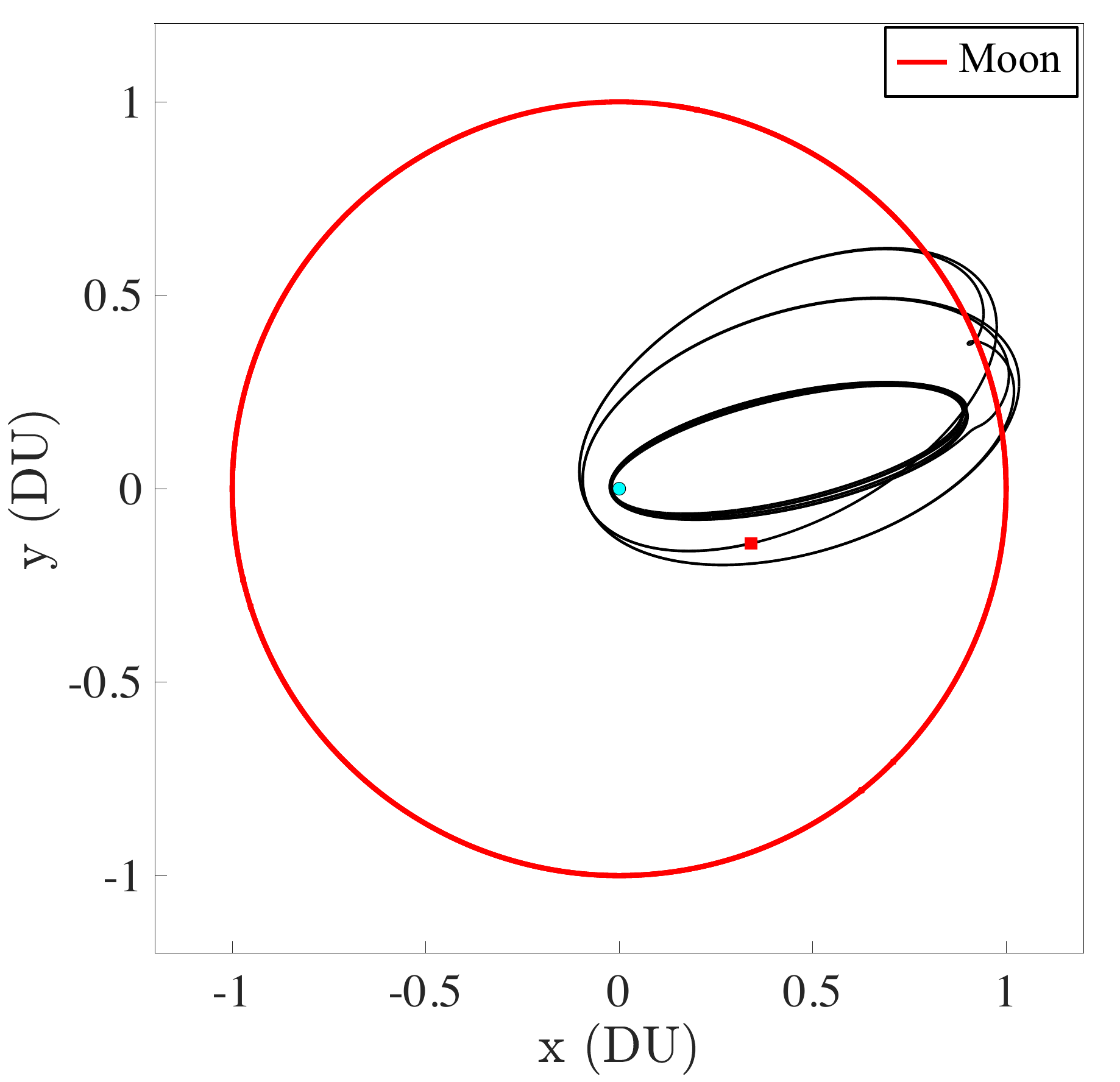}
\vspace{-0.2cm}
\caption{ \label{fig:3121heteroclinic} 3:1 to 2:1 MMR type 1 heteroclinic transfers plotted in rotating frame (left) and inertial frame (right) for $C=2.86, 2.70$, and 2.54 (from top to bottom). Near-3:1 and near-2:1 portions in yellow and cyan respectively in rotating frame along with transfer trajectory in dashed black. Earth (cyan circle) to scale; Moon (black circle) scaled 2x for clarity. Red square marks end state for displayed trajectory propagations.}
\end{centering}
\end{figure*}

For some additional examples of 3:1 to 2:1 type 1 transfers, Figure \ref{fig:3121heteroclinic} displays plots of such heteroclinic trajectories for $C = 2.86, 2.70$, and 2.54. 
The trajectories' near-3:1-orbit portions are shown in yellow and near-2:1 parts are colored cyan in the rotating frame plots on left. The increase of orbital semi-major axis is clearly visible in the inertial frame plots on right, all achieved naturally by leveraging the Moon's gravity. Indeed, similar type 1 heteroclinic transfers were computed for $C$ values as low as 2.44; however, at $C=2.52$, the heteroclinic trajectory's lowest perigee passes just 123 km above the Earth's surface, while at 2.50 it lies about 600 km below. Linearly interpolating these two values suggests that the heteroclinic trajectories will collide with the Earth's surface or pass below the 100 km von K\'arm\'an line for all $C \leq$ 2.517 or 2.519, respectively. Thus, we display plots for $C \geq 2.54$, which is one of the highest-energy non-colliding heteroclinics computed. 

Note that the near-3:1 part of the $C=2.54$ heteroclinic trajectory passes within about 890 km of the Earth's surface, suggesting that this and other heteroclinics at similar energies could be used to reach the Moon from low-Earth orbit (LEO) with just a single LEO-to-3:1 impulsive transfer maneuver. In fact, even portions of heteroclinics for $C<2.517$---whose lowest perigees lie below the Earth's surface---could be useful if targeted for direct injection by a launch vehicle, with the Moon's gravity then providing a free ``perigee raise maneuver''.

\begin{remark*} 

{While this study only considers ballistic transfer trajectories guided by true heteroclinic connections---which only occur between orbits at the same Jacobi constant $C$---similar non-ballistic transfers can also be found between unstable resonant orbits at different $C$ values using an impulsive maneuver. To find these, one first computes the two orbits' stable and unstable manifolds' intersection curves with the perigee section $\Sigma$, just as in the same-$C$ case. Then, intersections between these curves' $(x,y)$-plane projections will yield points where a single impulsive maneuver would allow the spacecraft to ``jump'' from the unstable manifold at the initial $C$ value to the stable manifold at the target $C$ value. }

{Note that while one still uses the perigee section $\Sigma$ here, manifold curves \emph{must} be projected into $(x,y)$ space before finding their intersections, not into $(g,a)$ space. This is because even if manifolds at different $C$ values intersect at perigee when projected into $(g,a)$ coordinates, the resulting $(g,a)$-space intersection point will not correspond to stable/unstable manifold points having the same position coordinates\footnote{If the manifolds have different $C$ values, then stable/unstable manifold points corresponding to a $(g,a)$-plane intersection at perigee will have the same $a$ but different eccentricities $e$. They will thus lie at differing distances $a(1-e)$ from the Earth.}, thus precluding any single-impulse transfer between the manifolds. Recall from Section \ref{heteroComputeSection} that the heteroclinic computation method described therein uses $(x,y)$ coordinates for the manifold intersection search, and would thus be largely applicable to the different-$C$ transfer computation case as well. }

\end{remark*}

\subsubsection{Times-of-Flight for 3:1 to 2:1 Heteroclinics} \label{tofSection}

\begin{figure}
\centering
\includegraphics[width=1\linewidth, trim=0.8cm 0cm 2.2cm 0cm, clip]{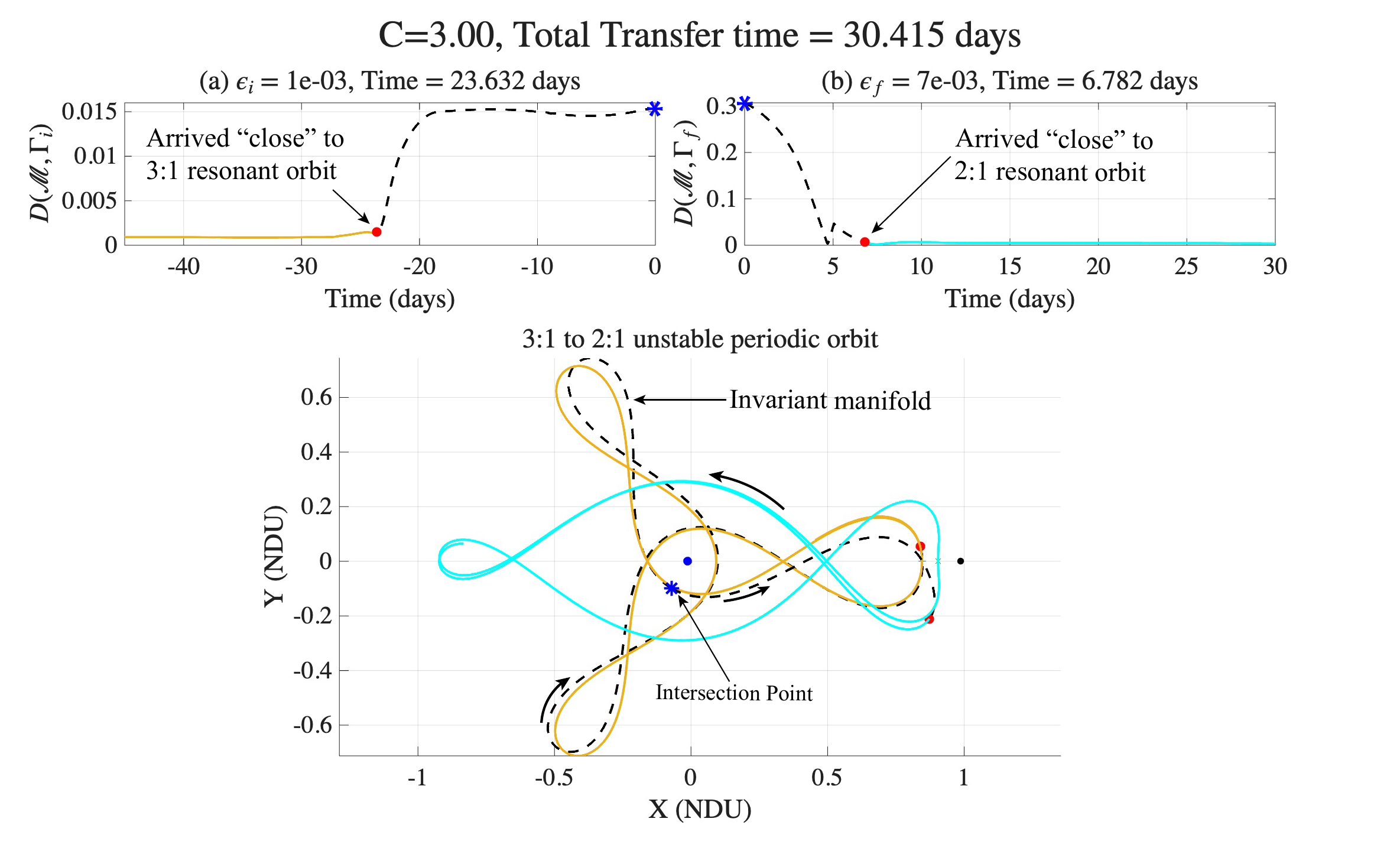}

\caption{\label{fig:transfer_time} Variation of distance metric with time for $C=3.00$ heteroclinic intersection point (denoted by blue star) propagated forwards and backwards towards 2:1 and 3:1 unstable periodic orbits, respectively. Red dots mark points before/after which the transfer trajectory is close enough (within tolerance) to the target orbit.}
\end{figure}

Since heteroclinic trajectories only converge asymptotically to their initial and final unstable periodic orbits, the transfer time-of-flight is infinite from a purely mathematical point of view. 
Operationally, though, a transfer can be considered to begin or end once it approaches sufficiently close to the initial or final periodic orbit in phase space \cite{braik2025aug}. As introduced in Section \ref{DistanceMetric}, in this work we use the generalized point-to-set distance metric of Equation \eqref{eq:point_set_distance} to define this notion of ``closeness'' to an orbit, and to determine the practical transfer times-of-flight. 
As is shown in Figure \ref{fig:transfer_time}, the minimum point-to-set distance metric between the propagated heteroclinic trajectory point and the final (initial) periodic orbit subsides to within very low tolerances when propagated forwards (backwards) in time.
These tolerances are heuristically determined by noting the time when variations and oscillations of the distance metric seem to subside, indicating that the manifold has converged closely to the desired orbit; we use different tolerances for transfers to/from different orbits. 
   
\begin{figure}
\begin{centering}
\includegraphics[width=0.8\columnwidth]{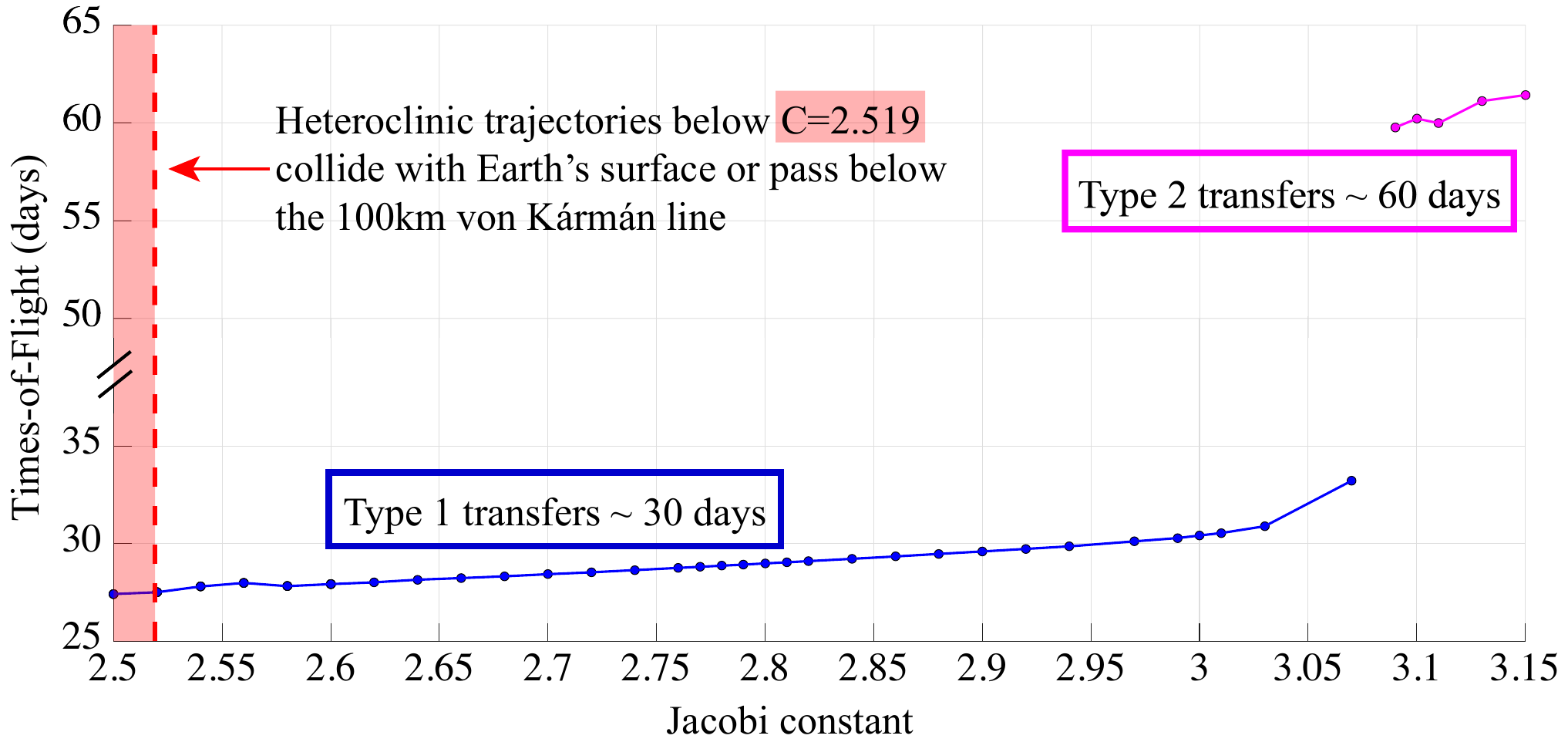}\\
\caption{ \label{fig:TOF} Variation of times-of-flight of shortest transfers from 3:1 to 2:1 unstable periodic orbits with Jacobi constant. Type 1 and Type 2 transfers are illustrated.
}
\end{centering}
\end{figure}

Taking a discrete set of Jacobi constants $C$ ranging from 2.50 to 3.15, and using the transfer time definition of Section \ref{DistanceMetric}, we found the minimum time-of-flight among all 3:1-to-2:1 heteroclinic transfers at each $C$ value. The resulting TOF vs $C$ plot is shown in Figure \ref{fig:TOF}. For $2.50\leq C\leq3.07$, the shortest (type 1) transfer trajectories take approximately 30 days. Such trajectories cease to exist for $C\geq3.09$; however, the longer duration type 2 transfer trajectories still exist, with times-of-flight of approximately 60 days. These transfer times depend on the heuristic tolerances used to define the beginning and end of each transfer. Nevertheless, these results indicate that for a given transfer type, the time-of-flight does not vary significantly with $C$; the main change occurs when the direct 3:1-to-2:1 transfer becomes available for $C<3.09$. Table ~\ref{Tab:31_21_transfer} summarizes transfer times and tolerances for few selected Jacobi constants $C=3.00$, 3.05, 3.10, and 3.15.

\begin{table}
\centering
\begin{tabular}{ | m{4em} | m{2cm}| m{2cm} |  m{2cm} | } 
  \hline
  Jacobi Constant & Transfer Time (days) & $\epsilon_{3:1}$ (NDU) & $\epsilon_{2:1}$ (NDU)\\ 
  \hline
  3.00  & 30.415 & $1\times10^{-3}$ & $7\times10^{-3}$\\ 
  \hline
  3.05 & 32.895 & $2\times10^{-3}$ & $1\times10^{-2}$ \\ 
  \hline
   3.10 & 60.212 & $1\times10^{-2}$ & $5\times10^{-3}$ \\ 
  \hline
   3.15 & 61.420 & $2\times10^{-2}$ & $4\times10^{-3}$ \\ 
  \hline
\end{tabular}
\caption{Summary of transfer times-of-flight and distance metric tolerances for $C=3.00,3.05,3.10,$ and 3.15.}
\label{Tab:31_21_transfer}
\end{table}

\subsection{The 4:1 Resonance and Barriers to Transport}

Having found that it is possible to transfer without $\Delta v$ from the 3:1 to 2:1 resonances given a sufficiently high energy level, it is natural to ask whether natural transfers from 4:1 to 3:1 may also be possible. If so, then one would be able to start in an even-lower 4:1 orbit and follow a chain of heteroclinics until the Moon without using propellant. 

\begin{figure}[t!]
\begin{centering}
\includegraphics[width=0.49\columnwidth]{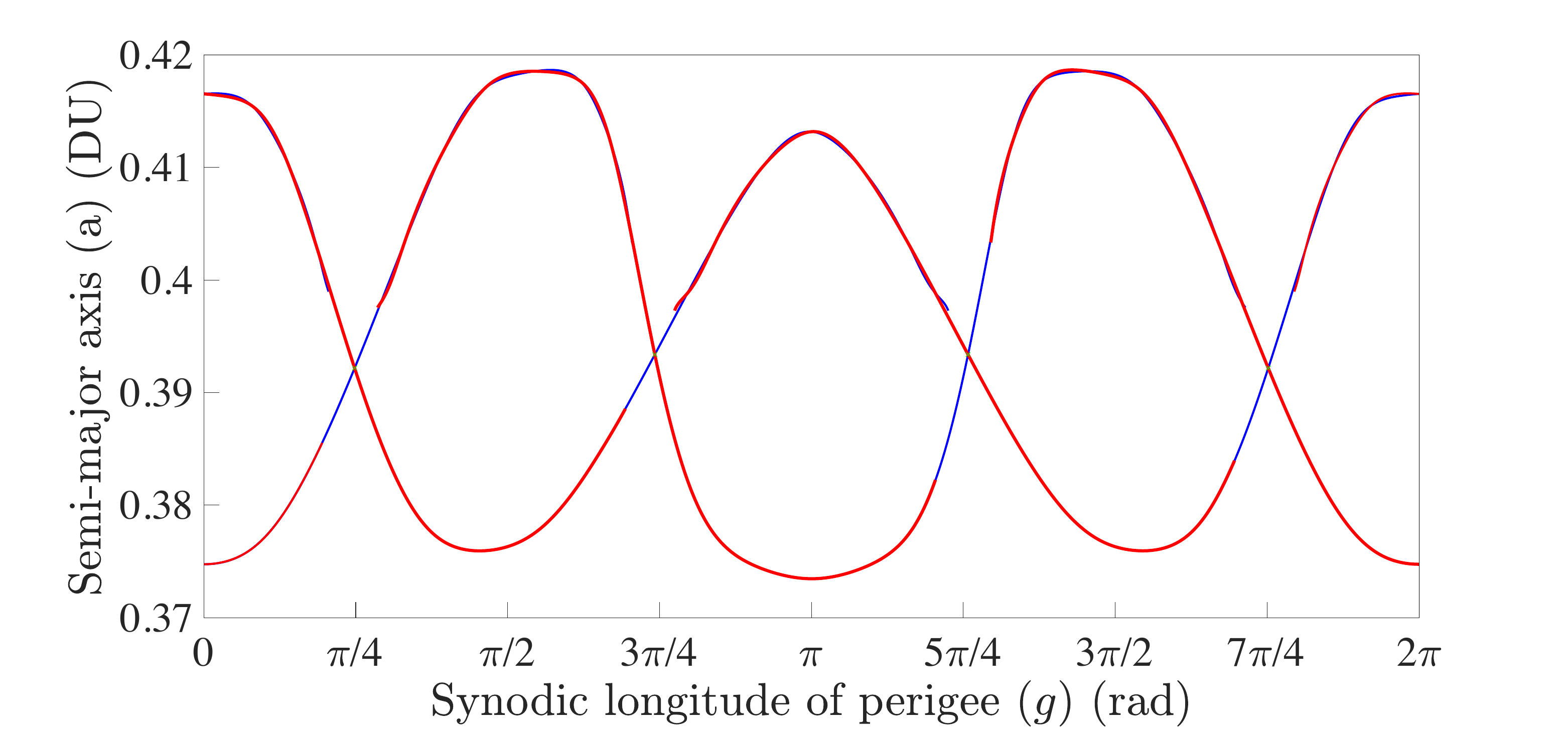}\\
\includegraphics[width=0.49\columnwidth]{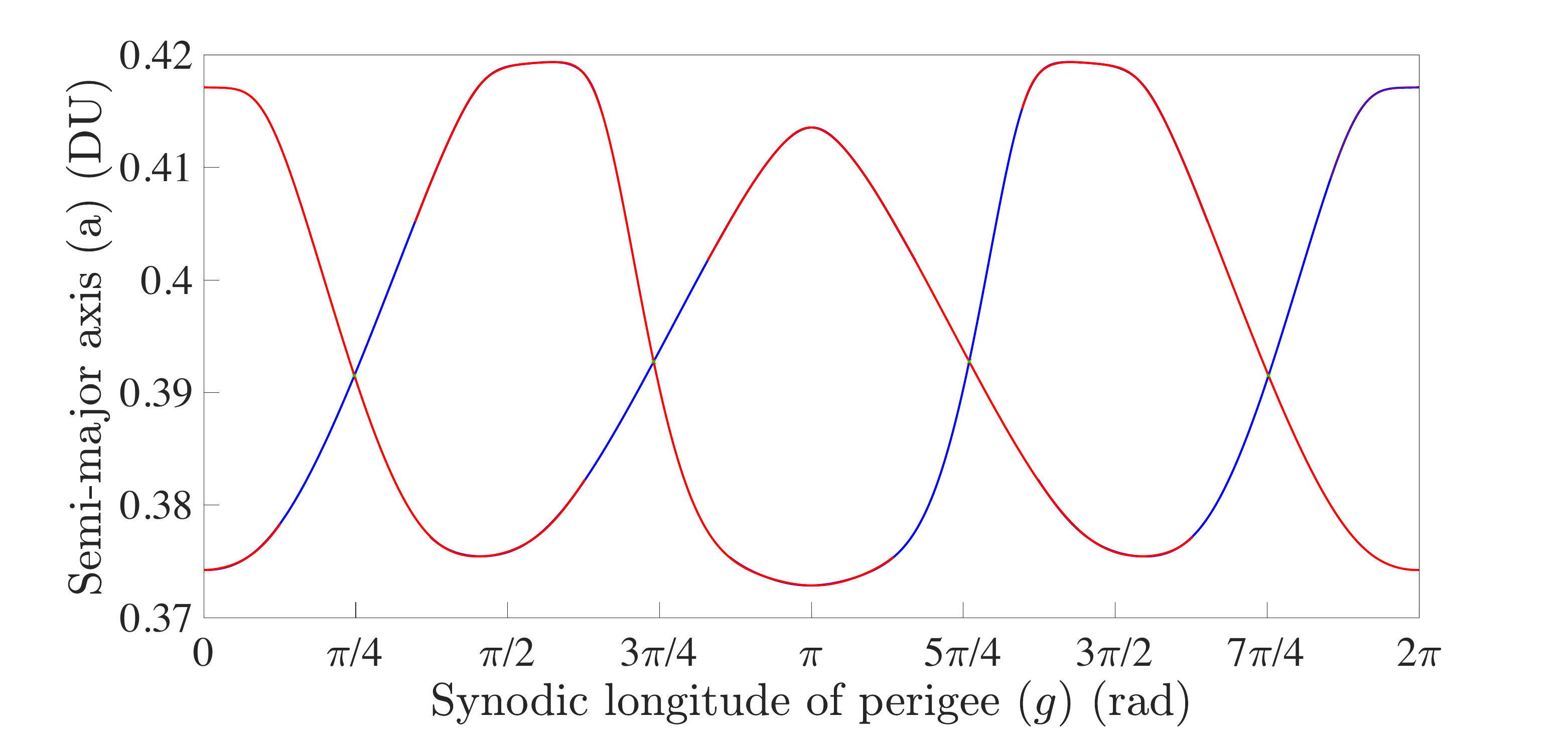}\\
\includegraphics[width=0.49\columnwidth]{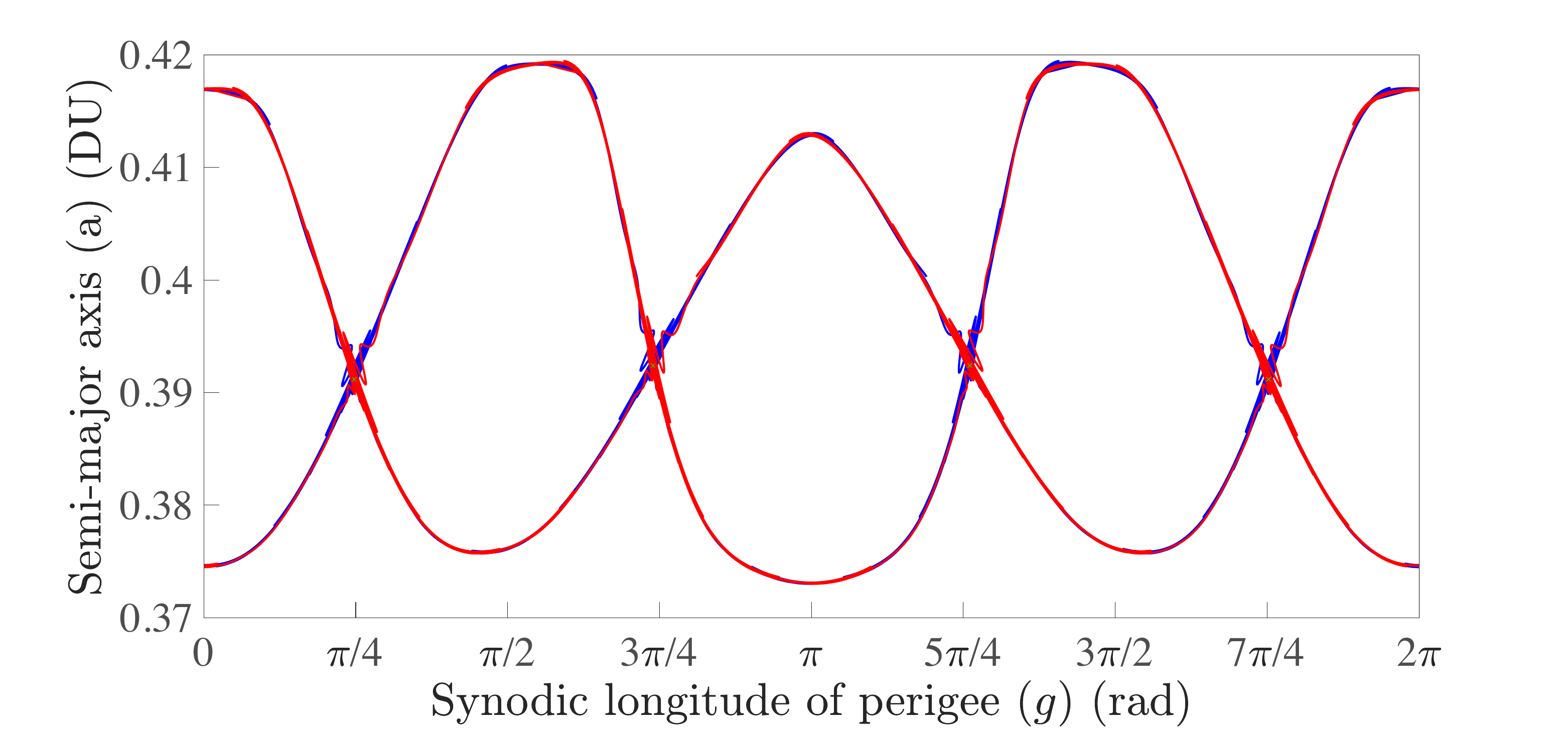}
\caption{ \label{fig:41manifolds} 4:1 stable/unstable manifolds (blue/red) on perigee section for $C = 3.15, 3.00$, and $C = 2.85$ (top to bottom). Note the lack of heteroclinic excursions to other resonances. 
}
\end{centering}
\end{figure}

In Figure \ref{fig:41manifolds}, we plot stable and unstable manifolds of 4:1 unstable resonant orbits for $C = 3.15, 3.00$, and 2.85 on our perigee Poincar\'e sections $\Sigma_C$. 
It is immediately visible that these manifolds are shaped very similarly to pendulum phase portrait separatrices, as would be the case in a region dominated by a single resonance without other MMRs overlapping. In fact, for $C = 3.15$ and 3.00, the 4:1 stable and unstable manifolds seem to almost lie on top of each other, very much like the case of an ideal pendulum whose stable/unstable manifolds exactly coincide. This strongly suggests that no heteroclinic can occur from these orbits to the 3:1 MMR or beyond; the same is in fact also visible for the 4:1 orbit at $C = 2.85$, whose stable and unstable manifolds intersect slightly more transversally than those of 3.00 and 3.15, but still clearly do not make significant excursions to other semi-major axis ($a$) values. Comparing these 4:1 manifolds to the 3:1 orbit manifolds computed earlier, we are clearly in a situation similar to the top two manifold plots of Figure \ref{fig:31manifolds} which had no heteroclinics to 2:1, rather than the case of Figure \ref{fig:3121manifolds} where heteroclinics first appeared.

\begin{figure}[!t]
\begin{centering}
\includegraphics[width=0.49\columnwidth]{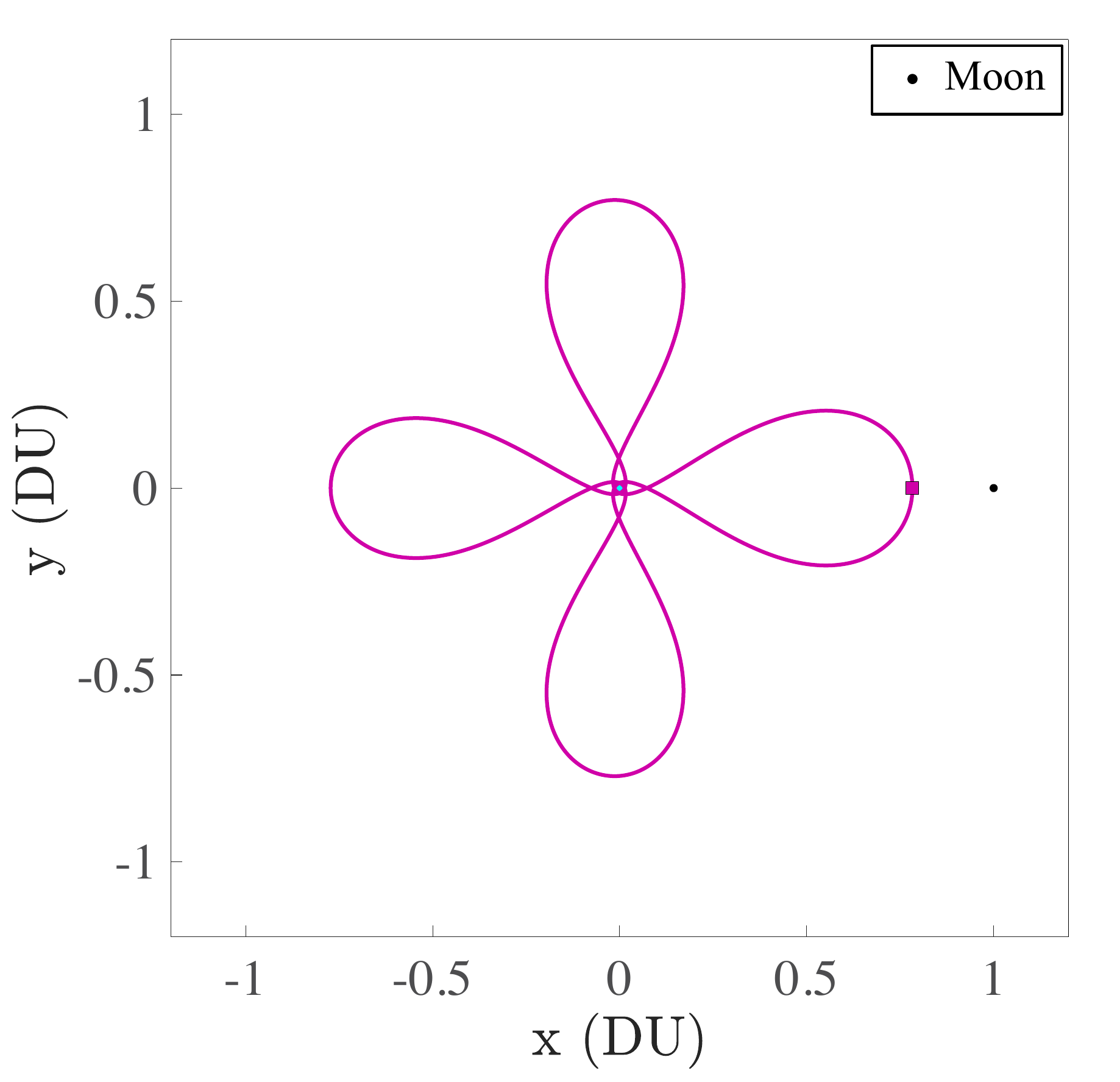}
\caption{ \label{fig:41orbitHighE} 4:1 unstable resonant orbit at $C = 2.85$, Earth (cyan circle at origin) drawn to scale.}
\end{centering}
\end{figure}


The above evidence suggests that 4:1 orbits with $C \geq 2.85$ do not possess heteroclinic connections to the 3:1 MMR. Although one might expect that decreasing $C$ further (i.e., increasing the energy) could yield a 4:1 orbit with such a connection, the $C = 2.85$ case already exhibits energy and eccentricity so high that its perigee lies below Earth’s surface (Fig.~\ref{fig:41orbitHighE}). Prograde 4:1 unstable orbits at even lower $C$ necessarily have higher eccentricities and thus also collide with Earth. Despite examining these Earth-impacting cases, we found no evidence of heteroclinics between 4:1 and 3:1 orbits at such low $C$ values. Overall, we conclude that no heteroclinic connections exist from any 4:1 orbit to any 3:1 orbit in the Earth–Moon system.

\begin{figure}[t!]
    \includegraphics[width=1\linewidth]{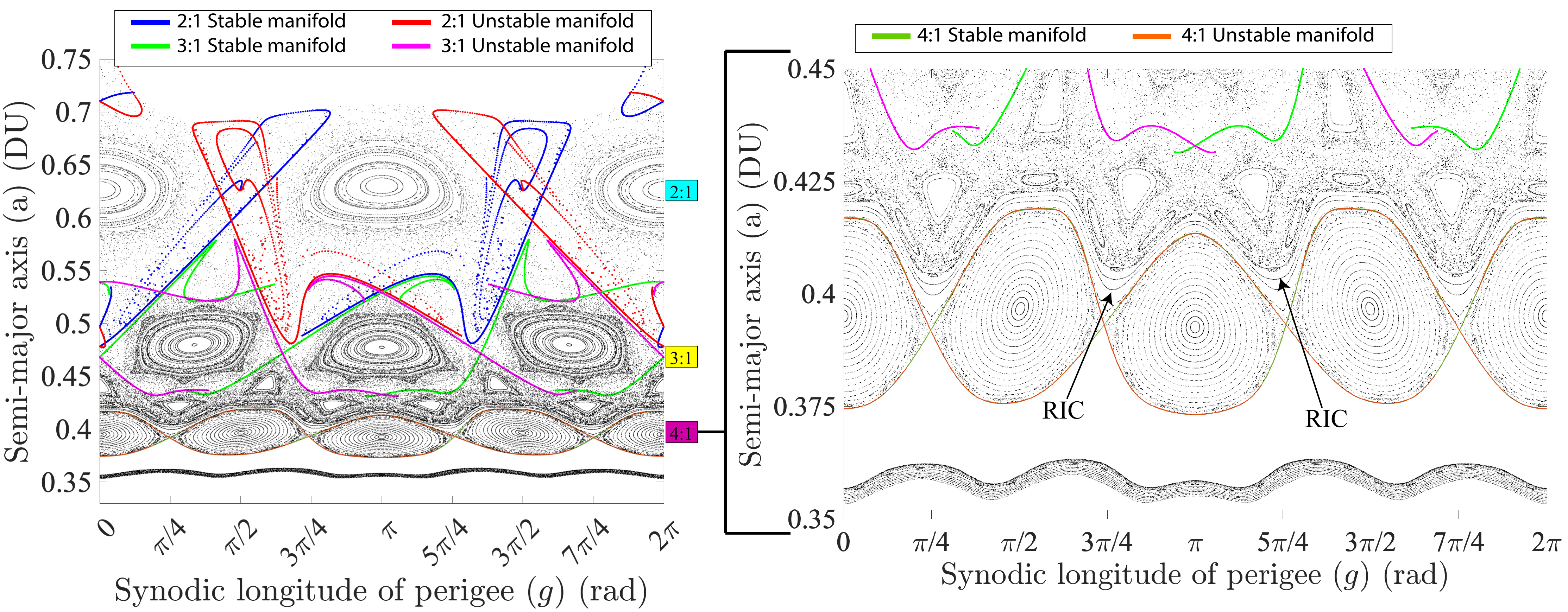}
    \caption{ 
    (Left) 2:1, 3:1, and 4:1 stable/unstable manifolds and Poincar\'e map points for $C=3.10$, with no connection observed between 3:1 and 4:1 resonances. 
    (Right) Zoomed-in plot for $C=3.10$ with potential rotational invariant circles (RICs) indicated, forming a barrier between 4:1 and 3:1 manifolds. 
    }
    \label{fig:4:1RIC}
\end{figure}

To understand why no heteroclinics from 4:1 to 3:1 appear to exist, we turn to the $C = 3.10$ case in Fig.~\ref{fig:3121manifolds}, which is redrawn in Fig.~\ref{fig:4:1RIC} with the 4:1 manifolds overlaid on the Poincar\'e map along with a zoomed view of the 4:1 MMR shown on the right. 
Here, continuous curves stretching horizontally across the entire plot, from $g=0$ to $2\pi$, seem to be present just above the 4:1 resonant island (the four ``eye'' shaped regions centered near $a = 0.40$ and encompassed by the 4:1 manifolds). These curves seem to be 1D KAM tori \citep{kamTutorial}, sometimes also referred to in the 2D-map case as rotational invariant circles (RICs) \citep{Meiss1992,RoSc2007,werner2022multiple}, of our PCR3BP Poincar\'e map. Such curves correspond to non-resonant Keplerian orbits which persist (after deformation) into the PCR3BP; they form barriers to transport between different resonances, as stable/unstable manifolds cannot intersect them. Thus, these potential RICs are clearly separating the 4:1 and 3:1 MMRs at the Jacobi constant value $C = 3.10$, preventing heteroclinics between them. Although less clearly visible, a similar curve seems to also potentially exist for $C = 3.05$ in Figure \ref{fig:3121manifolds}. For $C =3.00$ and below, a simple visual identification is not so easy, but we suspect that similar RICs most likely are the barrier preventing heteroclinics from 4:1 to 3:1 orbits from occurring over the entire range of $C$ values considered. 

\section{Connections with $L_1$ Tubes} \label{resToL1Section}

In the previous section, we computed stable/unstable manifolds of 4:1, 3:1, and 2:1 resonant orbits and demonstrated heteroclinic trajectories naturally connecting the 3:1 and 2:1 orbits together. However, also recall from Figure \ref{fig:3121manifolds} that the 2:1 orbits' stable/unstable manifolds make excursions towards $a=1$, corresponding to the Moon's orbital semi-major axis.
To better understand the mechanisms by which this occurs, we next investigate how these resonant orbit manifolds interact with other dynamical structures present near lunar $L_1$.

\begin{figure}[h!]
\centering
\begin{tabular}{c@{\hspace{5pc}}c}
\includegraphics[height=2in]{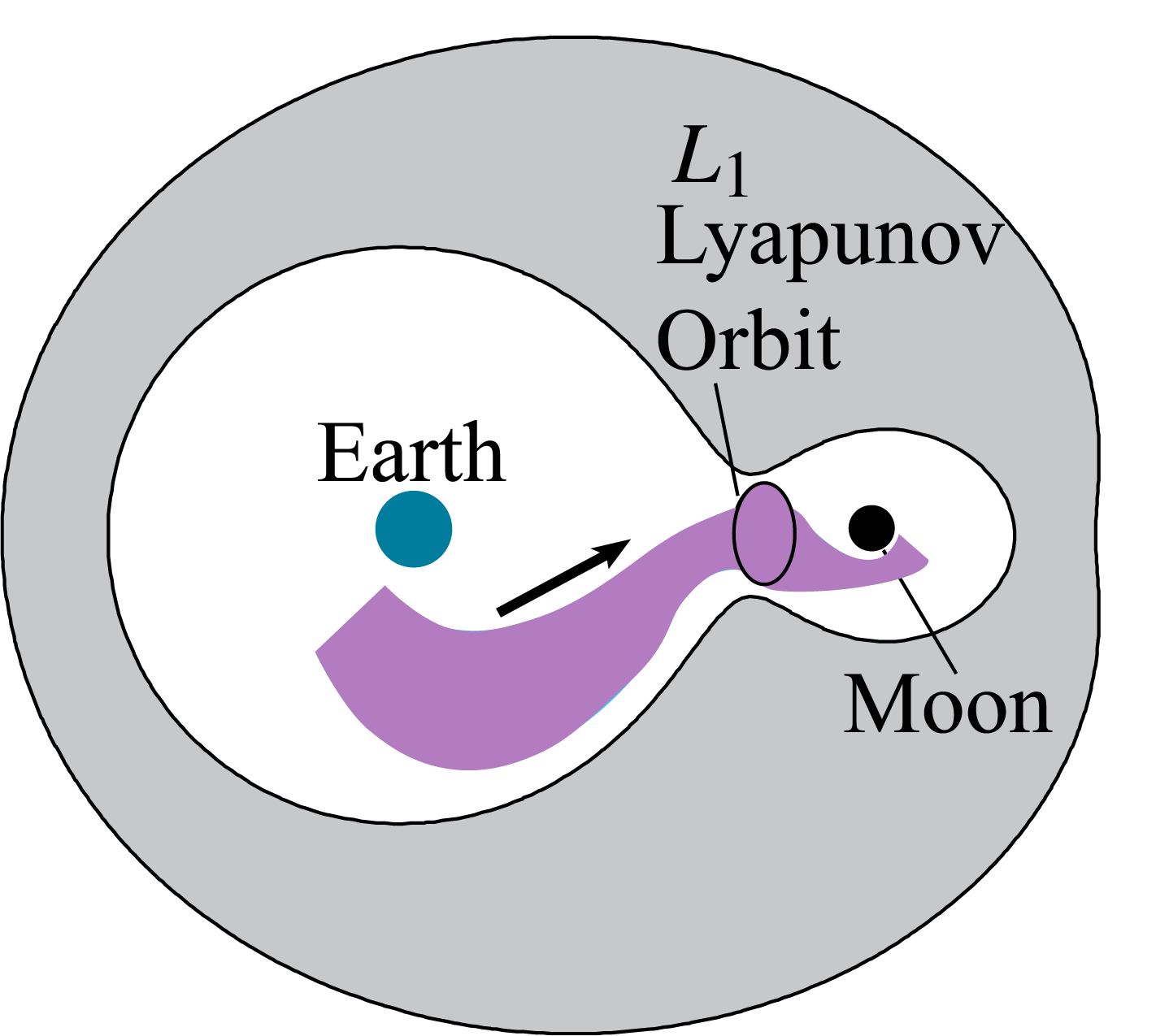}&
\includegraphics[height=2in]{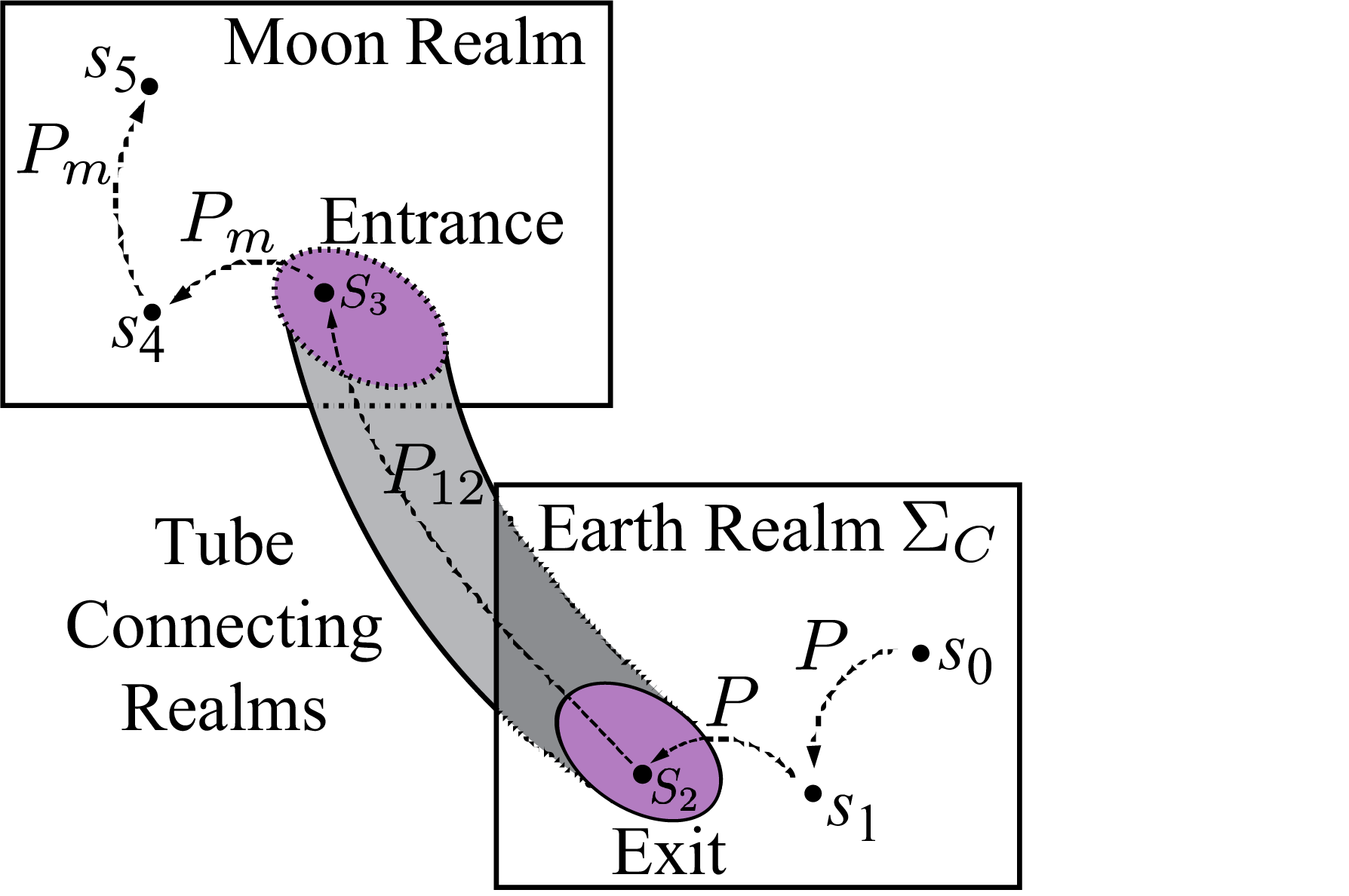}
\end{tabular}
\caption{\label{fig:tubes}Tube dynamics: leaving the Earth realm through an exit on the Poincar\'e section $\Sigma_C$, taking trajectories into the Moon realm.}
\end{figure}

As first noted in \citet{KoLoMaRo2000} and elaborated upon in \citet{KoLoMaRo2001} and \citet{Ross2003a}, intersections between ``lobes'' formed by stable/unstable manifolds of unstable MMR periodic orbits and ``tubes'' corresponding to $L_1$ planar Lyapunov orbit manifolds can be used to design low-energy trajectories between Earth orbits and lunar orbits; such a strategy was also recently studied by \citet{hiraiwa2024designing}. The first intersection of an $L_1$ Lyapunov orbit stable manifold on our 2D Poincar\'e section $\Sigma_C$, called its first Poincar\'e cut \citep{KoLoMaRo2000}, will have the topology of a (deformed) circle. The interior of this circle forms an ``exit'' from the Earth realm on $\Sigma_C$. All trajectories entering it will then emerge from an ``entrance'' into the Moon realm, as illustrated in Figure \ref{fig:tubes}.

\begin{figure}[h]
\centering
\includegraphics[width=0.6\linewidth]{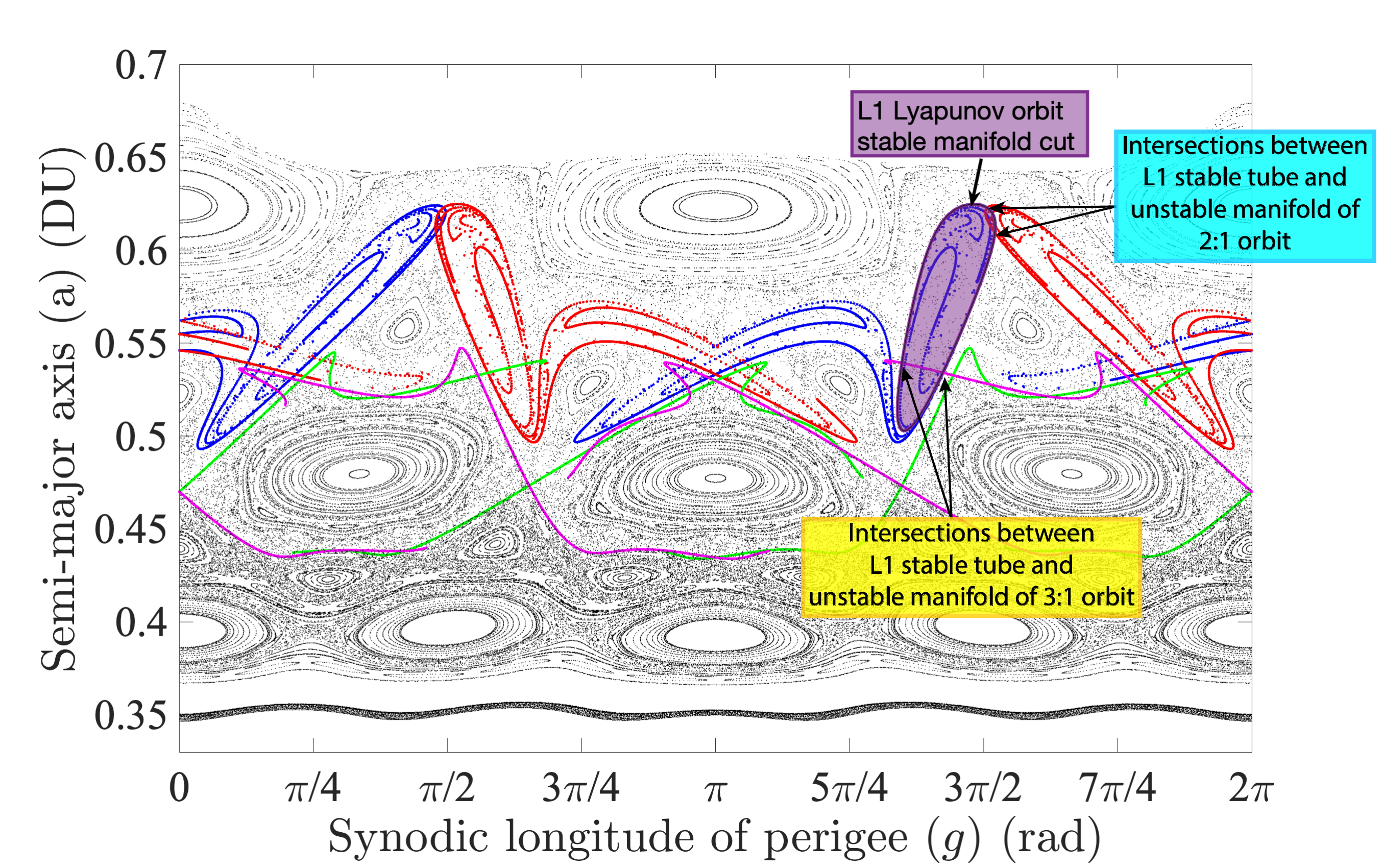}
\caption{\label{fig:cut}Poincar\'e section $\Sigma_C$ for $C=3.15$ with $L_1$ Lyapunov orbit stable manifold (the Earth realm ``exit'' as in Figure \ref{fig:tubes}) overlaying the 2:1 and 3:1 unstable resonant orbit stable/unstable manifolds.}
\end{figure}

\begin{figure}[t!]
    \centering
    \includegraphics[width=0.6\linewidth]{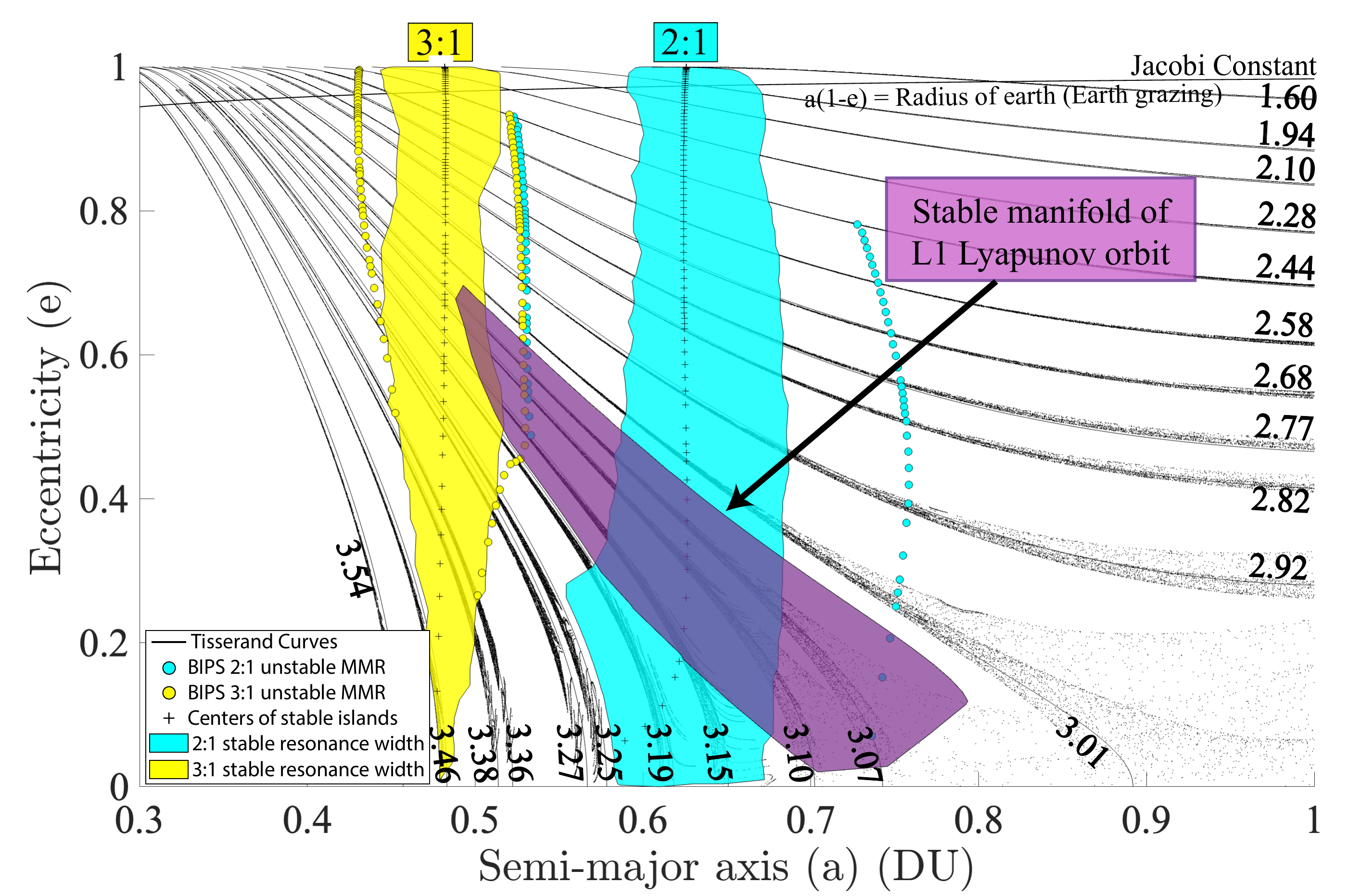}
    \caption{
    $(a,e)$ projections of $L_1$ Lyapunov orbit stable manifold points for $C=3.05$, 3.09, 3.11, 3.13, and 3.15. Overlaid on the PCR3BP resonance widths and regions calculated in \citet{rawat2025preprint}.
    }
    \label{fig:cut_ae}
\end{figure}

During numerical investigations, for a certain range of Jacobi constants, we discovered that the invariant manifolds of 2:1 unstable periodic orbits and the $L_{1}$ Lyapunov orbits' stable/unstable tubes seem to strongly interact with each other. One such case, at $C=3.15$, is shown in Figure \ref{fig:cut}, 
with the $L_1$ Lyapunov orbit stable manifold's first Poincar\'e cut highlighted. We found that the ``swirling'' of 2:1 manifolds seen inside the cut is a consequence of trajectories ``exiting'' the Earth realm through this cut on $\Sigma_C$, entering the Moon realm, and then re-emerging in the Earth realm during propagation. Similar behavior was seen for other Jacobi constants as well; this leads to numerical challenges similar to those encountered in other contexts, including atomic physics and chemistry \citep{JaFaUz1999}. These strong 2:1-$L_1$ manifold interactions can also be represented on the  $(a,e)$ semi-major axis vs eccentricity plane by calculating and plotting osculating $a$ and $e$ for all $L_1$ Lyapunov cut points across a range of $C$ values, as shown in Figure \ref{fig:cut_ae}. This quantifies the region of influence of the $L_1$ Lyapunov orbit stable manifold tube in the Earth realm and its overlap with the 2:1 and 3:1 resonance zones. 

\subsection{Heteroclinic Connections from 2:1 Resonant to $L_1$ Lyapunov Orbits}
\begin{figure}[t!]
    \includegraphics[width=0.55\linewidth]{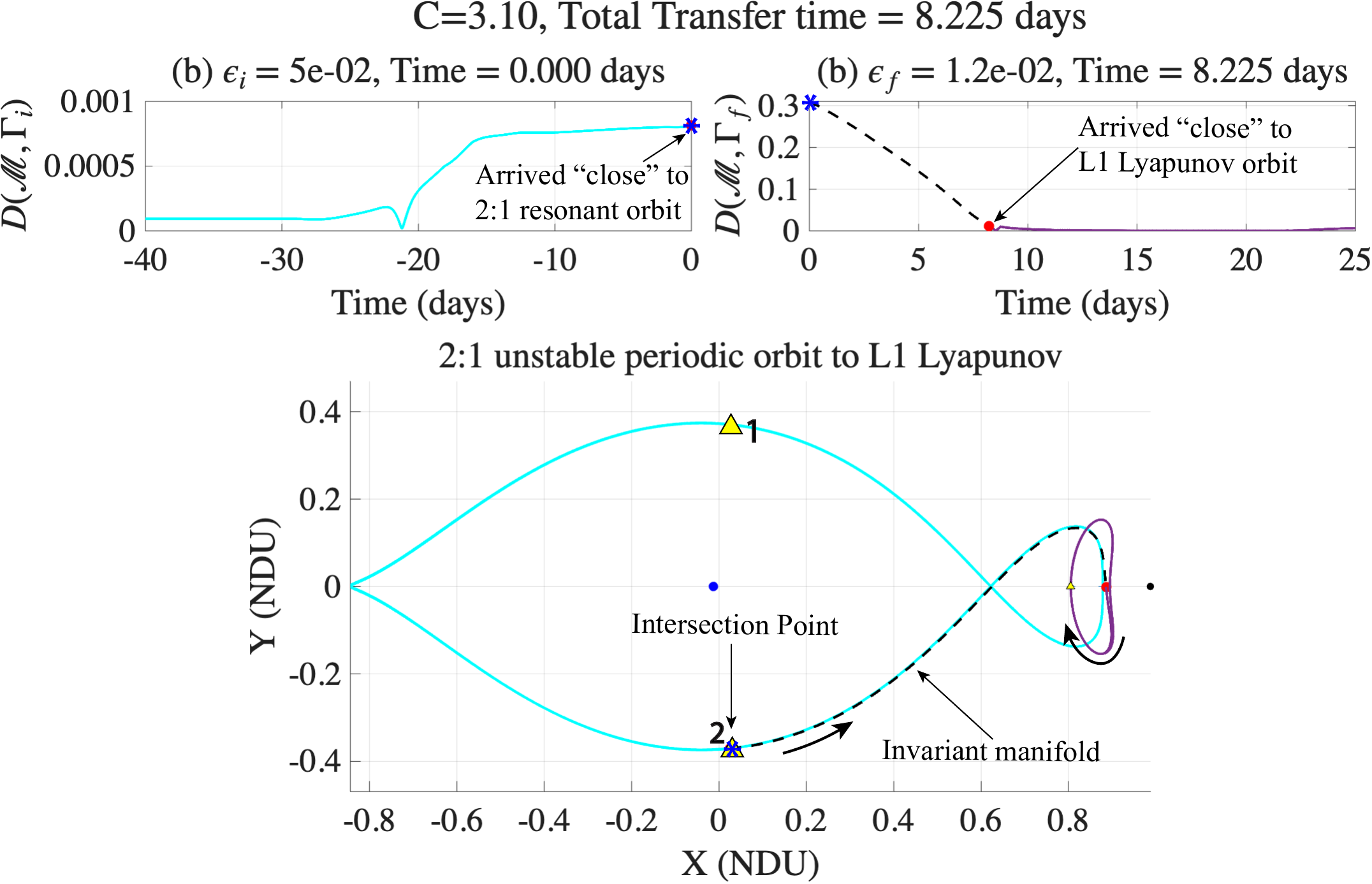}
    \includegraphics[width=0.45\linewidth]{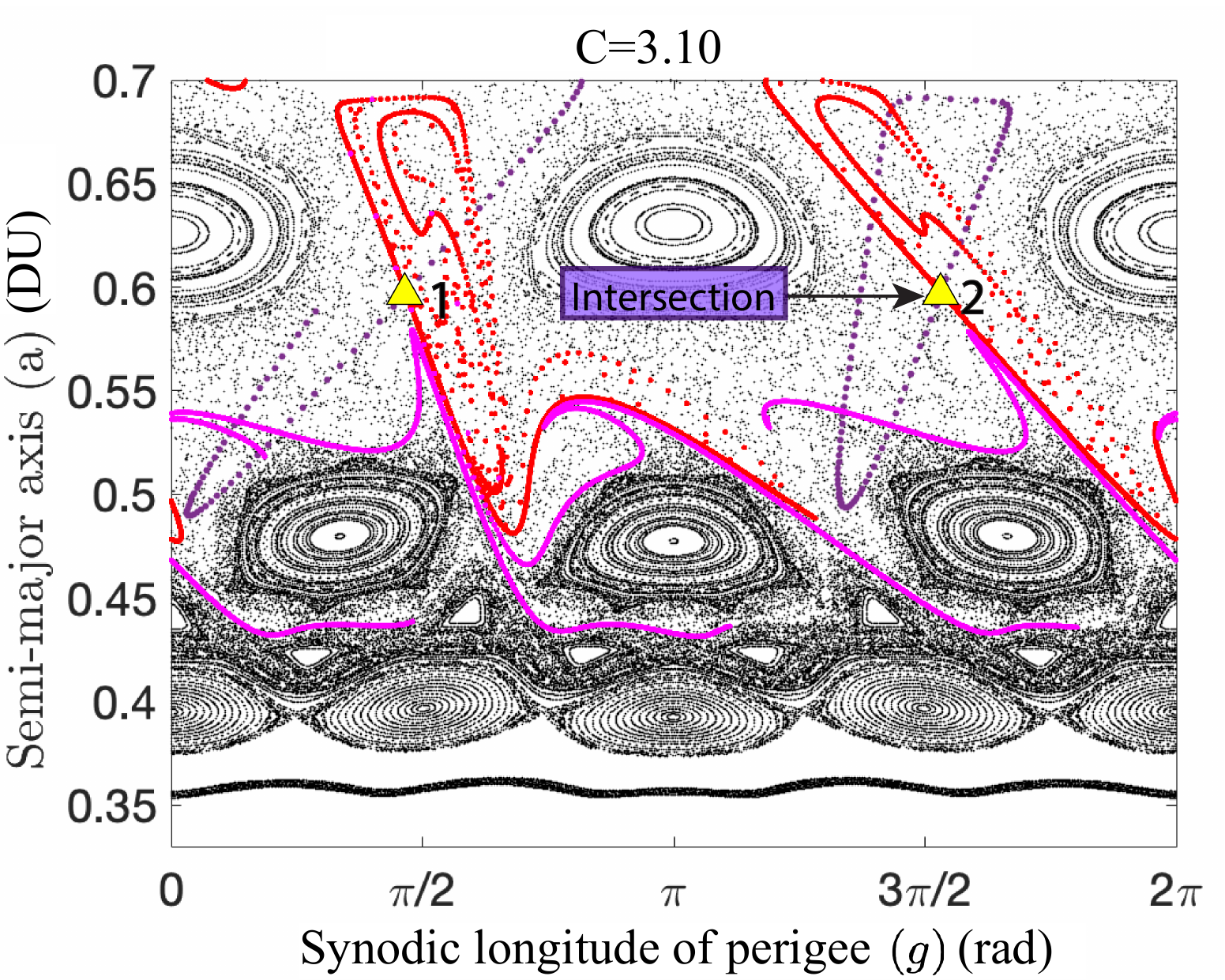}
    \caption{
    Shortest transfer from 2:1 unstable resonant periodic orbit to $L_1$ Lyapunov orbit, $C=3.10$.
    }
    \label{fig:21_L1_short}
\end{figure}

\begin{figure}[t!]
    \includegraphics[width=0.55\linewidth]{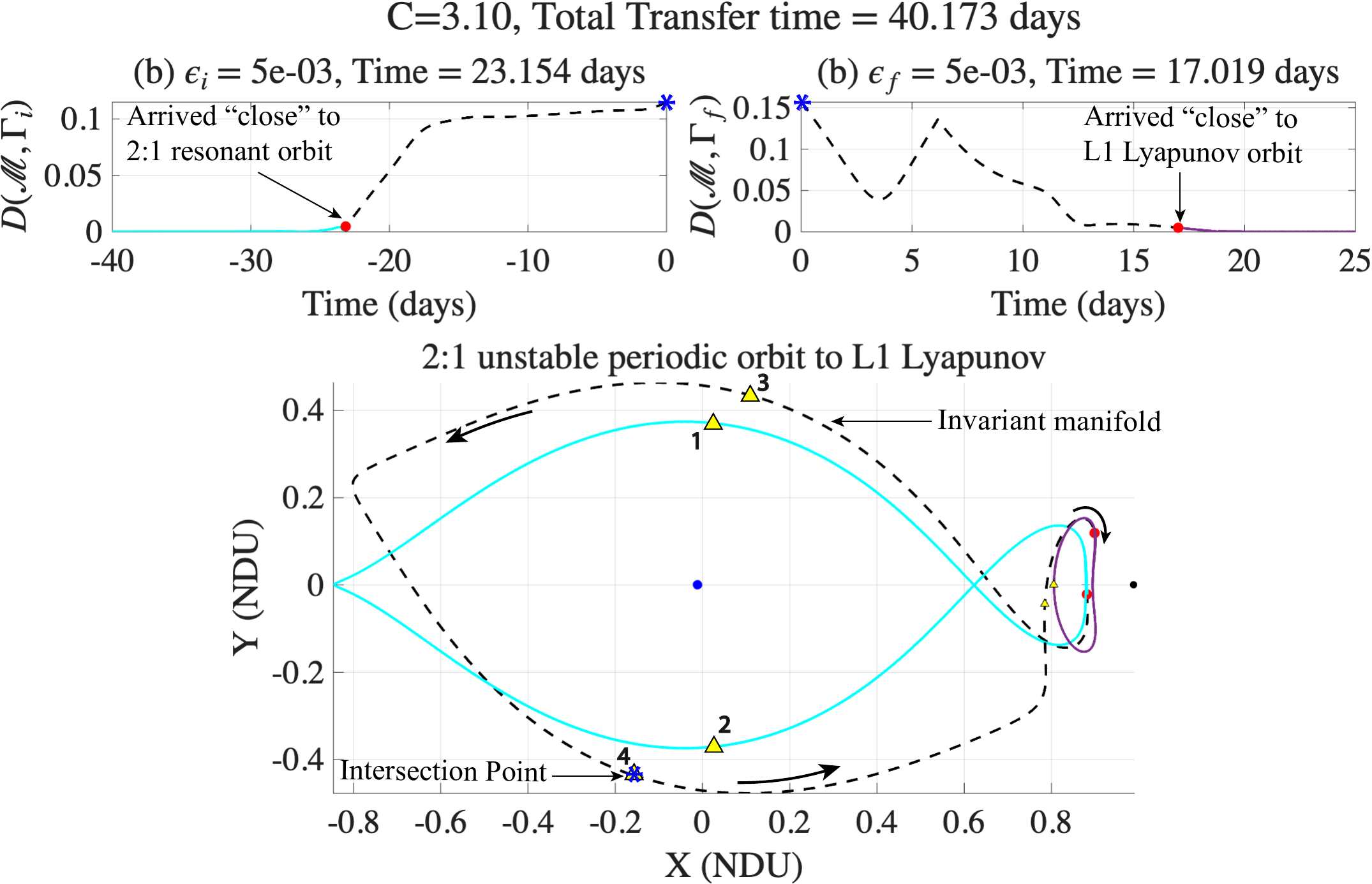}
    \includegraphics[width=0.45\linewidth]{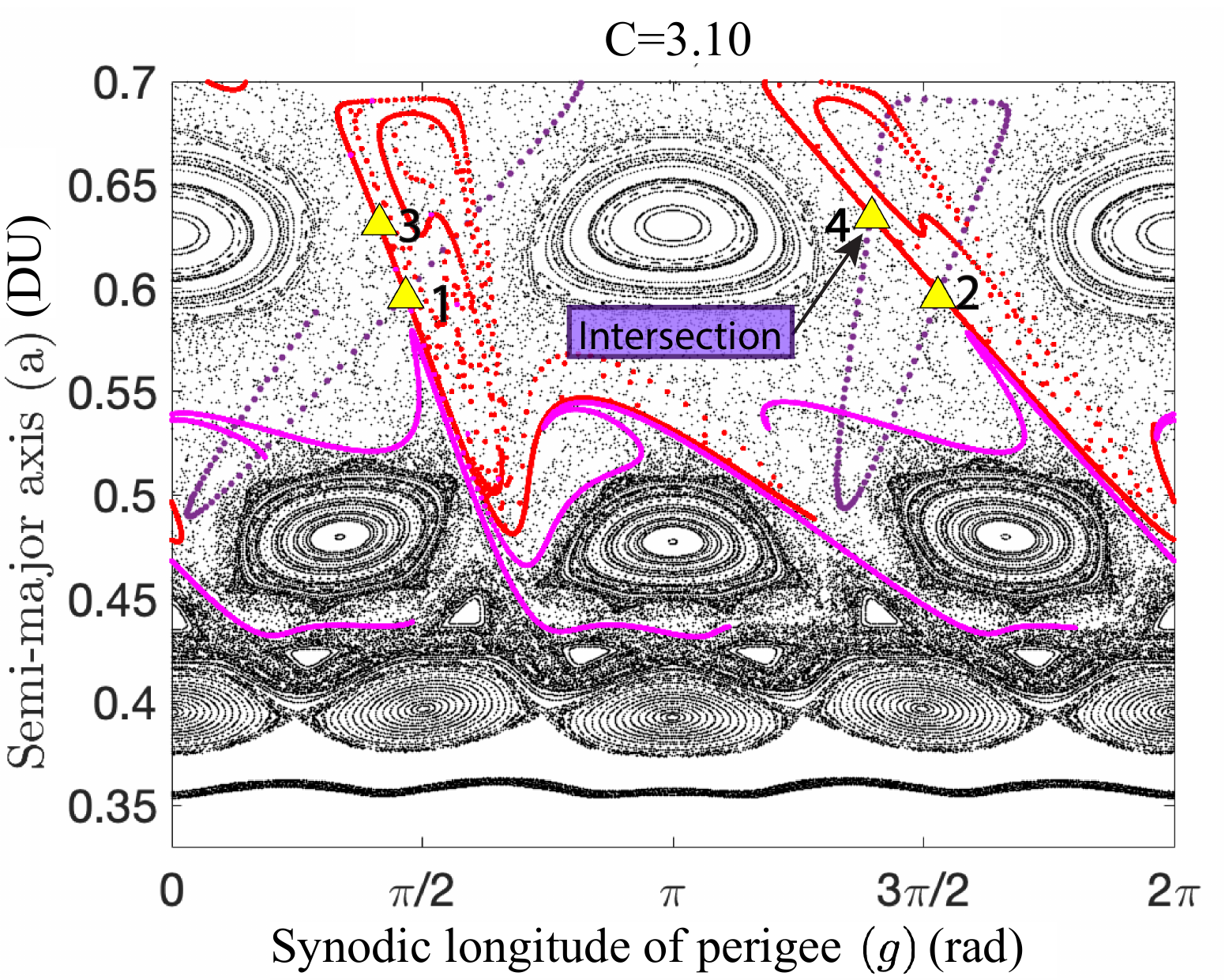}
    \caption{Longer transfer from 2:1 unstable resonant periodic orbit to $L_1$ Lyapunov orbit, $C=3.10$.
    }
    \label{fig:21_L1_long}
\end{figure}

It is visible in Figure \ref{fig:cut} that intersections occur between the unstable manifold of the 2:1 unstable resonant periodic orbit and the $L_1$ Lyapunov orbit's stable manifold tube, indicating the presence of a heteroclinic connection. Thus, we computed such heteroclinic trajectories from 2:1 unstable orbits to $L_1$ Lyapunov orbits for $C=3.05$ and $3.10$, using the tolerances given in Table \ref{Tab:L1_transfer} with the $(\mathbf{L},\mathbf{A})$-based distance metric of Section \ref{DistanceMetric} to define transfer times-of-flight (TOF). Two types of 2:1-to-L1 heteroclinics were found;  Figure \ref{fig:21_L1_short} displays the shortest transfer, with TOF $\approx 8.2$ days, for $C=3.10$; this trajectory corresponds to manifold intersections---denoted by 2 in the plot on right---that occur very close to the 2:1 orbit perigee points on $\Sigma_C$. If instead one selects and propagates a manifold intersection point farther away from the 2:1 orbit, a longer duration transfer is found---as shown in Figure \ref{fig:21_L1_long}, corresponding to the intersection point labeled as 4, again for $C=3.10$. This trajectory has a TOF of approximately $40.17$ days.

\subsection{Heteroclinic Connections from 3:1 Resonant to $L_1$ Lyapunov Orbits}
\begin{figure}[t!]
    \includegraphics[width=0.55\linewidth]{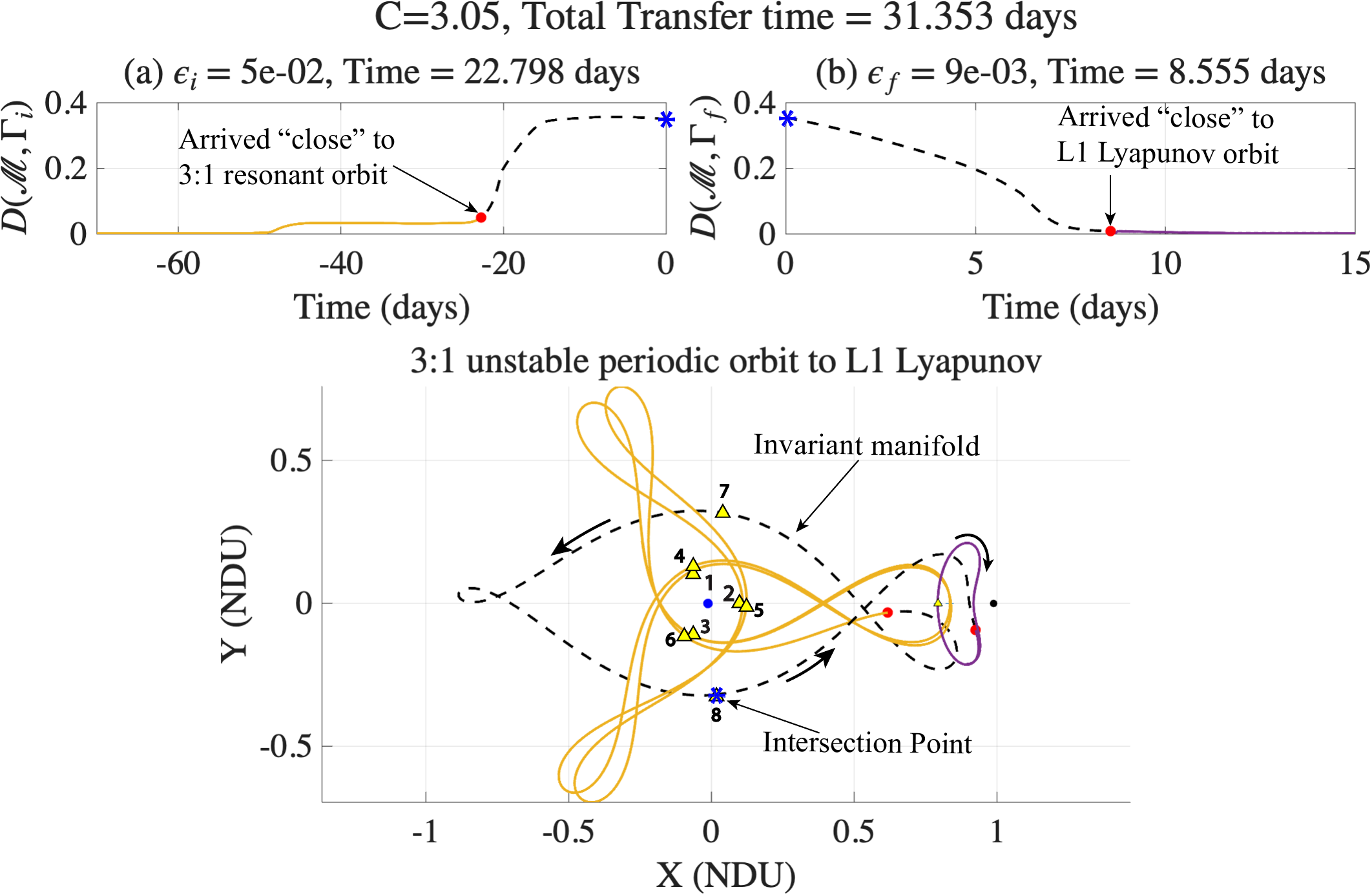}
    \includegraphics[width=0.45\linewidth]{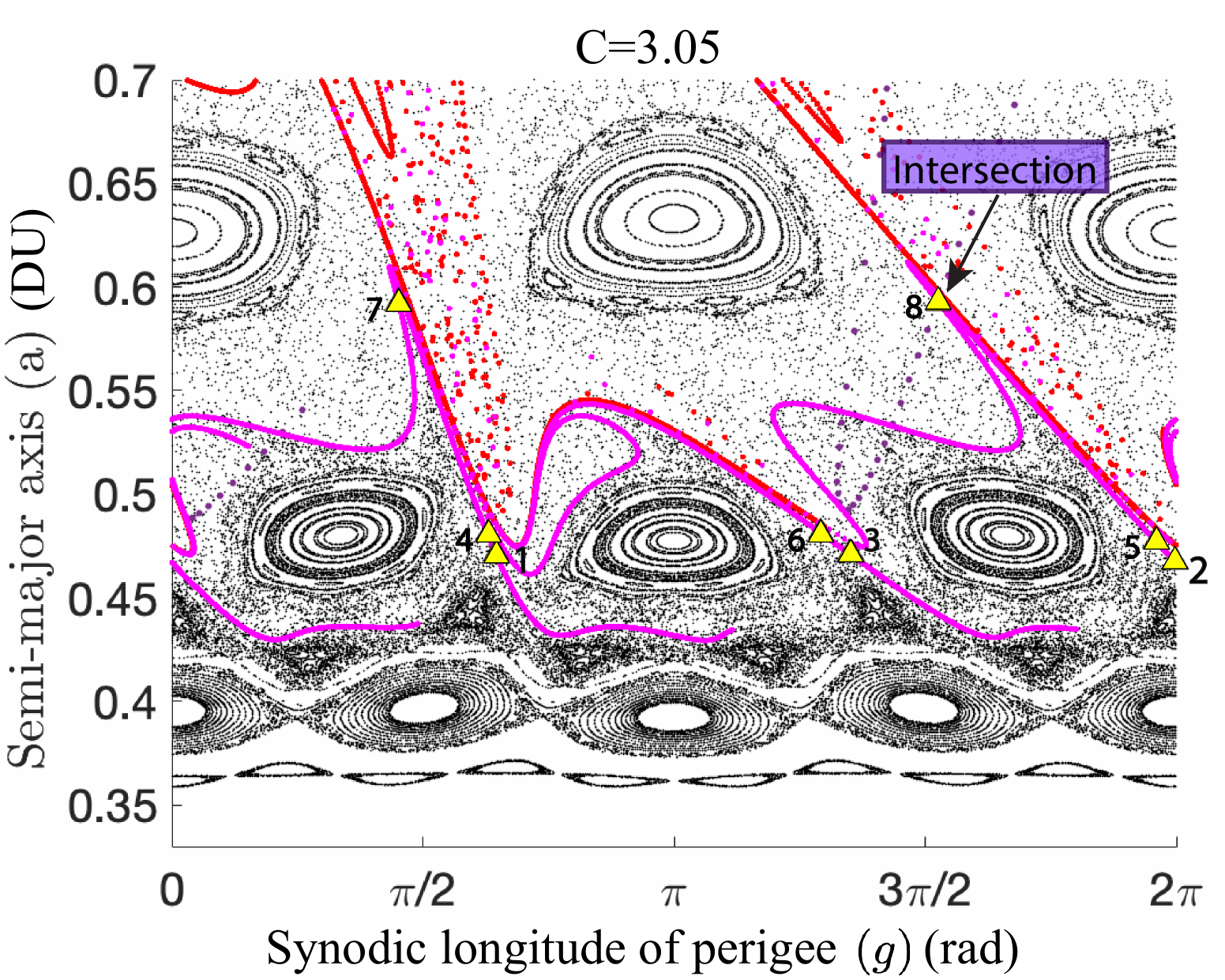}
    \caption{Shortest transfer from 3:1 unstable resonant periodic orbit to $L_1$ Lyapunov orbit, $C=3.05$.
    }
    \label{fig:31_L1_short}
\end{figure}

\begin{figure}[t!]
    \includegraphics[width=0.55\linewidth]{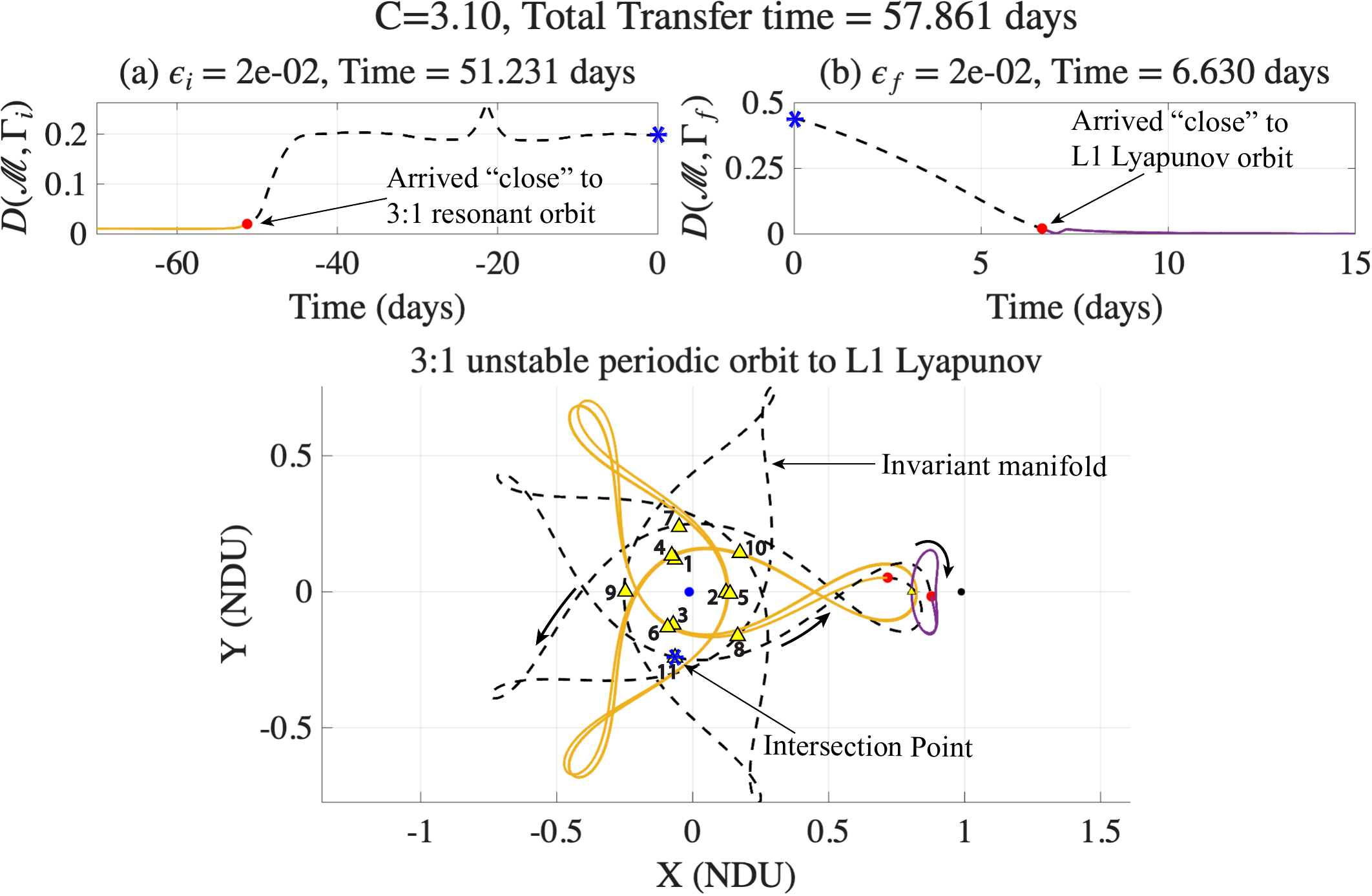}
    \includegraphics[width=0.45\linewidth]{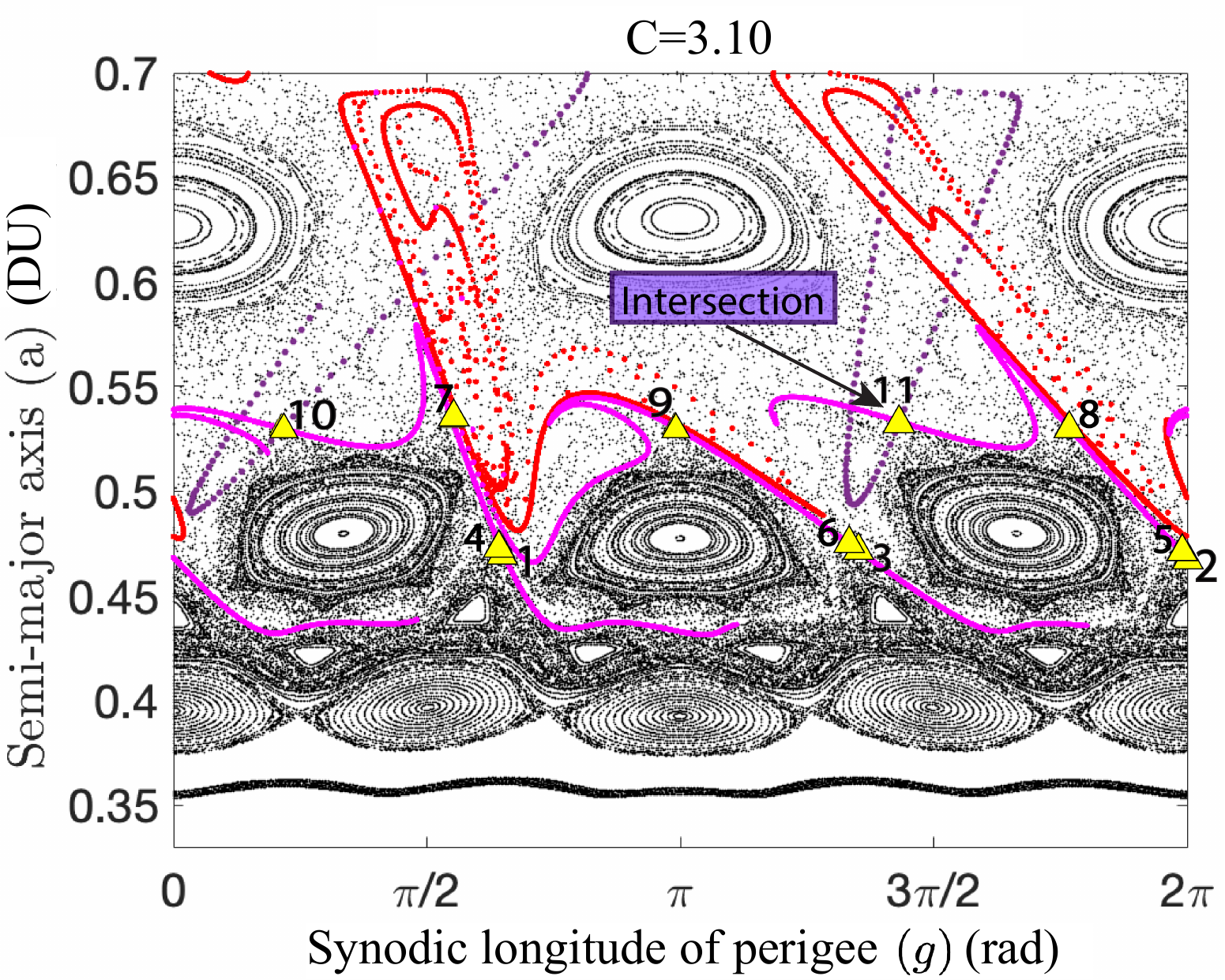}
    \caption{Longer transfer from 3:1 unstable resonant periodic orbit to $L_1$ Lyapunov orbit, $C=3.10$.
    }
    \label{fig:31_L1_long}
\end{figure}

Given that heteroclinic connections have been found between 3:1 and 2:1 resonant orbits as well as between 2:1 and $L_1$ planar Lyapunov orbits, mathematical ``shadowing'' lemmas \citep{fontichMartin2000, Arnold1967} imply that 3:1 to $L_1$ Lyapunov orbit heteroclinics must exist as well {(see Figure \ref{fig:cut})}. 
Indeed, in $(a,e)$ space, Figure \ref{fig:cut_ae} displays overlap between the $L_1$ Lyapunov orbit stable manifold cuts and the 3:1 chaotic (unstable) resonance zone over a wide range of Jacobi constants. Plotting 3:1 resonant orbit unstable manifolds and $L_1$ Lyapunov orbit stable manifolds on $\Sigma_C$ for $C=3.05$ and 3.10 in Figures \ref{fig:31_L1_short} and \ref{fig:31_L1_long}, respectively, intersections between these orbits' manifolds are immediately visible. 

Defining transfer times similarly to the 2:1 case using the tolerances of Table \ref{Tab:L1_transfer}, a short-TOF 3:1-to-$L_1$ transfer is found for $C=3.05$. Shown on the left of Figure \ref{fig:31_L1_short}, this heteroclinic trajectory takes approximately $31.35$ days, and occurs due to a 3:1-$L_1$ manifold intersection point---labeled as 8 in the right panel of the same figure---that exists very close to a 2:1 orbit point on $\Sigma_C$. In fact, it is clear from the trajectory plot that this transfer is mediated by the 2:1 resonance; initially, upon departure from the 3:1 orbit, the black dashed-line portion of the trajectory closely follows a type 1 3:1-to-2:1 transfer similar to that on the left of Figure \ref{fig:transfer_comb}. However, once the trajectory has closely approached the 2:1 periodic orbit---whose shape is clearly visible in much of the dashed-line trajectory---it then closely tracks a short 2:1-to-$L_1$ transfer similar to that of Figure \ref{fig:21_L1_short}. 

Recall from Section \ref{heteroclinicTrajsSection} that shorter-duration type 1 transfers between 3:1 and 2:1 orbits cease to exist for $C \geq 3.09$, with only longer type 2 transfers that traverse an intermediate 5:2 resonance remaining. Thus, for $C \geq 3.09$, changes in heteroclinics between 3:1 and $L_1$ Lyapunov orbits are to be expected as well. Figure \ref{fig:31_L1_long} shows such a heteroclinic trajectory for $C=3.10$ with a TOF of approximately 57.86 days---significantly longer than the 31.35 days taken for $C=3.05$. This transfer, corresponding to intersection point 11 from the right panel of Figure \ref{fig:31_L1_long}, has another clear difference from that of Figure \ref{fig:31_L1_short} as well: it no longer shadows a 2:1 orbit on its way to $L_1$. A close resemblance to the 5:2 resonant portion of 3:1-to-2:1 type 2 transfers (e.g. the right of Figure \ref{fig:transfer_comb}) is visible instead. However, rather than following the type 2 trajectory until 2:1, the transfer of Figure \ref{fig:31_L1_long} reaches $L_1$ directly from the 5:2 MMR, without the need for a 2:1 resonant segment. This change, from initially shadowing a short 3:1-to-2:1 type 1 transfer to following a 3:1-to-5:2 transfer before reaching $L_1$, explains the increased TOF required for $C \geq 3.09$.

In summary, short-duration transfers from 3:1 to $L_1$ Lyapunov orbits exist when $C\leq 3.09$, while longer-TOF transfers exist both for higher $C$ and for $C \leq 3.09$; note that one can thus choose either heteroclinic type for $C \leq 3.09$. 
As one would expect given the larger change in semi-major axis required, heteroclinic transfers from 3:1 to $L_1$ Lyapunov orbits take longer than those starting at 2:1 orbits. However, the TOF for a particular type of transfer (choice of heteroclinic intersection point) remains essentially the same when varied across Jacobi constants, with the key change once again occurring when type 1 3:1 to 2:1 transfers become available. The transfer times for both types of heteroclinics are summarized in Table \ref{Tab:L1_transfer}. 

\begin{table}
\centering
\begin{tabular}{ | m{5cm} | m{2.2cm} | m{2.2cm}| m{2.2cm} |  m{2.2cm} | } 
  \hline
  Transfer Type & Jacobi Constant & Transfer Time (days) & $\epsilon_{21/31}$ (NDU) & $\epsilon_{L1}$ (NDU)\\ 
  \hline
  \multirow{2}{4cm}{2:1 to $L_1$ Lyapunov (short)} & 3.05  & 8.112 & $5 \times 10^{-2}$ & $1 \times 10^{-2}$\\\cline{2-5}
  &3.10 & 8.225 & $5 \times 10^{-2}$ & $1.2 \times 10^{-2}$ \\ 
  \hline
   \multirow{2}{4cm}{2:1 to $L_1$ Lyapunov (long)} & 3.05 & 42.971 & $7 \times 10^{-3}$ & $5 \times 10^{-3}$ \\\cline{2-5}
  & 3.10 & 40.173 & $5 \times 10^{-3}$ & $5 \times 10^{-3}$ \\ 
  \hline
  3:1 to $L_1$ Lyapunov (short) & 3.05  & 31.353 & $5 \times 10^{-2}$ & $9 \times 10^{-3}$\\ 
  \hline
  3:1 to $L_1$ Lyapunov (long) & 3.10 & 57.861 & $2 \times 10^{-2}$ & $2 \times 10^{-2}$ \\ 
  \hline
\end{tabular}
\caption{Summary of transfer times-of-flight and tolerances for transfers from resonant orbits to $L_1$ Lyapunov orbits, $C=3.05, 3.10$.}
\label{Tab:L1_transfer}
\end{table}

\begin{remark*}

{Figures~\ref{fig:31_L1_short} and \ref{fig:31_L1_long} show an apparent multi-loop in the 3{:}1 unstable resonant orbit. This arises from our tolerance choice: we adopt the first dip of the distance metric (top-left panel) as the stopping criterion, indicating that the unstable manifold has come sufficiently close to the 3{:}1 resonant orbit. In the trajectory view, the red dot occurs near perigee~6; in the map view, this aligns closely with the semi-major axis of the 3{:}1 unstable resonant orbit (near perigees~1--3), and is thus considered operationally “close enough.” Choosing the \emph{second} dip (a tighter tolerance) would lessen the residual mismatch but would also extend the time of flight by approximately 30~days, which is unnecessary for the transfer under consideration.}
    
\end{remark*}

\section{Conclusions}

In this study, we demonstrated that mean-motion resonances play a major role in the dynamical structure of cislunar space, generating zero-$\Delta v$ heteroclinic pathways that traverse much of the region. After computing and characterizing families of unstable and stable 4:1, 3:1, and 2:1 resonant periodic orbits, in which the 2:1 family was found to have a very unique and interesting structure, we computed stable and unstable manifolds of the prograde unstable resonant orbits. By using a perigee Poincar\'e section 
represented using osculating orbital elements, we were able to effectively visualize the resulting manifolds in a manner that allowed us to identify and interpret the observed phenomena in line with Hamiltonian perturbation theory.  Heteroclinics between 2:1 and 3:1 unstable resonant orbits were found and precisely computed for Jacobi constants below 3.15, whereas it was also shown that the 4:1 MMR does not have any natural heteroclinics with the 3:1 MMR for any energy. Heteroclinics connecting 2:1 and 3:1 orbits to lunar $L_1$ planar Lyapunov orbits were also demonstrated, calculated, and characterized. By introducing a generalized distance metric, we further quantified effective times-of-flight for all heteroclinic transfers, finding them to be reasonable for real missions. Collectively, these results establish
the 3:1 mean-motion resonance as truly being a ``gateway to the Moon'' for lunar mission design. 

We also found that the resonant orbit and heteroclinic structures of the Earth–Moon PCR3BP can differ markedly from those of previously studied systems with much smaller $\mu$ values (e.g., outer planet–moon or star–planet CR3BPs), which tend to follow perturbation-theory dynamics more closely.
Thus, a CR3BP-based analysis of the Earth--Moon system is shown to be crucial to understanding the resonant dynamics of cislunar space. 
Our recent work \citep{rawat2025preprint}, which leverages the manifolds computed here, further shows that CR3BP-based analysis is essential: comparisons with a perturbed Keplerian semi-analytical approach reveal that the latter severely underestimates the regions of influence and effects of the 2:1 and 3:1 resonances.


While this study and \citet{rawat2025preprint} focused only on the planar Earth–Moon CR3BP, many spacecraft in practice follow orbits inclined relative to the Earth--Moon orbital plane. This is the case for TESS and IBEX, which occupy inclined stable 2:1 and 3:1 resonant orbits. Extending the present work to the spatial CR3BP therefore represents the most significant future direction for this research. Such an extension will require new tools to address the increased dimensionality of the system, since even after reduction to a Poincar\'e map on an energy submanifold, the spatial CR3BP remains a 4D symplectic map. Beyond this spatial generalization, it will also be of interest to examine other, less prominent MMRs such as 5:2 or 7:2, as well as resonances exterior to the Moon; some initial results on exterior MMRs are reported in \citet{rawatKumar2025}. Finally, the CR3BP does not model the  effects of lunar eccentricity and solar gravity, which can be significant; {while there exist theoretical reasons---e.g., KAM \citep{kamTutorial} and Fenichel \citep{fenichel1971persistence} theory---to expect that similar resonant orbits and heteroclinics will persist in models including such perturbations,} studies of MMR dynamics in higher-fidelity models remain a key avenue for future investigation. Considerable work remains to fully characterize the network of propellant-free heteroclinic highways that future missions may exploit throughout cislunar space.



\section*{Acknowledgments}

B. Kumar was supported in part by the National Science Foundation (NSF) under a Mathematical Sciences Postdoctoral Research Fellowship award no. DMS-2202994, and in part by the US Air Force Office of Scientific Research (AFOSR) under Award No. FA8655-24-1-7012. This research was carried out in part at the Jet Propulsion Laboratory, California Institute of Technology, under a contract with the National Aeronautics and Space Administration (80NM0018D0004). A. Rawat, A.J. Rosengren and S.D. Ross acknowledge support from the Air Force Office of Scientific Research (AFOSR) under Grant No. FA9550-24-1-0194.

\bibliographystyle{elsarticle-harv}
\bibliography{references_combined}

\section{Appendix}
\subsection{Alternative Distance Metrics}
\label{Alt_dist}
The general distance metric can be defined with any quantity that uniquely defines an orbit or trajectory. Osculating equinoctial elements $(p,L,f,g,h,k)$ uniquely define an orbit as
\begin{equation}
\begin{array}{rcl@{\quad}rcl}
  p &=& a(1-e^2),               & L &=& \Omega + \omega + \theta, \\[0.5em]
  f &=& e\cos(\omega+\Omega),   & h &=& \tan{\tfrac{i}{2}}\cos\Omega, \\[0.5em]
  g &=& e\sin(\omega+\Omega),   & k &=& \tan{\tfrac{i}{2}}\sin\Omega.
\end{array}
\end{equation}
The point-to-point distance metric can thus be defined similarly to Eq. \ref{eq:D(a,b)} using equinoctial elements as,
\begin{equation}
\label{eq:equinoctial}
    D(o_a,o_b) = \sqrt{||p_{o_a} - p_{o_b}||^2 + ||f_{o_a} - f_{o_b}||^2 + ||g_{o_a} - g_{o_b}||^2 + ||h_{o_a} - h_{o_b}||^2 + ||k_{o_a} - k_{o_b}||^2},
\end{equation}
with this metric being used in Eq.~\ref{eq:point_set_distance} to define the point-to-set (state-to-orbit) distance. 
The disadvantage of using equinoctial elements is that these are scalar quantities, and when the spatial problem is considered, $h$ and $k$ become undefined for inclination equal to $\pi$. Nevertheless, using the metric of Eq.~\eqref{eq:equinoctial}, the various heteroclinic transfer times remain nearly the same as those from Sections~\ref{tofSection} and \ref{resToL1Section}. Tolerances need to be adjusted accordingly based on the minimization of oscillation of the distance metric.

\begin{table}[hbt!]
\centering
\begin{tabular}{ | m{4em} | m{3cm}| m{2cm} |  m{2cm} | } 
  \hline
  Jacobi Constant & Transfer Time (days) & $\epsilon_{31}$ (NDU) & $\epsilon_{21}$ (NDU)\\ 
  \hline
  3.00  & 30.549 & $2 \times 10^{-2}$ & $6 \times 10^{-4}$\\ 
  \hline
  3.05 & 31.466 & $2 \times 10^{-2}$ & $5 \times 10^{-4}$ \\ 
  \hline
   3.10 & 57.533 & $2 \times 10^{-2}$ & $1.5 \times 10^{-2}$ \\ 
  \hline
   3.15 & 59.460 & $2 \times 10^{-2}$ & $1.5 \times 10^{-2}$ \\ 
  \hline
\end{tabular}
\caption{Summary of 3:1 to 2:1 transfer times and tolerances for $C=3.00$ - $3.15$ using equinoctial element-based metric.}
\label{Tab:31_21_transfer_equi}
\end{table}

Another choice could be to use the synodic Cartesian state vector, as it also uniquely defines the osculating orbit. The point-to-point distance metric is then simply
\begin{equation}
\label{eq:state}
    D(o_a,o_b) = \sqrt{||x_{o_a} - x_{o_b}||^2 + ||y_{o_a} - y_{o_b}||^2 + ||{p}_{x,o_a} - {p}_{x,o_b}||^2 + ||{p}_{y,o_a} - {p}_{y,o_b}||^2}.
\end{equation}
The disadvantage of using this state vector is that it changes very rapidly when compared to equinoctial elements or angular momentum and Laplace vector. As a result, the associated point-to-set (state-to-orbit) distance metric exhibits pronounced persistent oscillations before stabilizing within a prescribed tolerance, thereby extending the effective time-of-flight.

\begin{table}[htb!]
\centering
\begin{tabular}{ | m{5cm} | m{2.2cm} | m{2.2cm}| m{2.2cm} |  m{2.2cm} | } 
  \hline
  Transfer Type & Jacobi Constant & Transfer Time (days) & $\epsilon_{21/31}$ (NDU) & $\epsilon_{L1}$ (NDU)\\ 
  \hline
  \multirow{2}{4cm}{2:1 to $L_1$ Lyapunov (short)} & 3.05  & 8.851 & $2 \times 10^{-3}$ & $8 \times 10^{-3}$\\\cline{2-5}
  &3.10 & 8.225 & $1 \times 10^{-3}$ & $1.2 \times 10^{-3}$ \\ 
  \hline
   \multirow{2}{4cm}{2:1 to $L_1$ Lyapunov (long)} & 3.05 & 36.975 & $7 \times 10^{-3}$ & $2 \times 10^{-2}$ \\\cline{2-5}
  & 3.10 & 35.533 & $5 \times 10^{-3}$ & $2 \times 10^{-2}$ \\ 
  \hline
  3:1 to $L_1$ Lyapunov (short) & 3.05  & 31.766 & $3.5 \times 10^{-2}$ & $2 \times 10^{-4}$\\ 
  \hline
  3:1 to $L_1$ Lyapunov (long) & 3.10 & 57.874 & $2 \times 10^{-2}$ & $2 \times 10^{-4}$ \\ 
  \hline
\end{tabular}
\caption{Summary of transfer times and tolerances for $C=3.05,3.10$ for transfers from resonant orbits to $L_1$ Lyapunov orbit using equinoctial element-based metric.}
\label{Tab:L1_transfer_equi}
\end{table}

\subsection{Initial Conditions for Heteroclinic Transfers and Resonant orbits}
\label{IC}

{Cartesian coordinates for stable/unstable manifold points corresponding to heteroclinic transfers are provided in Table \ref{Tab:IC_heteroclinic}. In the case of 3:1 to 2:1 transfers, intersection points between unstable and stable manifolds were computed using the highly-accurate method of Section \ref{heteroComputeSection}; the resulting intersections are denoted by $\mathbf{x}$ in the table. Figures \ref{fig:transfer_comb}, \ref{fig:3121heteroclinic}, \& \ref{fig:TOF} and  Tables \ref{Tab:31_21_transfer} \& \ref{Tab:31_21_transfer_equi} can be reproduced by propagating $\mathbf{x}$ forwards and backwards.}

{For heteroclinics to Lyapunov orbits, the manifold intersections were not computed as precisely, as the methods of Section \ref{heteroComputeSection} rely on the assumption of a non-circular topology for manifold intersections with the Poincar\'e section $\Sigma$---a condition that does not hold for the stable and unstable manifolds of Lyapunov orbits. Thus, we approximate the intersection point by identifying the pair of computed stable and unstable manifold points, $\mathbf{x_+}$ and $\mathbf{x_-}$ respectively, that are closest together in the $(g,a)$ plane. These points provide a reasonable estimate for the actual manifold intersection. Figures \ref{fig:21_L1_short}, \ref{fig:21_L1_long}, \ref{fig:31_L1_short}, and \ref{fig:31_L1_long}, as well as Tables \ref{Tab:L1_transfer} and \ref{Tab:L1_transfer_equi}, can be reproduced by propagating $\mathbf{x_+}$ forward and $\mathbf{x_-}$ backward. }

\begin{table}[htb!]
\centering
\begin{tabular}{ | m{1.9cm}  | m{1.2cm} | m{1cm} | m{2.8cm}| m{2.8cm} |  m{2.8cm} | m{2.8cm} | } 
  \hline
  Transfer & $C$ & Point type & $x$ & $y$ & $\dot x$ & $\dot y$\\ 
  \hline
     \multirow{6}{4cm}{\textbf{3:1 to 2:1 \\ (Type 1)}} & 2.54 & \textbf{x}   & -0.02495763718 & -0.02013755597 & 7.56003664579 & -4.80802086624 \\\cline{2-7}
     & 2.70 & $\mathbf{x}$  & -0.03022725673 & 0.042561712914 & -5.82508502316 & -2.47401118974  \\\cline{2-7} 
      & 2.86 & $\mathbf{x}$  & 0.06136961660 & -0.00485126788 & 0.322456526264 & 4.886777873750  \\\cline{2-7} 
        & 3.00 & $\mathbf{x}$ & -0.07187159167 & -0.09858383084 & 3.22059393928 & -1.95100061370  \\\cline{2-7} 
        & 3.05 & $\mathbf{x}$ & 0.029351172720 & 0.320801654785 & -1.77006016626 & 0.228990735501  \\
  \hline
     \multirow{2}{4cm}{\textbf{3:1 to 2:1 \\ (Type 2)}} & 3.10 & $\mathbf{x}$ & 0.173726091985 & 0.150836454723 & -1.44182292916 & 1.77676712380  \\\cline{2-7} 
          & 3.15 & $\mathbf{x}$ & 0.203387734653 & 0.165418603215 & -1.25080421783 & 1.62978185751  \\

  \hline
        \multirow{4}{4cm}{\textbf{2:1 to $L_1$ \\ (short)}} & \multirow{2}{4cm}{3.05} & $\mathbf{x_+}$ & 0.0366500307265 & -0.34165173319 & 1.66146760213 & 0.237319550248  \\\cline{3-7} 
        &  & $\mathbf{x_-}$ & 0.0202388267797 & -0.32661071935 & 1.7522136554 & 0.173763948854  \\\cline{2-7}
          & \multirow{2}{4cm}{3.10} & $\mathbf{x_+}$ & 0.0255766502798 & -0.37136396824 & 1.52674168622 & 0.155103195482  \\\cline{3-7} 
          &  & $\mathbf{x_-}$ & 0.0252310397037 & -0.37105702307 &  1.5283544673 & 0.153971946205  \\
 \hline
        \multirow{4}{4cm}{\textbf{2:1 to $L_1$ \\ (long)}} & \multirow{2}{4cm}{3.05} & $\mathbf{x_+}$ & -0.14132039003 & -0.38970297519 & 1.32702940649 & -0.43985328225  \\\cline{3-7} 
        &  & $\mathbf{x_-}$ & -0.14191807532 & -0.37724673884 & 1.36549493788 & -0.46971075927  \\\cline{2-7}
          & \multirow{2}{4cm}{3.10} & $\mathbf{x_+}$ & -0.16039196073 & -0.43439108894 & 1.13516561771 & -0.38738942941  \\\cline{3-7} 
          &  & $\mathbf{x_-}$ & -0.16459697997 & -0.43399693818 & 1.12834267293 & -0.39634328833  \\

\hline
     \multirow{2}{4cm}{\textbf{3:1 to $L_1$ \\ (short)}} & \multirow{2}{4cm}{3.05} & $\mathbf{x_+}$ & 0.0123138303884 & -0.31908512499 & 1.79730258732 & 0.137800064457  \\\cline{3-7} 
        &  & $\mathbf{x_-}$ & 0.0141669888515 & -0.32035243046 & 1.78927012305 & 0.146992008865  \\\cline{3-7}
\hline
     \multirow{2}{4cm}{\textbf{3:1 to $L_1$ \\ (long)}} & \multirow{2}{4cm}{3.10} & $\mathbf{x_+}$ & -0.06441165789 & -0.24077029190 & 2.18597985368 & -0.47448403335  \\\cline{3-7} 
        &  & $\mathbf{x_-}$ & -0.06447093170 & -0.24079902449 & 2.18559541702 & -0.47488195526  \\\cline{3-7} 
\hline
  
\end{tabular}
\caption{Initial conditions for heteroclinic transfers at various Jacobi constant ($C$) values. Points labeled $\mathbf{x}$ are accurately-computed manifold intersections. Points labeled $\mathbf{x_+}$ and $\mathbf{x_-}$ correspond respectively to the stable \& unstable manifold points which are closest to each other, and can be taken as reasonable approximations for the true intersection point $\mathbf{x}$.}
\label{Tab:IC_heteroclinic}
\end{table}

{Finally, initial conditions for the resonant periodic orbits shown in Figures \ref{fig:41orbit}, \ref{fig:31orbit}, \ref{fig:highe21}, and \ref{fig:41orbitHighE} are also provided in Table \ref{Tab:IC_orbit}. These were also used to help define and compute the times-of-flight for various heteroclinic transfers.}

\begin{table}[htb!]
\centering
\begin{tabular}{ | m{3cm}  | m{2cm} | m{3.5cm}| m{3.5cm} | } 
  \hline
  Resonant orbit (unstable) & Jacobi Constant  & $x$ & $\dot y$\\ 
  \hline
     \multirow{2}{4cm}{\textbf{4:1}} & 2.85  & 0.770658112628  & -0.616166626758 \\\cline{2-4}
     & 3.15  & 0.737385941470  &  -0.355888785284 \\\cline{2-4} 

    \hline
     \multirow{6}{4cm}{\textbf{3:1}} 
     & 2.54  & 0.901330167142  & -0.846224994375\\\cline{2-4}
     & 2.70  & 0.888102851739  & -0.725919236562  \\\cline{2-4}   
      & 2.86  & 0.868787031779 & -0.584481100346  \\\cline{2-4} 
      & 3.00  & 0.845409258968  & -0.434953051104  \\\cline{2-4} 
        & 3.05 & 0.834875967543  & -0.37200499129  \\\cline{2-4} 
         & 3.10 & 0.822429022871 & -0.300987128481  \\\cline{2-4} 
          & 3.15 & 0.806804382814  & -0.218228896847  \\\cline{2-4} 
  \hline
         \multirow{7}{4cm}{\textbf{2:1}} 
         & 2.54  & 0.964310487216 & -1.20233213851 \\\cline{2-4}
         & 2.628  & 0.959722043439  & -1.090831532610 \\\cline{2-4}   
     & 2.70  & 0.954308268237 & -0.989690975727  \\\cline{2-4}   
      & 2.86  & 0.934036118105 & -0.743014738588  \\\cline{2-4} 
      & 3.00  & 0.904862874177 & -0.515850946602  \\\cline{2-4} 
        & 3.05 & 0.892104772476 & -0.429597033392  \\\cline{2-4} 
         & 3.10 & 0.878280334961  & -0.334629543419  \\\cline{2-4} 
          & 3.15 & 0.864401165205  & -0.219058827351  \\\cline{2-4} 
\hline
  
\end{tabular}
\caption{Initial conditions for resonant periodic orbits (all with $y=0$, $ \dot x=0$.)}
\label{Tab:IC_orbit}
\end{table}

\end{document}